\documentclass[12pt,preprint]{aastex}

\begin{document}

\title{The Velocity Dispersion Profile of the Remote Dwarf Spheroidal Galaxy Leo~I: A Tidal Hit and Run?}

\author{Mario Mateo\altaffilmark{1}, Edward W. Olszewski\altaffilmark{2}, and Matthew G. Walker\altaffilmark{3}}
\altaffiltext{1}{Department of Astronomy, University of Michigan, 830 Dennison Building, Ann Arbor, MI 48109-1042; {\tt mmateo@umich.edu}}
\altaffiltext{2}{Steward Observatory, The University of Arizona, Tucson, AZ 85721; {\tt edo@as.arizona.edu}}
\altaffiltext{3}{Department of Astronomy, University of Michigan, 830 Dennison Building, Ann Arbor, MI 48109-1042; {\tt mgwalker@umich.edu}}

\begin{abstract}

We present new kinematic results for a sample of 387 stars located in
and around the Milky Way satellite dwarf spheroidal galaxy Leo~I.
These spectra were obtained with the Hectochelle multi-object echelle
spectrograph on the MMT, and cover the MgI/Mgb lines at about 5200\AA.
Based on 297 repeat measurements of 108 stars, we estimate the mean
velocity error ($1\sigma$) of our sample to be 2.4 km/s, with a
systematic error of $\leq 1$ km/s.  Combined with earlier results, we
identify a final sample of 328 Leo~I red giant members, from which we
measure a mean heliocentric radial velocity of $282.9 \pm 0.5$ km/s,
and a mean radial velocity dispersion of $9.2 \pm 0.4$ km/s for Leo~I.
The dispersion profile of Leo~I is essentially flat from the center of
the galaxy to beyond its classical `tidal' radius, a result that is
unaffected by contamination from field stars or binaries within our
kinematic sample.  We have fit the profile to a variety of equilibrium
dynamical models and can strongly rule out models where mass follows
light. Two-component Sersic+NFW models with tangentially anisotropic
velocity distributions fit the dispersion profile well, with isotropic
models ruled out at a 95\%\ confidence level.  Within the projected
radius sampled by our data ($\sim$ 1040 pc), the mass and V-band
mass-to-light ratio of Leo~I estimated from equilibrium models are in
the ranges 5-7 $\times 10^7 M_\odot$ and 9-14 (solar units),
respectively.  We find that Leo~I members located outside a `break
radius' at $R_b \sim 400$ arcsec (500 pc) exhibit significant velocity
anisotropy, whereas stars interior of this radius appear consistent
with an isotropic velocity distribution.  We propose a heuristic model
in which the break radius represents the location of the tidal radius
of Leo~I at perigalacticon of a highly elliptical orbit.  Our scenario
can account for the complex star formation history of Leo~I, the
presence of population segregation within the galaxy, and Leo~I's
large outward velocity from the Milky Way.  Within the framework of
our model, the lack of extended tidal arms in Leo~I -- both
perpendicular to and along the line of sight -- suggests the galaxy
has experienced only one perigalactic passage with the Milky Way;
thus, Leo~I may have been injected into its present orbit by a third
body a few Gyr before perigalacticon.  We discuss the plausibility of
this idea within the context of hierarchical models and conclude that
such an interaction is entirely possible.  We also report the possible
detection of a distinct kinematic structure in the Leo~I field at
about a $2\sigma$ significance level.

\end{abstract}


\section{Introduction}

An understanding of how external and internal dynamical effects
combine to produce the observed kinematic properties of dwarf galaxies
is fundamental to probing the nature of these systems.  Since dwarfs
may represent the smallest dark-matter (DM) halos that have survived
to the present epoch -- or simply the smallest that have retained
baryons (Ferrara and Tolstoy 2000; Benson et al. 2002; Grebel and
Gallagher 2004; Susa and Umemura 2004) -- they represent an important
link to broader structure formation models. As we learn more about the
dynamical state of the Galactic halo, however, is has become clear
that complex interactions of the components within the halo --
including local dwarfs -- may be common (Taylor and Babul 2004, 2005;
Coleman et al. 2004).  Detailed $n$-body models hint at the frequency
of interactions (Moore et al. 1999; Kravtsov et al. 2004), while
hybrid $n$-body+hydrodynamical models reveal the richness of the
baryonic phenomena -- episodic star formation, gas streaming -- that
may result (Mayer et al. 2001a,b, 2005, 2007; Kravtsov et al. 2004).
Observational evidence in the halo of the Milky Way (Sagittarius:
Ibata et al. 1994, Majewski et al. 2003, Belokurov et al. 2006; The
Magellanic Stream: Putman et al. 1998, Connors et al. 2006; The Fornax
dwarf spheroidal galaxy: Coleman and Da Costa 2004, 2005, Olszewski et
al. 2006; The halo of M31: Ibata et al. 2001; Brown et al. 2006)
broadly confirms model expectations that tides/encounters strongly
alter the structural features of satellite dwarf systems.

To describe the halo and its interaction history fully, we need to
know the nature of the DM subhalos.  Nearby dwarfs and their DM halos
represent the best local proxies for the original DM seeds that built
up the Milky Way (Read et al. 2006a).  But to probe the satellites and
study their DM halos in detail, we need to account for the effects of
interactions, since these can strongly alter our conclusions regarding
the properties of the halos (Klessen and Kroupa 1998; Fleck and Kuhn
2003; Metz et al. 2007).  Moreover, understanding the tidal history of
dwarfs depends on the knowledge of the total mass and structure of the
DM halo of the Milky Way in which these interactions occur.  The
problem is deliciously interconnected and devilishly hard to solve,
particularly in the absence of a secure description of the small halos
comprising the Milky Way's extended halo.  We have embarked on an
effort to address part of this riddle by systematically studying the
kinematic properties of many of satellites of the Milky Way (Walker et
al. 2007a,b).  The present paper focuses on the Leo~I dwarf spheroidal
(dSph) galaxy.

Two general cases arise when we consider the dynamical state of a
galaxy such as Leo~I\footnote{We restrict ourselves here to cases in
which Newtonian gravity applies.  We will address the interpretation
of dSph kinematics in cases assuming non-Newtonian gravity in a later
paper.}.  First, a galaxy can be considered to be in dynamical
equilibrium if it is sufficiently isolated from outside
perturbations. This case is well-understood theoretically though it
does admit to various degeneracies (e.g. anisotropy and the mass
distribution) and other complications (deprojections of possibly
non-spherical mass or tracer distributions).  Nonetheless, one can,
with enough data, close in on the mass distribution from the
kinematics and distribution of baryonic tracers.  However, given the
hierarchical nature of galaxy formation noted above, the second, more
realistic expectation is that galaxies, particularly dwarf satellites,
can never be in true dynamical equilibrium.  The question then becomes
one of degree: How far out of equilibrium is a given galaxy and how do
deviations from equilibrium affect our interpretation of its
kinematics?

The motions of stars in a satellite galaxy are governed by its
underlying mass distribution and the external potential in which the
galaxy orbits.  To complicate matters, the external potential may be
asymmetric, or significantly time-dependent, particularly for orbits
with periods that span a significant fraction of the age over which
the parent galaxy formed.  But even if the external potential is
highly symmetric, a dwarf galaxy in a non-circular orbit will
necessarily experience a time-variable potential as it traces its
orbital path (Pinchardo et al. 2005).  Clearly, many factors dictate
the tidal history of any given galaxy, and, by implication, the degree
to which a given system's kinematics are affected by tidal influences.

We know that in extreme cases these processes can have a
transformative effect, converting dwarf galaxies into streams that
encircle their parent galaxies (Ibata et al. 1994, 2001; Majewski et
al. 2003).  Whether less severe effects have been seen in other
systems remains a matter of debate (e.g., Ursa Minor:
Mart\'inez-Delgado et al. 2001, Palma et al. 2003, Mu\~noz et
al. 2005; Carina: Kuhn et al. 1996, Monelli et al. 2004;
Majewski et al. 2005).  As a population, local dwarfs exhibit trends,
such as decreasing mass-to-light (M/L) ratio with increasing
Galactocentric distance ($R_G$) for Milky Way and M31 satellites
(Mateo et al. 1998a; hereafter M98) and correlations of stellar ages
in dwarf systems as a function of $R_G$ (van den Bergh 1994; Mayer et
al. 2007), that provide circumstantial evidence tides do influence
satellite properties to varying degrees.

Since our goal is to characterize the DM halos dwarf galaxies, it is
clearly advantageous to try to identify galaxies that best exemplify
systems in dynamical equilibrium.  It is for this reason we (and
others: M98; Sohn et al. 2007; Koch et al. 2007) have chosen to focus
on the Leo~I dSph galaxy.  Located 255 kpc from the Milky Way (Caputo
et al. 1999; Held et al. 2001; Bellazzini et al. 2004), Leo~I seems to
be a good candidate as a truly isolated dwarf that is close enough for
detailed kinematic study.  This reasoning motivated M98 to obtain
radial velocities for 33 red giants in Leo~I from which they derived a
central velocity dispersion.  Under the assumptions of dynamical
equilibrium and mass follows light, they argued that Leo~I does
contain a DM halo of comparable mass to those inferred in other,
closer dSph galaxies.  This result implied that tides are not the sole
cause of the observed kinematic properties of dSph galaxies.

However, Leo~I has two puzzling properties that suggest it may not be
the ideal, dynamically-isolated system we would hope it to be.  First,
the heliocentric radial velocity, $v_h$ of Leo~I is extreme (282.9
km/s; see Section 4.3.1).  Relative to the Galactic Center rest frame,
it has a recessional velocity of 174.9 km/s as observed from the Sun,
very high for a remote satellite of the Milky Way.  Indeed, Leo~I can
singlehandedly inflate the inferred mass of our Galaxy by nearly a
factor of five compared to dynamical analyses of kinematic samples
that exclude it (Zaritsky et al.  1989; Fich and Tremaine 1991;
Kochanek 1996; Wilkinson and Evans 1999; Sakamoto et al.  2003).  But
the large velocity also has an important implication for Leo~I itself.
Standard cold DM (CDM) hierarchical formation models of the Milky Way
tend to produce outward-streaming dwarfs at late times as the smaller
galaxies begin to interact strongly with the dominant, massive central
halo (Taylor et al. 2005) or other dark halos (Taylor and Babul 2004,
2005; Sales et al. 2007a,b).  Is the high systemic velocity of Leo~I
direct evidence of a strong past interaction with the Milky Way or,
possibly, some third body?

The second enigma is that Leo~I has a complex star-formation (SF)
history (Gallart et al. 1999b; Hernandez et al. 2000; Dolphin 2002),
with the latest significant episode ending about 1 Gyr ago.  Within
the framework of the tidal-stirring models of Mayer et al. (2001a,b,
2005), periods of enhanced star formation represent times when the
gas in a dwarf is compressed toward the center of its DM halo during
strong interactions with the parent halo.  Does the complex star
formation history of Leo~I reveal evidence of such an event --
presumably an interaction with the Milky Way -- about 1 Gyr ago?

This paper presents new high-precision kinematic observations of over
300 individual member red giants in the Leo~I dSph galaxy,
significantly extending both the quantity, quality, and spatial
coverage of earlier samples.  With these new data it is possible to
address whether Leo~I represents an example of an isolated,
equilibrium system, or if its dynamics are significantly influenced by
external tidal effects.

\section{Observations and Data Reduction}

The principal data of this paper consist of new radial velocity
measurements derived from spectra obtained using the multi-fiber
Hectochelle spectrograph on the MMT telescope (Szentgyorgyi et
al. 1998; Szentgyorgyi 2006).  The observations were carried out
during two nights dedicated to this project (March 31/April 1 and
April 2/3, 2005), and during parts of six separate nights of two
Hectochelle `queue runs' in April, 2006 and March/April 2007 (see
Table 1 for details).  Seven different fiber configurations were used,
denoted chronologically as `c1' through `c7'.  We purposely allowed
for considerable target overlap in the various configurations to allow
us to use repeat measurements to quantify the velocity uncertainties
(Section 3.2).  The spatial distribution of the targets is shown in
Figure~\ref{figs:xieta}, along with all of the Leo~I red giant (RG)
candidates.

Candidates stars to be observed with Hectochelle were selected from
the ($I$, $V$--$I$) color-magnitude diagram (CMD) of Leo~I derived
from CCD data obtained with the Hiltner 2.4m Telescope at MDM
Observatory in February, 2004, and for the outermost candidates, from
data we obtained with the 90Prime camera on the Steward 2.3m telescope
at Kitt Peak in Feb 2006 (Williams et al. 2004).  The locations of the
all candidate stars in the CMD are shown in Figure~\ref{figs:cmd}.
The photometric calibration was carried out using 22 stars in the
field that have calibrated photometry from M98.  The rms scatter of
the adopted transformations is about 0.02 mag in both $I$ and $(V-I)$.
The astrometry for these stars was determined by transforming of the
CCD coordinates to standard coordinates, $(\xi,\eta)$ using 200-400
USNOB/NOMAD (Monet et al. 2003) coordinates of stars in the field.

We chose regions devoid of detectable stars in our deep imaging to be
used for spectroscopic 'sky' measurements.  These sky fiber locations
span the full radial extent of the target fibers relative to the
center of Leo~I.  Additional 'sky' fibers were also randomly assigned
by the Hectochelle fiber robot software at the time of observation to
utilize otherwise idle fibers.  Since these could, by chance, land on
an astronomical source, no automatically-assigned sky fibers were used
in our analysis.  We also observed the radial velocity standard
HD171232, as standards in the SA57 field (Stefanik et al.  2006).  The
Hectochelle fiber robot software generally assigned different fibers
to these stars on separate visits.  Further details of our Leo~I and
standard-star observations are provided in Table~1.

Hectochelle employs two $2048 \times 4096$ CCDs, illuminated by 244
fibers and read out through 4 amplifiers (2 on each CCD). Fibers are
assigned to targets to avoid mechanical overlaps and collisions, while
maximizing user-specified priorities, and can be positioned anywhere
within a 1-degree diameter field.  The fibers terminate at the focal
plane of the collimator of an echelle spectrograph located on an
optical bench near the telescope.  To increase the packing factor on
the CCD, the fibers are mounted in a zigzag pattern at the slit plane.
Thus, consecutive spectra are offset in both the spatial and
dispersion directions from one another.  The entire pattern also
exhibits slight curvature along the `spatial' axis due to anamorphic
effects in the spectrograph optics.  To avoid order overlap, an
interference filter was used to isolate the wavelength region of
interest.  We observed our targets through the `RV52' filter which
covers the region of the Mg~I/Mgb features around 5200\AA. Further
technical information on Hectochelle is available at (Szentgyorgyi
2006) and from the user's manual\footnote{{\tt
http://cfa-www.harvard.edu/mmti/hectospec/Hectochelle\_Observers\_Manual.pdf}}.

IRAF\footnote{IRAF was distributed by the National Optical Astronomy
Observatories, which are operated by the Association of Universities
for Research in Astronomy, Inc., under cooperative agreement with the
National Science Foundation.}  routines were used to process the raw
images, extract and wavelength-calibrate the individual spectra, and
to measure the final heliocentric velocities.  The procedures are
broadly similar to ones we have employed in the past for single-object
slit spectroscopy (e.g. Vogt et al. 1998) and for multi-fiber
spectroscopy of Southern dwarf galaxies (Walker et al. 2007a).  We
briefly summarize them here.

All data were processed by first subtracting the overscan, trimming
the images, then combining the data from the two amplifiers for both
CCDs.  Individual exposures were combined to form single deep images
and to remove the brightest point-like cosmic rays in the raw data
using a conservative sigma-clipping algorithm.  Hectochelle had a
source of significant scattered light within the spectrograph chamber
during the 2005 run (configurations c1 and c2 in Table 1).  This
problem made it impossible to flatfield these data or to perform
background subtraction.  We estimate the scattered light cost us about
one magnitude in depth compared to the 2006/2007 observations, all of
which were unaffected by this problem.

For the 2006 and 2007 observations (configurations c3 through c7 in
Table 1), we determined relative fiber throughputs from observations
of the twilight sky.  Quartz lamp spectra were inadequate for this as
the light sources on the telescope do not illuminate the fibers
uniformly.  After extraction, wavelength calibration (see below) and
dividing by the relative throughputs, we then combined all the sky
fiber spectra to make a master sky for each configuration.  This
combined sky spectrum was subtracted this from the spectrum of each
target and each individual sky spectrum.  The master sky spectrum was
typically produced by averaging 40-60 individual spectra, so it adds
little variance to the final sky-subtracted spectra.  Overall, this
procedure appears to have worked well: Individual sky fibers had
little mean residual flux after subtraction (typically $\leq 5$\%\ of
the original sky flux).  In contrast to the 2005 data,
cross-correlations of both target and sky fibers rarely produced a
signal at the expected velocity of scattered sunlight at the time of
observations (see Figure~\ref{figs:ccf1}).

We established the locations of the individual spectra on the
detectors by tracing and extracting spectra of quartz lamps obtained
just before or after each set of observations for a given
configuration.  The spectral traces from the quartz images were then
shifted spatially {\it en masse} to best match the fainter spectra of
the corresponding target exposures.  The shifts of the quartz spectra
to the target spectra were stable to about 0.01 pixel, as determined
by comparing results for individual target exposures of a given
configuration.  ThAr calibration emission spectra were extracted using
the same traces used for the corresponding target spectra.  A
dispersion solution for each arc spectrum was determined by fitting
the centroids of 35-40 lines with known wavelengths to a fifth-order
polynomial.  The fits had a typical RMS scatter of about 0.3-0.5 km/s.
The resulting dispersion solutions were then applied to all spectra,
producing wavelength-calibrated, one-dimensional spectra for every
fiber.  The spectra are defined from 5150\AA\ to 5300\AA, with an
effective dispersion of 0.01 \AA/pix ($R \sim 25,000$).  Spectra of
the standard stars were processed in essentially the same way except
that these were sufficiently bright to be traced without need of a
quartz exposure.

\section{Velocity Measurements}

\subsection{Velocity Standards}

We constructed a high signal-to-noise master template spectrum from
individual spectra of radial velocity standard stars observed during
the 2005 MMT run (see Table 1).  We chose HD171232 to act as a
`master' template, then used the Fourier cross-correlation routine
{\bf fxcor} in the {\it rv} package of IRAF to measure relative shifts
of this template and all the standard-star spectra.  Our results for
each standard-star observation spectrum are listed in Table 2, based
on an assumed heliocentric velocity of HD171232 of $V_{h,HD171232} =
-37.3 \pm 0.8$ km/s (Udry et al. 1999).  We also list the heliocentric
velocities for the standard stars from SA57 (Stefanik et al. 2006).
The individual standard-star spectra were Doppler-shifted by their
observed velocity shift relative to HD171232, then summed to make a
master template with a mean signal-to-noise ratio of about 400:1 per
resolution element.  At this stage, the final template was tied to the
velocity scale defined by HD171232 given the Udry et al. (1999)
velocity.

A problem became evident when we found that the SA57 standards
(Stefanik et al. 2006) from all runs exhibit an offset of
$\Delta_{SA57} = -3.4 \pm 1.3$ km/s when tied to HD171232 (see Table
2).  We suspect this offset is not associated with the SA57 standards
for the following reasons.  First, the mean heliocentric velocity of
sky fiber spectra with a Tonry-Davis (1979) index, $R_{TD}$, greater
than 2.8 (see section 3.2) was $-4.1 \pm 0.8$ km/s, consistent with
$\Delta_{SA57}$, though not with zero.  Second, during the 2006 run,
three sets of twilight exposures were obtained.  These data gave a
mean heliocentric velocity of $-3.4 \pm 0.2$ km/s for the twilight
spectra using the HD171232 velocity zeropoint; again consistent with
$\Delta_{SA57}$ but not with zero.  Third, velocity measurements from
other configurations where the spectra of faint stars are overwhelmed
by moonlight were found to be offset by $-3.7$ km/s using the HD171232
zeropoint.  Finally, Udry et al. (1999) remark that the velocity of
HD171232 appears to 'drift', a result we appear to confirm.

We can reconcile HD171232, the SA57 standards and the night-sky and
twilight spectra by simply shifting our adopted velocity scale by
$-\Delta_{SA57} = +3.4$ km/s.  We did this by amending the
heliocentric velocity in the master template spectrum by $+3.4$ km/s,
effectively adopting $v_{helio} = -33.9$ km/s for HD171232.  With this
change, the heliocentric twilight and night-sky velocities all
averaged to within $\pm 1$ km/s of zero, and the velocities of the
SA57 standards came into excellent agreement to their published values
(Stefanik et al. 2006; Table 2).  We believe that our final
heliocentric velocity zeropoint is systematically accurate to $\leq 1$
km/s.

\subsection{Velocities of Leo~I Candidates}

We used {\bf fxcor} in IRAF to measure the velocities of Leo~I
candidates relative to the template.  Typical cross-correlation
functions are shown in Figure~\ref{figs:ccf1}.  Tonry and Davis (1979)
defined a parameter, $R_{TD}$, which measures the height of the
cross-correlation peak relative to the amplitude of the noise in the
cross correlation function near the peak.  The {\bf fxcor} task
reports $R_{TD}$ for all spectra for which it estimates a velocity.
All of the spectra used here have $R_{TD} \geq 2.8$.  This cutoff
represents the value of $R_{TD}$ where it became difficult to identify
consistently a correlation peak, and where quantitative comparisons of
independent measurements of individual stars revealed that our
velocity measurements were becoming unreliable (see below).

Some statistics for our MMT/Hectochelle fiber observations are
provided in Table 3.  Of the 749 fibers assigned to astronomical
(i.e., non-sky) targets in the five Leo~I configurations, we obtained
543 spectra (72\%) of 371 different targets that produced
cross-correlations with $R_{TD} \geq 2.8$.  About 59\%\ of our
assigned fibers (440 spectra) produced good velocities for the final
sample of 312 likely Leo~I members observed with Hectochelle
(membership is quite clear-cut in the case of Leo~I as described
below).  If we add the 33 targets from M98 (we justify this below),
our final sample consists of 387 stars, 328 of which are likely Leo~I
members (17 of the M98 stars were reobserved with Hectochelle).  A
total of 108 stars (297 spectra) were observed multiply, with 51 stars
twice, 38 three times, 15 four times, and 3 five times. One star
(number 359 in Table 5) was observed on six separate occasions.  Table
4 gives a complete listing of all the repeat measurements within the
combined MMT/Keck dataset.

We have used the repeat observations to assess the quality of our
velocity measurements.  Figure~\ref{figs:mmtdiffs} plots the histogram
of the velocity differences in our dataset, along with plots of the
velocity differences as a function of position, velocity, and
brightness.  There are no significant correlations apparent, nor do we
see evidence that different configurations are offset relative to one
another.

Repeat measurements can also be used to estimate the individual
velocity errors as described by Walker et al. (2006a).  We assume a
relation between the velocity error, $\sigma_i$, and $R_{TD,i}$ for
star $i$, of the form
$$\sigma_i^2 (R_{TD,i}) = \left({{\alpha}\over{(1 +
R_{TD,i})^x}}\right)^2 + \sigma_0^2.$$ The best-fit parameters
$\alpha$, $x$ and $\sigma_0$ are determined via a least-squares
minimization process (Walker et al. 2006a; 2007) from which we
obtained $\alpha = 2.60$ km/s, $x = 0.16$, and $\sigma_0 = 0.14$ km/s.
In calculating these parameters, we did not vet the sample in any
way. Astrophysical sources of velocity variations (e.g. atmospheric or
binary motions) will contribute to our estimates of the individual
velocity errors.  Our data do indeed reveal evidence of possible
binaries (Section 3.4), but neither the magnitude of the velocity
variations nor the frequency of detectable binaries significantly
alter our results (Section 4.1).  The distribution of $\sigma(R_{TD})$
from repeat measurements, is quite flat, probably owing to systematic
run-to-run velocity errors at the 1-2 km/s level.  These represent a
negligible contribution to our final error budget since, even with a
catalog of 300+ Leo~I members, our kinematic results are still
dominated by sampling uncertainties (Sections 3.3, 4.2).

For the stars with multiple observations, we used these error
estimates to determine the mean velocity for each star. We assumed
that the $n$ multiple, independent velocity measurements
$\{v_1,...,v_n\}$ of a star with true velocity $u$ follow a Gaussian
distribution centered on $u$.  From maximum likelihood statistics
(Rice 1995), the estimate of $u$ is given by the weighted mean: $\hat
{u}=\sum_{i=1}^n(v_i\sigma_i^{-2})/\sum_{i=1}^n(\sigma_i^{-2})$.  The
velocity range that includes $u$ with probability $1-\gamma$ is then
given by $\hat{u} \pm n^{1/2}t_{n-1}(\gamma/2)S$, where $t_{n-1}$ is
the $t$ distribution with $n-1$ degrees of freedom and $S^2 \equiv
(n-1)^{-1}\sum_{i=1}^n(v_i-\hat{u})^2$.  For each star with multiple
measurements, we calculate $\hat{u}$ and its $1\sigma$ ($\gamma=0.32$)
confidence interval.  Figure~\ref{figs:hists} is a histogram of the
the velocity differences of repeat measurements ($\Delta_2 \equiv v_i
- \langle v\rangle$) plotted in bins of km/s and in units of the
velocity error, $\sigma$ ($\Delta_3 \equiv \Delta_2/\left(\sqrt{2}\
\langle \sigma\rangle\right)$.  The Gaussian profile in the lower
histogram is the expected distribution for $n = 2$ and $\sigma_i =
\sigma_j$, which is reasonably valid for about 70\%\ of the cases
plotted in the histogram.

The final adopted velocities for 387 stars in our Leo~I dataset are
listed in Table 5, excluding velocities from sky fibers and for any
spectra with $R_{TD} < 2.8$.  This table includes results for stars
from M98, adopting the velocity uncertainties tabulated in that paper.
Table 5 also lists positions and photometric data for each star and
the configurations used to determine their velocities.  For stars with
a single observations (and a known value of $R_{TD,i}$), velocity
uncertainties are $\sigma_{TD,i}$ from the equation above.  For stars
with multiple observations, the velocity uncertainties are obtained
from the 68\%\ ($1\sigma$) confidence limits as described in the
previous paragraph.

Sohn et al. (2007; hereafter S07) carried out an independent kinematic
study of Leo~I based on near-IR spectra of resolution $R \sim 3000$
obtained with DEIMOS on the Keck telescope.  Using a 0.5 arcsec
matching radius, we find 26 stars from S07 in common to our sample.
Figure~\ref{figs:sohndiffs} summarizes the velocity differences,
$\Delta$ in the same format as Figure~\ref{figs:mmtdiffs}.  The
lower-left panel of Figure~\ref{figs:sohndiffs} reveals a strong
systematic trend of $\Delta$ with velocity.  The coefficients of the
fitted line are given in the figure caption.  The correlation
coefficient indicates that a better linear fit could occur by chance
less than 0.02\%\ of the time for a sample this size.  The slope of
this trend ($-$0.31 km/s {\it per} km/s) is a significant contributor
to the width of the histogram of $\Delta$ in the upper-left panel of
Figure~\ref{figs:sohndiffs}.  The standard deviation of $\Delta$ about
the fitted line is 3.5 km/s while the standard deviation of the data
in the histogram is 4.8 km/s.  We find no significant correlation of
velocity difference between our data and that of S07 as a function of
position or $I$-band brightness.

Since the comparisons of our MMT and Keck datasets reveal no
systematic trends, we tentatively conclude that the source of the
systematic trend lies in S07 measurements.  Many authors have noted
special complications with velocity measurements using near-IR spectra
due to telluric contamination (a summary of this is given in Walker et
al. 2007).  If the near-IR spectra are indeed the principal source of
the trend in Figure~\ref{figs:sohndiffs}, the S07 velocities are
offset by about $+$10 km/s at the low end of the Leo~I velocity range
(approximately $250$--$310$ km/s on the MMT velocity scale) and by
about $-$2 km/s at the upper end of the range.  Given the
complications and uncertainties involved with dealing with this trend,
the unusual spatial coverage of targets observed by S07 (due to
constraints associated with DEIMOS slit masks), and the modest number
of new Leo~I members we would gain from the S07 sample (an additional
15\%\ at most to our sample), we have chosen not to include the S07
results in our analysis.

Koch et al. (2007; hereafter K07) have also recently published a
kinematic study of Leo~I based also on moderate-dispersion, near-IR
spectroscopy.  Using a 5 arcsec matching radius (K07 report their
coordinates to a precision of only 1 arcsec) and requiring 10\%\
photometric matching, we find 17 stars in common to the two samples.
Figure~\ref{figs:kochdiffs} summarizes the velocity differences,
$\Delta$, plotted in the same format as Figure~\ref{figs:mmtdiffs}.
Apart from a systemic offset (see upper left panel), we find no
significant trend in the comparison of these velocities, nor as a
function of stellar brightness.  In part because of this systematic
offset, but mostly since the typical velocity uncertainties of the K07
sample are about twice those of our new measurements, we have chosen
not to merge the datasets here.
 
\begin{figure}
  \begin{center}
    \plotone{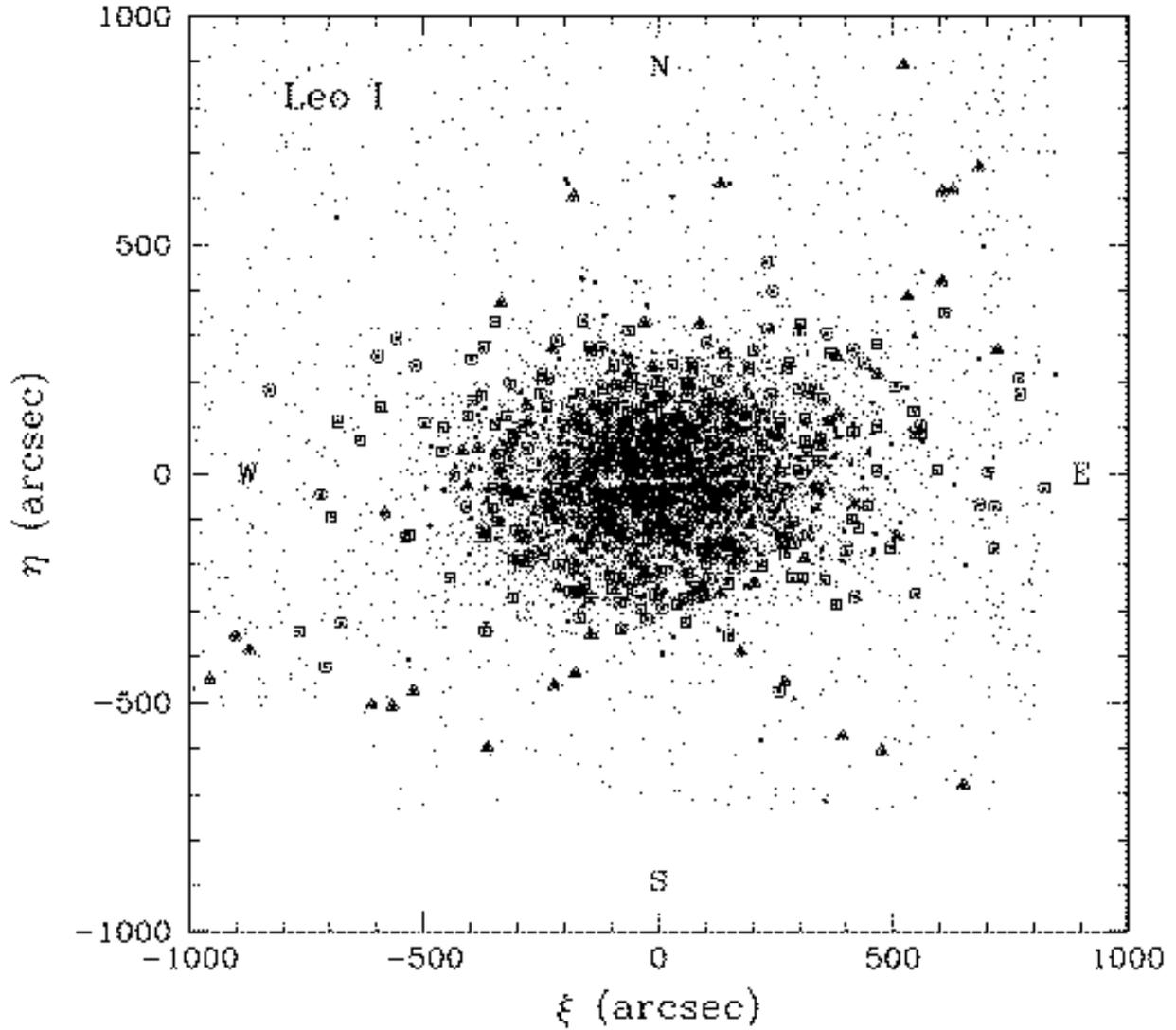}
      \caption{\label{figs:xieta} The spatial distribution of stars in
      our Leo~I color-magnitude diagram (Figure~\ref{figs:cmd}).
      Candidate Leo~I red giants are denoted as small filled black
      squares (see Figure~\ref{figs:cmd} for the candidate selection
      definition).  Stars with MMT/Hectochelle spectroscopy from
      2005-2007 are shown as open squares (kinematic members) and open
      triangles (kinematic non-members).  Open circles denote stars
      also observed spectroscopically with HIRES on Keck by M98.  The
      standard coordinates adopt the center of Leo~I at
      $(\alpha,\delta)_{2000.0} = $(10:08:27, +12:18:30) (Mateo 1998)
      as the tangent point.  Each Hectochelle field (see Table~1)
      spans an area much larger than the full extent of the region
      plotted here.}
  \end{center}
\end{figure}

\begin{figure}
  \epsscale{0.80}
  \begin{center}
    \plotone{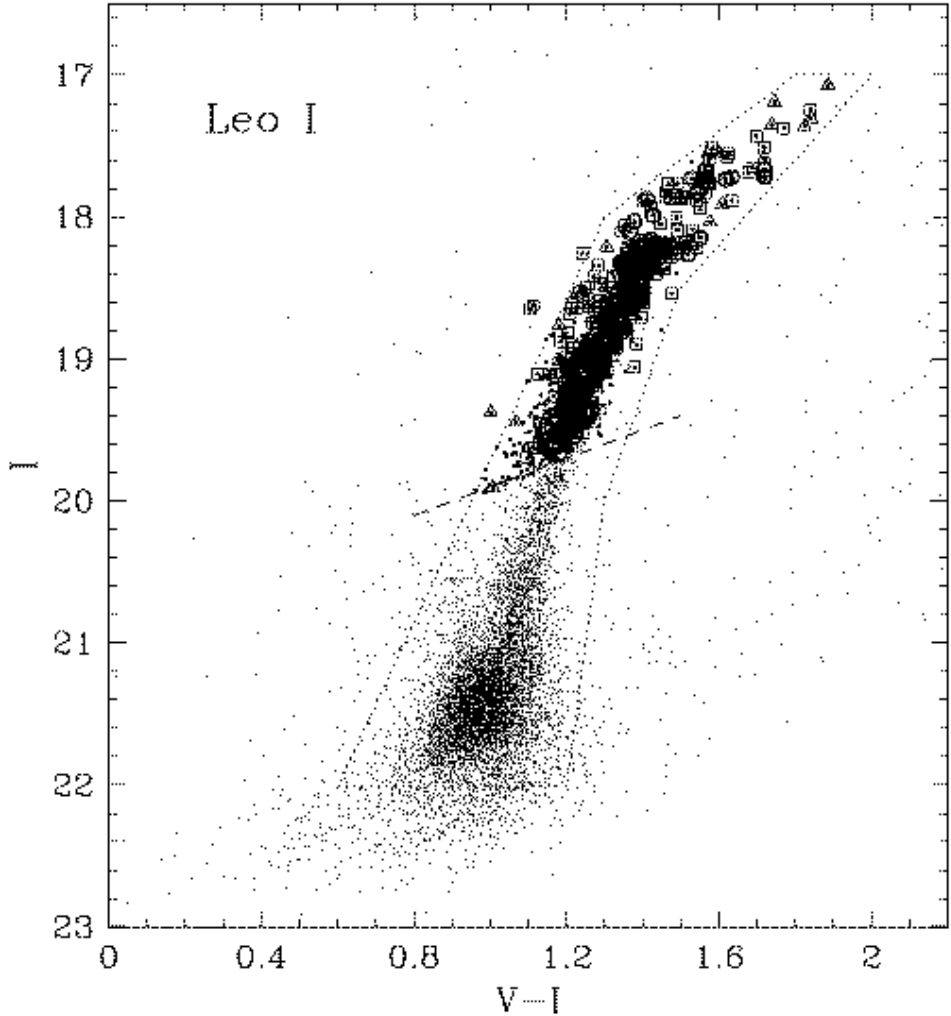}
    \caption{\label{figs:cmd} The calibrated Leo~I color-magnitude
    diagram based on our new 90Prime photometry.  The selection region
    for spectroscopic candidates can be discerned from the
    distribution of the small filled squares, while open squares and
    open triangles denote stars we have observed with Hectochelle.
    Open circles denote stars from the Keck/HIRES sample of M98.  The
    dotted lines enclose the stars used to measure the structural
    parameters of Leo~I (Section 4.3.2).  Though difficult to see here
    because of crowding, about 75\%\ of the stars we have observed
    with Hectochelle lie along the RGB, below the base of the extended
    AGB at $I \sim 18.1$.  The dashed line denotes our effective selection
    limit at $V = 20.9$.  These data have been transformed to the
    photometric scale of M98 to a precision of about 0.02 mag.}
  \end{center}
\end{figure}

\begin{figure}
  \begin{center}
    \plotone{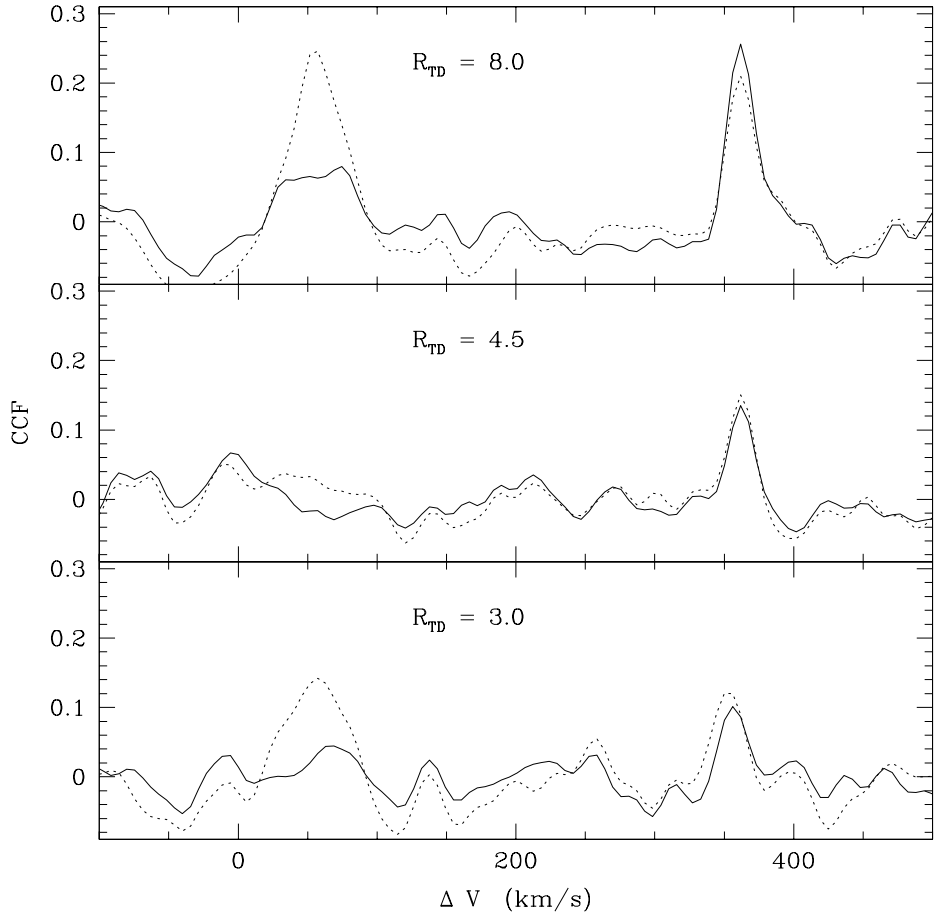}
      \caption{\label{figs:ccf1} Raw Hectochelle cross-correlation
      functions from the IRAF task {\tt fxcor} are plotted here in the
      velocity range $-100$ to $+500$ km/s.  Results for three Leo~I
      stars observed during 2006 or 2007 are shown.  The Tonry-Davis
      $R_{TD}$ (1979) values for these cases span much of the range
      exhibited by our kinematic sample (see Figure~\ref{figs:ccf2}).
      Dotted lines correspond to cross-correlations for spectra for
      which we did not do sky subtraction, while the solid lines are
      for sky-subtracted spectra.  The broad, low-velocity peak in two
      of the non-sky-subtracted cases (top and bottom panels) is due
      to scattered moonlight (the middle panel is for a spectrum
      obtained in dark conditions).  The velocity scale of these plots
      has not been corrected for the heliocentric velocity of the
      template.}
  \end{center}
\end{figure}

Figure~\ref{figs:rpa} is a plot of all velocities measured from our
sample for spectra with $R_{TD} \geq 2.8$, as a function of radial
distance from the adopted center of Leo~I (Mateo 1998; see
Figure~\ref{figs:xieta}).  The results in Table 5 are plotted as
filled points, while velocities measured from sky fibers (but still
with $R_{TD} \geq 2.8$) are shown as open symbols.  There is a clear
concentration of stars around the mean systemic heliocentric radial
velocity of Leo~I (282.9 km/s, Section 4.3.1; Zaritsky et al. 1989;
M98).  The stars in this velocity range also exhibit a clear
concentration toward the center of Leo~I.  When we also consider the
expected velocity distribution of field stars in this direction (Robin
et al. 2003), it is clear that all stars with $V_{helio}$ in the range
250-320 km/s are highly probable Leo~I members (the range 200-400 km/s
would have identified precisely the same sample of likely members).
Stars with velocities outside this range are uniformly distributed
spatially in the field, consistent with non-membership (see
Figure~\ref{figs:xieta}).

Sky fibers with $R_{TD} \geq 2.8$ tend to exhibit a mean velocity of
about $-19.7 \pm 0.8$ km/s when plotted in Figure~\ref{figs:rpa}
because {\tt fxcor} assumes the target is an astronomical object and
applies a heliocentric correction.  For our Leo~I observations, the
heliocentric correction for a Leo~I field was always about $-19 \pm 2$
km/s.  Although only non-sky fiber results are plotted in
Figure~\ref{figs:rpa}, many velocities cluster around $-19$ km/s.
These are likely `false positives' where the target stars were too
faint to produce usable spectra, but for which the sky velocity could
be measured.  We somewhat arbitrarily define false positives as cases
where we observe a heliocentric velocity in the range $-10$ to $-28$
km/s and $R_{TD} \leq 4.2$.  There are 14 such cases in
Figure~\ref{figs:rpa}, all of which are noted in Table 5.  Each has
data from only the 2005 dataset (which we could not sky-subtract).
Moreover, many false positives based on 2005 spectra alone turned out
to have well-determined, non-sky velocities obtained from spectra from
the 2006 or 2007 runs, or from the M98 sample.  The converse of this
effect -- sky spectra that scatter into the acceptance range for Leo~I
members (250-320 km/s) -- never occurred in multiply-observed data
with 2005 observations.  Given the narrow distribution of velocities
we observe in the sky fibers, such scatter is statistically extremely
unlikely.

\subsection{Velocity Dispersion Profiles}

Merritt and Saha (1993) and Wang et al. (2005) have described
non-parametric approaches that produce dispersion profiles without
binning.  These methods are marginally appropriate for the present
Leo~I data set because the sample size ($N = 328$ members) remains
rather small.  We will describe a non-parametric analysis for Leo~I
and other dSph galaxies in a separate paper (Walker et al. 2007, in
preparation).

For now, we adopt here the more standard approach of using binned
profiles in our analysis.  The bins in our profiles contain (nearly)
equal numbers of stars as dictated by the sample size and number of
bins, $N_{bin}$.  Profiles for $N_{bin} = 15, 20$ and 25 are shown in
Figure~\ref{figs:leoibin}.  The horizontal `error bars' in the
profiles show the standard deviation in $R$ for the stars in each bin.
A common problem with binning is the possibility that false structures
may be produced in the profiles.  Walker et al. (2006b) have
investigated this problem and find that bins contain $10$ or more
stars seem to be systematically stable to within the calculated
Poisson uncertainties.  From top to bottom in
Figure~\ref{figs:leoibin}, the three binning options correspond to
21/22 ($N_{bin} = 15$), 19/20 ($N_{bin} = 20$) and 13/14 ($N_{bin} =
25$) stars per bin.  There are no significant features in any of the
profiles that are not visible in the other (this remains true if we
offset the bins), so these binning options appear to be fairly robust.
Since our aim is to model the Leo~I dispersion profile with simple
dynamical models, we will take advantage of the higher S/N per bin of
the $N_{bin} = 15$ profile and use this one exclusively in our
subsequent analysis.  The smaller, downward error bars on the
dispersions in Figure~\ref{figs:leoibin} are based on the method
described by Kleyna et al. (2004), while the larger, symmetric error
bars are calculated using the method described by Walker et
al. (2006a, 2007a).

\subsection{Temporal Stability}

With the inclusion of the M98 results, the subset of stars in Table 5
with multiple observations spans slightly over 11 years of temporal
coverage.  The possibility of kinematic variability in our sample is
suggested by the outliers apparent in the lower panel of
Figure~\ref{figs:hists}.  To explore this further, we produced
Figure~\ref{figs:chi2}, a plot of the reduced chi-squared,
$\chi_\nu^2$ for all stars with multiple measurements for $\nu$
degrees of freedom.  The lines show the values of $\chi^2_\nu$
corresponding to a 0.5\%\ probability of exceeding the value of
$\chi^2_\nu$ by chance for $\nu = 1$ (2 observations) to 4.  Seven
stars, including five Leo~I members, exhibit values of $\chi^2_\nu$
that suggest they may be binaries.  In every case, the stars have only
two observations and exhibit $\Delta V \leq 13.3$ km/s.  These stars
are noted in Tables 4 and 5.

The two stars with the smallest heliocentric velocities have
$\chi^2_\nu \sim 9$, and, in both cases, the observations were
obtained only a few days apart.  As neither are members of Leo~I, they
can plausibly have short orbital periods.  In the other cases (all
Leo~I members; see Tables 4 and 5), the time intervals between
observations range from as little as 1 to at most 10 years.  For a
sample of 102 stars with repeat measurements, we might reasonably
expect 1-2 stars to exceed the 0.5\%\ line in Figure~\ref{figs:chi2}
by chance.  The presence of five outliers suggests we have detected at
least a few physical binaries in Leo~I.

We can look for more subtle evidence of binarity by splitting the
sample of multiply-observed stars into subsamples with different time
intervals between the individual observations.  One subsample consists
of stars observed multiply in a single run.  We take these to have a
time interval between observations, $\Delta t$, of zero since
plausible red giant binaries in Leo~I must have orbital periods long
compared to the length of a single run.  The second subgroup contains
stars observed either one or two years apart (that is, during
different MMT/Hectochelle runs; see Table 1) $\Delta t \sim 1$-2 yr.
Finally, stars observed with the MMT in 2005-07 and during the 1996
Keck observations published by M98 have $\Delta t$ of 9 or 10 years
and comprise a final subgroup (there are no Keck repeats from the 2007
MMT run).  For the $\Delta t = 0$ subgroup we find a mean velocity
difference of $\langle \Delta v_{(0\ yr)}\rangle = 0.58 \pm 0.37$
km/s, with rms of $\sigma_{(0\ yr)} = 3.4$ km/s, and $N=87$.  For the
$\Delta t = 1$ yr subgroup the values are $\langle \Delta_{(1\
yr)}\rangle = 0.10 \pm 0.37$, $\sigma_{(1\ yr)} = 4.2$ km/s, and $N =
138$.  Finally, for the long-interval subgroup we find $\langle
\Delta_{(10\ yr)}\rangle = -0.29 \pm 0.66$, $\sigma_{(10\ yr)} = 3.8$
km/s, and $N = 33$.  These values are consistent with no discernable
change in $\sigma_{(\Delta t)}$ as time interval increases.

\section{The Dynamics of Leo~I}

In this section, we explore two cases that serve as frameworks to
interpret our new kinematic data for Leo~I.  The first corresponds to
the case where the stellar and dark components of Leo~I are in
dynamical equilibrium, while the second explores the possibility that
some or none of the dynamical components that comprise Leo~I are in
dynamical equilibrium.  For convenience, various parameters for Leo~I
that we use or derive in this section are summarized in Table~6.

\subsection{Binary Stars}

Before we can discuss dynamical models for Leo~I, we must address the
potential contamination by spectroscopic binaries.  In general,
orbital motions in binaries will enhance the velocity dispersion of a
kinematic sample.  We found above that our dataset reveals at most
seven possible binaries (5 Leo~I members) among the 108 stars (84
Leo~I members) with multiple observations.  Are these binaries present
in sufficiently large numbers and with sufficiently large velocity
excursions to significantly affect our interpretation of the galaxy's
kinematics?

Red giants in Leo~I that reside in binaries with periods up to a few
hundred years will exhibit maximum velocity amplitudes comparable to
the internal dispersion we measure for the galaxy ($\sim 9$ km/s).
Projection (inclination) effects, a spread in binary mass ratios, and
a range of orbital separations all tend to reduce the velocity
amplitudes one actually observes in a realistic sample. A hint that
suggests binaries are unimportant in our Leo~I data is our evidence
(Section 3.4 above) that the sample dispersion does not change with
increasing time baseline interval.  Our observations are consistent
with a population of binaries that contributes a `dispersion'
comparable to or smaller than the mean measurement errors, about 2--3
km/s.

Previous studies have addressed the issue of binary contamination in
kinematic samples of dSph galaxies (Hargreaves et al 1996; Olszewski
et al. 1996), taking into account plausible period, inclination and
mass distributions for binary populations that amount to some
fraction, $f_b$, of the total sample.  Here, $f_b$ is defined here as
the total number of apparently single stars that are actually
unresolved binaries, divided by the total number of apparently single
stars.  The simulations of Olszewski et al. (1996) for the case of $N
= 17$ and a population with an intrinsic dispersion of about 7 km/s
are, remarkably, appropriate for {\it single bins} in our Leo~I
dispersion profile (Figure~\ref{figs:leoibin}).  From their results
(Table 9 of Olszewski et al. 1996) we find that even for an extreme
case with $f_b = 0.7$ and all binaries distributed within the shortest
period (highest velocity amplitude) range of 0.5-100 yrs, a
single-epoch measurement will typically overestimate the true
dispersion by only about 10\%, with the 95\%\ confidence interval
ranging between about 4.5-10 km/s for an assumed sample dispersion of 7
km/s.

In Leo~I, we observe a lower limit to the binary frequency of $f_b
\geq 5/84 \sim 0.06$ for a set of observations that was sensitive to
binaries a period range of a few days to few decades.  The lack of an
increase in dispersion with time baseline suggests there is no
significant population of longer-period binaries lurking under the
radar.  Thus, for Leo~I the likely binary frequency appears to be much
lower than the 70\%\ frequency assumed in the simulation from
Olszewski et al. (1996) we cited above.  We conclude that binaries
have inflated the dispersions of single bins in
Figure~\ref{figs:leoibin} by at most $\sim 10$\%.  Given that the
typical error bars on these individual dispersions are around 15-20\%,
the effects of binaries are negligible for any given bin.  To be safe,
we use the larger symmetric error bars from Walker et al. (2006a,
2007a; see Figure~\ref{figs:leoibin}) to calculate goodness-of-fit in
all subsequent analyses.

\subsection{Equilibrium Models}

Figure~\ref{figs:leoibinfits} shows a comparison of the velocity
dispersion profile of Leo~I (for the $N_{bin} = 15$ profile) with an
isothermal model ($\sigma = 9.2 \pm 0.4$ km/s; see Section. 4.3.1 and
Table 6), a single-component King (1966) dynamical model, and a
two-component Sersic+NFW (Sersic 1968; Navarro et al. 1997; {\L}okas
2002) model.  The core radii of the isothermal and King models is
taken to be that of the visible stellar distribution ($R_{core} = 245$
pc for an assumed distance of $255$ kpc) and a concentration parameter
of $c \equiv \log(R_{tidal}/R_{core}) = 0.6$ (Irwin and Hatzidimitriou
1995; hereafter IH95).  The Sersic profile used with the NFW model is
assumed to be concentric with the DM halo.  We followed the recipe of
{\L}okas (2002), adopting a Sersic profile index of $m_S = 0.6$ and a
Sersic radius of $r_{Sersic} = 370 \pm 30$ pc (see
Figure~\ref{figs:leoishape}).  We assumed $M/L = 1.0$ for the visible
matter, implying $M_{visible} = 5.6 \pm 1.8 \times 10^6 M_\odot$).
Models with anisotropy parameter, $\beta$ (see Binney and Tremaine
1987), ranging from 0.0 (isotropic) to $-3.0$ (moderately tangentially
aniostropic) are compared to the observations in
Figure~\ref{figs:leoibinfits}.  For a given model, $\beta$ is
constant.

Because the observed dispersion profile of Leo~I is so flat, the
isothermal sphere (top panel of Figure~\ref{figs:leoibinfits})
provides a good fit to the kinematic data.  For our assumption that
the mass distribution has the same core radius as the visible matter,
the central density for the isothermal case is $0.23 \pm 0.04\ M_\odot {\rm
pc}^{-3}$ (Richstone and Tremaine 1986; M98).  The baryonic central
density is considerably lower, $\rho_b \sim 0.02$-0.05 $M_\odot {\rm
pc}^{-3}$ for $\left(M/L\right)_{V,baryons} = 0.3$-0.7 (M98; below),
where we have converted the projected surface density assuming a King
profile with parameters from IH95 corrected to an adopted distance of
255 kpc.  The mass of the best-fitting spherical, isothermal sphere
out to 1040 pc ($\sim 840$ arcsec, the location of the outermost
kinematic member of Leo~I in our sample) is $5.2 \pm 1.2 \times 10^7
M_\odot$, implying $M/L = 9.3 \pm 4.0$ (in Solar units) interior to
this radius.  The projected mass density of the isotropic model does
not resemble the visible mass density (see
Figure~\ref{figs:leoishape}), but this just means that mass does not
follow light.  In particular, $\rho_{DM} >> \rho_{vis}$ everywhere in
Leo~I, with the DM distribution considerably more extended than the
visible matter.

It has long been known that King (1966) models provide a good fit to
the visible matter distribution in Leo~I and other dSph
galaxies(IH95).  However, a single-component King model in which mass
follows light (middle panel of Figure~\ref{figs:leoibinfits}) fails
spectacularly to fit the observed dispersion profile. For any King
model that fits the light distribution, the predicted dispersion
begins to decrease steadily outside the core radius, falling well
below the observed profile, reaching zero at the tidal radius (by
design, of course).  Through its failure to fit the kinematics of
Leo~I, the King model also implies that, in equilibrium, the mass
distribution of the galaxy must be considerably more extended than
that of the visible matter.

A two-component Sersic+NFW model does considerably better at
accounting for the dispersion profile and the visible matter
distribution, particularly if we allow for some (radially constant)
kinematic anisotropy (lower panel, Figure~\ref{figs:leoibinfits}).
The isotropic case, $\beta = 0$, cannot simultaneously fit the inner
and outer parts of the dispersion profile for any assumed Leo~I mass.
The best fit is for $\beta = -1.5$ and $M_{vir} = 7 \pm 1
\times 10^8 M_\odot$, where $M_{vir}$ is the mass interior to the
virial radius (defined here as the radius where $\rho(R_{vir}) = 200
\rho_{crit}$).  The 95\%\ confidence interval on $\beta$ ranges from
$-0.4$ to $-3.2$.  The best-fit Sersic+NFW model implies a mass of
$8.1 \pm 2.0 \times 10^7 M_\odot$ interior to 1040 pc, the radius of
the outermost Leo~I member in our kinematic sample.  This mass is
about 60\%\ higher than the mass obtained from the isothermal model
(above; see Table 6).  At this radius, $M/L = 14.4 \pm 5.8$ (in Solar
units) for the Sersic+NFW case.  


\subsection{Are Equilibrium Models Valid for Leo~I?}

The large heliocentric velocity of Leo~I and its unusual star
formation history suggest that the galaxy may have experienced a
strong encounter with the Milky Way in the not-too-distant past.  Here
we explore the interpretation of our new kinematic results in the
context of such an interaction.

\subsubsection{The Radial Velocity of Leo~I}

From the entire sample of 328 Leo~I members, we find a weighted ($w_i
= 1/\sigma_i^2$) mean heliocentric velocity of $282.9 \pm 0.5$ km/s
and a sample dispersion of $\sigma = 9.2 \pm 0.4$ km/s.  For an
assumed motion of the Local Standard of Rest (LSR) of 220 km/s toward
an apex at $(l,b) = (90,0)$ and a peculiar solar motion relative to
the LSR of 16.6 km/s toward $(l,b) = (53,25)$, Leo~I has a velocity of
$174.9 \pm 0.5$ km/s for an observer located at the Sun but stationary
with respect to the Galactic Center (we refer to this as the
'Galactostationary', or 'GS', reference frame).  The angular
separation of the Sun and Galactic Center as seen from Leo~I is only
1.7 deg, so the GS radial velocity of Leo~I must be very close to the
Galactocentric radial velocity component of Leo~I for any tangential
velocity that keeps Leo~I bound to the Milky Way.  Leo~I is fairly
compact on the sky, so the sample dispersion is unaffected by this
change of reference frame.  Hence, $\sigma_{GS} = \sigma_{helio} = 9.2
\pm 0.4$ km/s.

The large outward velocity of Leo~I has long been problematic and
puzzling.  Analyses of the total mass of the Milky Way using halo
tracers (e.g.  Zaritsky et al. 1989; Kochanek 1996; Wilkinson and
Evans 1999; Sakamoto et al. 2003) are often strongly affected by the
inclusion of Leo~I. Taylor et al. (2005) pointed out that a system
with kinematics similar to Leo~I is rarely seen in CDM simulations of
the formation and late-time evolution of the Local Group.  Byrd et
al. (1994) suggested that Leo~I is unbound to the Milky Way (but bound
to the Local Group), postulating that the dwarf originated closer to
the Andromeda galaxy than the Milky Way.  In this scenario, Leo~I
follows a hyperbolic trajectory relative to our Galaxy, exceeding the
local escape velocity of the Milky Way at its current position.

Whether bound or unbound, the sign of Leo~I's radial velocity means
the galaxy was much closer to the Galactic Center in the past.  In the
preferred model of Byrd et al. (1994), $R_{peri, LeoI} = 70$ kpc,
similar to the distances of the closest present-day dSph galaxies from
the Galactic Center.  Orbits for which Leo~I is bound to the Milky Way
imply even smaller perigalactica.  This result is remarkable in that
it implies Leo~I entered the `no-fly zone' ($R \leq 60$-70 kpc; Mayer et
al.  2001a,b) of the Milky Way, a volume in which we find no
internally bound dSph systems (Mateo 1998; Grillmair 2006; Belokurov
et al. 2006).  If Leo~I passed through this no-fly zone, it would have
experienced tidal forces of sufficient strength that, given enough
time, are able to destroy dwarf systems of comparable luminosity.

To explore this further, Dr. C. Pryor kindly calculated a series of
simple bound orbits for a point-like Leo~I model in a logarithmic
Galactic potential.  These models suggest that Leo~I passed closest to
the Galactic Center 0.5-2 Gyr ago for $v_{tan} \leq 100$ km/s in a
Galactocentric rest frame.  One such model, based on the orbital pole
suggested by S07, is shown in Cartesian projection in
Figure~\ref{figs:leoiorbit}, while $R(t)$ is plotted in
Figure~\ref{figs:leoirorbit}.  Within the framework of tidal stirring
models (Mayer et al. 2001a,b; 2005), this most recent close passage to
the Galactic Center would have been responsible for Leo~I's latest
significant episode of star formation (which ended about 1 Gyr ago;
Gallart et al. 1999b; Hernandez et al. 2000; Dolphin 2002), and
possibly may have stripped the galaxy of gas, completing its
transformation into a spheroidal system.  The important issue this
raises is whether tidal interactions have left an imprint on Leo~I.
If so, what is the nature of the imprint and how does it affect our
dynamical analysis?

\subsubsection{Tidal Imprints in Leo~I:  Structural Properties}

It is well established that encounters between dwarf spheroidal
galaxies and the Milky Way can raise strong tides in the smaller
systems (Oh et al. 1995; Piatek and Pryor 1995; Read et al. 2006b;
Klimentowski et al. 2006).  One result is that debris from the dwarf
is spread along the galaxy's past and future orbit.  In the case of
Leo~I, its large distance and likely highly eccentric orbit (see
below) would cause tidally-stripped matter -- which we refer to
loosely as `tidal arms' -- to project closely onto the main body of
the dwarf.

The extent and structure of tidal arms depends sensitively on the
dwarf mass.  At one extreme, Kuhn and Miller (1989), Kuhn et
al. (1996), Klessen and Kroupa (1998) and Fleck and Kuhn (2003), among
others, have explored the effects of tides on interacting dwarfs that
contain no dark matter at all.  In such cases, prominent tidal arms
are produced that create clear structural and line-of-sight kinematic
signatures (e.g. Klessen and Kroupa 1998; Klessen et al. 2003; Read et
al. 2006b).  For example, systems with significant tangential
velocities may produce tidal extensions visible along the projected
orbital path on the sky. These may be seen as S-distortions in the
projected stellar distribution (Odenkirchen et al. 2002; Grillmair and
Dionatos 2006), or as `breaks' in the light profiles (Read et
al. 2006b).  If the tangential velocity is small compared to the radial
velocity component (as is likely the case for Leo~I), the geometry of
the orbit may cause arms to project mostly onto the main body of the
disrupting dwarf.  For this case, an observer may fail to notice
striking structural anomalies (Read et al 2006b), even though
significant line-of-sight extension is present.  For Leo~I,
observations of the red giant branch and red clump (Gallart et
al. 1999a; Bellazzini et al. 2004) and of RR~Lyr stars (Held et
al. 2000, 2001) reveal no compelling evidence that a significant
fraction of the galaxy extends more than $\pm 15$\%\ (or about $\pm$
40 kpc) from its main body.  Similar observations have ruled out
significant line-of-sight extensions in other local dwarfs (Klessen et
al. 2003).

We conclude that if Leo~I had a close encounter with the MW (say, as
implied by the orbit shown in Figures~\ref{figs:leoiorbit} and
\ref{figs:leoirorbit}), then the lack of observable depth in Leo~I and
the absence of {\it prominent} structural anomalies require the
presence of dark matter in Leo~I.  Otherwise, the galaxy would have
been disrupted or greatly distorted in the encounter given its deep
incursion into the Milky Way's `no-fly zone' (Mayer et al. 2001b).
But the presence of DM does not mean Leo~I would have been {\it
unaffected} by the encounter, just that tidal effects may be
considerably subdued compared to the no-DM cases considered above.

To search for subtle structural tidal features, we have used our
photometric observations to try to identify distortions in the
projected geometry of Leo~I.  Figure~\ref{figs:greycontour} shows
contours fit to a greyscale representation of the star counts of 12630
stars chosen within the boundary shown in the Leo~I CMD
(Figure~\ref{figs:cmd}), and smoothed with a Gaussian kernel with
$\sigma = 2$ arcmin.  By selecting stars from the CMD, we reduce field
star contamination to about 1-2\%\ of the sample.  The contours are
independent fits to different isopleths of the stellar distribution
for ellipses with semi-major axes ranging from 2-12 arcmin, in 1
arcmin intervals.  These contours reveal no striking distortions or
asymmetries, though possibly a slight centroid shift to the NE with
decreasing surface brightness.  

Figure~\ref{figs:centers} quantifies this by plotting the locations of
the fitted ellipse centers for successive contours.  The ellipse
centroids shift about 10 arcsec to the NE as we map fainter contours,
consistent with the visual impression from
Figure~\ref{figs:greycontour}.  This shift is quite modest: the
centroid offsets in Figure~\ref{figs:centers} are never larger than
about 5\%\ of the King core radius of Leo~I (IH95).
Figure~\ref{figs:leoishape} plots some of the other parameters we
derive from the fitted ellipse contours.  Neither the position angles
of the fitted ellipses nor their ellipticities vary more than
1-2$\sigma$ about the observed means, though the trend in ellipticity
(increasing outward) appears to be systematic.  The properties of the
contours in Figure~\ref{figs:greycontour} do not rule out weak tidal
features in Leo~I, but they do indicate that Leo~I lacks prominent
tidal distortions in its projected structure.  Models of dSph galaxies
that lack DM but are optimally aligned to conceal structural
anomalies, still tend to exhibit larger centroid shifts and isophot
variations than we observe in Leo~I (Klessen and Kroupa 1998; Klessen
and Zhao 2002; Fleck and Kuhn 2003; Read et al. 2006b).

\subsubsection{Tidal Imprints in Leo~I: Kinematic Properties}

Tides should produce observable kinematic signatures, including
large-scale streaming motions that mimic rotation (Piatek and Pryor
1995; Oh et al. 1995; Klessen and Zhao 2002) or that produce
distinctive, rising dispersion profiles (Read et al. 2006b).  This
pseudo-rotation signal may be seen as a coherent velocity gradient
across the galaxy roughly oriented along the projected orbital path.
A persistent motivation for exploring dSph models that lack DM has
been to determine if tidal effects alone can account for the observed
dispersions and dispersion profiles.  We have already shown that the
mildly distorted structural properties of Leo~I imply the galaxy
contains a significant DM content.  Our discussion here aims to
explore the extent to which tides may have affected the kinematics of
Leo~I given the presence of a DM halo.

To search for streaming, we calculate the mean velocity differences,
$\Delta v$, within of our entire sample of kinematic members of Leo~I
on either side of a bisector passing through the galaxy center (see
Walker et al. 2006a).  We repeat this for a range of bisectors, each
oriented at different position angles, $\theta$, producing a function
$\Delta v(\theta)$ (Figure~\ref{figs:deltav}; see Figure~\ref{figs:xieta2} for a graphical
description of this procedure).  When we apply this test to the full
kinematic sample of Leo~I members, we find a maximum $\Delta v$ of
$1.8 \pm 1.2$ km/s.  Monte-Carlo simulations (described in detail
below) indicate that this observed maximum value of $\Delta v$ occur
by chance about 30\%\ of the time when we consider all 328 Leo~I
kinematic members in our dataset.  Thus, we find no compelling
evidence for a velocity gradient in Leo~I from the full kinematic
sample or from the stars located within.

We can split the Leo~I sample into two radial groups corresponding to
an `inner' subsample ($R \leq 400$ arcsec; $N = 264$) and an `outer'
subsample ($R > 400$ arcsec; $N = 64$).  The motivation for this
division comes from Figure~\ref{figs:leoibreak}.  All stars (members
and nonmembers) in the inner subsample are uniformly distributed about
the center of Leo~I.  In the outer subsample, non-members remain
uniformly distributed, but the kinematic members appear elongated
along a position angle of about $90/270$ deg.  This effect does not
evidently result from any selection effect in our sample or fiber
assignments.  

The spatial differences of the inner/outer subsamples can also be
illustrated by comparing the frequency of members and nonmembers as a
function of position angle.  Figure~\ref{figs:memfrac} shows that
along a position angle of about $90 \pm 10$ deg, a significantly
higher proportion of members make up the full outer kinematic
subsample.  For the inner subsample, no such trend is seen.  This behavior
is evident from inspection in the spatial distribution of kinematic
members and non-members plotted in Figure~\ref{figs:xieta}, more
plainly illustrated in Figure~\ref{figs:xieta2}.  We conclude that
stars of the outer sample of Leo~I follow a distribution that is
elongated along an axis that is similar to, but possibly slightly
offset from the axis corresponding to the maximum velocity gradient of
Leo~I members in the outer subsample.

We have applied the streaming test described above for Leo~I members
in both the inner and outer subsamples separately
(Figure~\ref{figs:deltav}).  The inner subsample alone still shows no
convincing evidence of streaming.  However, the further out we sample
Leo~I members, the stronger the streaming signal becomes in terms of
$\Delta v$ (Figure~\ref{figs:deltav}) and, generally, significance
(see below).  Moreover, the bahavior of $\Delta v(\theta)$ is highly
coherent, consistent with streaming motion along the PA $\sim$ 90/270
axis.  We have fit the outer subsample (Leo~I members only) to a
linear velocity gradient model as a function of position angle.  The
strongest velocity gradient corresponds to $PA \sim 108 \pm 10$ deg.
The fitted slope is $-0.34 \pm 0.15$ km/s/arcmin or $-0.0046 \pm
0.0020$ km/s/pc (for $D = 255$ kpc).  The sign of the gradient is such
that outer subsample Leo~I members to the west of the center of Leo~I
have on average positive velocities relative to the systemic velocity
of the galaxy, while outer subsample members to the east have on
average negative relative velocities (see Figure~\ref{figs:xieta2}).

To determine the significance of this result, we ran Monte-Carlo
simulations where we assigned the observed velocities to permutations
of the stellar positions (each MC experiment consisted of 10000
trials).  We then calculated $\Delta v$ as above for each simulated
sample.  The probabilities of seeing a value of $\Delta v$ as large or
larger than the observed maximum $\Delta v$ at {\it any position
angle} are given in Figure~\ref{figs:deltav}.  The observed maximum
value of $\Delta v$ is exceeded about 10\%\ of the time for our
simulations of the inner subsample (top panel,
Figure~\ref{figs:deltav}), while for the outer subsamples (for $R >
400,\ 455$, and 600 arcsec), the probability of exceeding $\Delta
v_{max}$ is (0.03, 0.006, 0.014), respectively.  If we add the
requirement that the simulations exhibit the coherence apparent in
Figure~\ref{figs:deltav}, then virtually none of the simulations
($\leq 5$ out of 30000) for the three outer subsamples.  We conclude
that Leo~I exhibits a statistically highly significant velocity
gradient along an axis very close to its apparent major axis, but only
among stars with projected radii $\geq 400$ arcsec.  We shall refer to
this radial distance at which the galaxy's kinematics change as the
`break radius', $R_b$, of Leo~I.

\subsubsection{Tidal Imprints in Leo~I: Population Segregation}

Our data also reveal evidence of a change in the stellar populations
of Leo~I at the break radius.  To illustrate
this, we have taken our photometry from Figure~\ref{figs:cmd} and
plotted it for the inner and outer subsample regions separately
(Figure~\ref{figs:cmd2}).  Four regions in the CMD were identified that
correspond to red giant branch stars (RGB; Region 1), asymptotic giant
branch stars (AGB; Region 2), field stars (Region 3), and blue-loop
stars (Region 4).  Regions 1 and 2 correspond, roughly, to our
kinematic selection region (see Figure 2).  Table~7 lists the numbers
of stars in each region from direct counts in the CMD, along with
kinematic results for stars in each CMD region where available.

Consider first the RGB and AGB regions (Regions 1 and 2, respectively;
the data used here are listed in Table 7).  The ratio of RGB
candidates in the inner and outer samples is $732/91 = 8.0$, and
$72/18 = 4.0$ for the AGB candidates.  If we only consider the
kinematic sample and assign membership to stars based on their
velocities, the RGB kinematic members in the inner and outer samples
becomes $209/52 = 4.0$, while for the AGB stars the ratio is $42/1 =
42$.  That is, the intermediate-age population traced by the AGB stars
(Gallart et al. 1999a; Hernandez et al. 2000; Dolphin 2002) is almost
exclusively located within with the break radius, $R_b = 400$ arcsec.
RGB stars, which arise from both the intermediate and older
populations, extend over both the inner and outer regions.

This analysis underscores the value of the kinematic data to search
for population gradients.  Field stars, almost all certain non-members
of Leo~I based on their location in the CMD (Region 3 of
Figure~\ref{figs:cmd2}) exhibit a ratio of
${{N_{outer}}\over{N_{inner}}} = 2.1 \pm 0.5$, consistent with the
ratio of the areas of the inner and outer regions
(${A_{outer}\over{A_{inner}}} = 2.6$).  This contamination explains
the relatively small ratio of inner/outer AGB stars in the
non-kinematic sample.  Blue-loop stars (Region 4 in
Figure~\ref{figs:cmd2}) exhibit an inner/outer ratio of about 6.0,
and, like the AGB, are intermediate-age stars (Dohm-Palmer and
Skillman 2002).  The actual ratio of inner/outer blue-loop stars is
likely considerably larger since non-members preferentially
contaminate the outer sample.

These points are presented graphically in
Figure~\ref{figs:rcumulative} where we plot cumulative radial
distributions of various subsets of stars from the Leo~I CMD
(Figure~\ref{figs:cmd2}).  The left panel shows that, among
radial-velocity members, stars selected in the AGB region of
Figure~\ref{figs:cmd2} (Region 2) are more centrally distributed than
the RGB stars (Region 1). This holds even if we only consider
kinematically-selected RGB stars (which are biased in radius due to
fiber restrictions and science aims), or the full sample of RGB
candidates (thick and thin solid lines in
Figure~\ref{figs:rcumulative}, respectively) noting that the counts of
photometrically-selected RGB candidates are essentially complete at
all radii. Virtually all (44 of 45) of the AGB members are inside
$R_b$ despite the fact that we observed 10 stars from this region in
the CMD outside the break radius.  A KS test comparing the observed
radial distributions of the AGB and RGB kinematic members reveals a
low probability of $< 0.01$\% that the two are drawn from a common
parent distribution.

The right panel of Figure~\ref{figs:rcumulative} addresses the
possible segregation of AGB and blue-loop stars from the older stars
in Leo~I and each other.  The cumulative distribution of
photometrically-selected AGB stars exhibits a change in slope near
$R_b$.  We know from the kinematic sample of AGB stars that this is
where field contamination begins to dominate those counts.  The
blue-loop stars show a similar slope change, but at about $0.75 R_b$.
If this break in slope is due to contamination by blue disk and halo
stars, and background quasars and galaxies, the radial profile
suggests that the blue-loop stars may be even more centrally
concentrated than the AGB in Leo~I.  A KS test suggests that there is
a 0.1\%\ probability the radial profiles of the AGB and blue-loop
photometric candidates are drawn from the same parent distribution,
suggesting some difference in the distributions of these two
populations.

We conclude that the Leo~I stellar populations segregate by age such
that the younger populations (AGB and blue-loop stars) are
preferentially near the galaxy center relative to the older (RGB)
population.  Moreover, this segregation appears to occur at the break
radius, $R_b = 400$ arcsec, where we see kinematic segregation
(Figure~\ref{figs:deltav}) and (possibly) structural changes in Leo~I
(Figure~\ref{figs:leoishape}).

\subsection{A Heuristic Model for Leo~I}

We present here a descriptive model that aims to account for all the
features we have identified in our study of Leo~I, as well as many
long-standing enigmas of this galaxy.  We assume Leo~I is bound to the
Milky Way and is currently on a highly elliptical orbit;
Figure~\ref{figs:leoiorbit} shows a representative example.  The
period for the orbit in Figure~\ref{figs:leoiorbit} is about 5.5 Gyr;
one orbit (defined here as one complete cycle in $R$) is shown in
Figure~\ref{figs:leoirorbit}.  The other orbits that we consider in this
discussion have similar periods.  It is important to appreciate that
these orbital periods are sufficiently long that it may not be valid
to assume a static Milky Way potential or that Leo~I has remained
isolated from other objects in the halo.  We begin our
discussion considering only the effects of the last perigalactic
passage of Leo~I, about 1 Gyr ago for all cases considered here.

One aspect of an elliptical orbit that is well known (see King 1962;
Allen and Richstone 1988; Read et al. 2006b; Choi et al. 2007) is that
the instantaneous tidal radius of a dwarf galaxy in such an orbit
varies with Galactocentric distance.  When closest to the Galactic
Center, for example, Leo~I's true tidal radius (in the Roche sense)
will be smallest, while far from the Galactic Center, the tidal radius
will be large.  For any assumed orbit, we can calculate the tidal
radius crudely in a two-body approximation as $R_t = D (M_{Leo\ I}/2
M_{MW})^{(1/3)}$ (we assume $M_{MW} >> M_{Leo\ I}$; King 1962).  We
can then ask, at what perigalactic distance does the minimum tidal
radius, $R_{t,min}$, equal the kinematic break radius, $R_b$, that we
identified in Section 4.3.3?  Figure~\ref{figs:tidalperi} is a plot of
$R_{t,min}$ as a function of perigalacticon for orbits of Leo~I in a
logarithmic Milky Way potential with total mass (to the Virial radius)
of $10^{12} M_\odot$ for a range of assumed Leo~I mass, and for
various assumed present-day Galactocentric tangential velocities for
Leo~I. Further details of these orbits are listed in Table~8.  Our
assumption that Leo~I is bound to the Milky Way implies that the
galaxy's present tangential velocity is less than about 100 km/s, so
only results from models with $v_{tan} \leq 100$ km/s (Table 8) are
plotted in Figure~\ref{figs:tidalperi}.  The horizontal dashed line in
Figure~\ref{figs:tidalperi} corresponds to $R_b = 500$ pc, equal to
the physical size of the observed break radius ($R_b = 400$ arcsec) at
the current distance of Leo~I.

For $M_{Leo\ I} = 5 \times 10^8 M_\odot$, we find that Leo~I had to
pass within about 4 kpc from the Galactic Center for $R_{t,min} =
R_b$.  Lower masses imply larger perigalactica, up to $\sim 20$ kpc
for $M_{Leo\ I} = 2 \times 10^7 M_\odot$.  The latter mass, the lowest
considered in Figure~\ref{figs:tidalperi}, corresponds to a global
mass-to-light ratio of $\sim 4$ for Leo~I and is reasonable for a
purely baryonic case.  For a lower limit of $M/L = 0.8$ (appropriate
for the comparatively young central populations of Leo~I; see M98s),
we estimate that perigalacticon may have been as large as about 35 kpc.

One implication of this scenario is that many stars initially bound to
Leo~I prior to perigalactic passage, would have found themselves
outside the galaxy's tidal radius as it passed by the Milky Way.
Stars with radially-outbound orbits relative to the center of the
dwarf would have begun to drift away at a relative speed comparable to
the internal velocity dispersion, and in the process initiating the
formation of tidal arms (Piatek and Pryor 1995; Oh et al. 1995; S07).
Over the time interval, $t_p$, since perigalacticon, stars will drift
a distance of order $\sigma t_p$ from the center of Leo~I, or about
10-30 kpc for $\sigma = 10$ km/s for the orbits listed in Table~8.  As
Leo~I continues to orbit away from the Milky Way, its tidal radius
grows so that, at present, it is about 20 kpc.  Consequently, some of
the stars that were formally unbound at perigalacticon will be
recaptured, but because the stars on rapid, radial orbits remain
unbound, the velocity distribution will become progressively more
tangentially anisotropic in the outer parts of the galaxy.

The key point is that for a satellite in an elliptical orbit,
variations in $R_t$ will preferentially affect the kinematics of the
outermost stars of the system, while leaving the kinematics internal
to this radius comparatively unaffected (this just reiterates the
conclusions of Piatek and Pryor 1995; Oh et al. 1995).  We contend
that our observation of a radius at which the internal kinematics of
Leo~I change suddenly may represent this incursion of $R_t$ into the
main body of Leo~I.  This behavior is reflected in some of the models
of Mayer et al. (2001a,b; 2005) though these authors did not explore
orbits quite as eccentric as we are considering for Leo~I.

It is reasonable to suppose that this simple model can plausibly
account for the fairly mild structural anomalies we see in Leo~I,
including the small shift in its photocenter as a function of surface
brightness and the elongation of the distribution of kinematic members
at large radii.  The timing of the last perigalactic passage is also
very similar to the age of the end of the last prolonged burst in the
star-formation history of Leo~I (Gallart et al. 1999b; Hernandez et
al. 2000; Dolphin 2002), while the spatial segregation of the stellar
populations suggest that this event was largely confined to the inner
regions of the galaxy.  Hybrid $n$-body/hydro models (such as Mayer et
al. 2005) predict that tidal stirring can result in a strong gaseous
inflow in dwarf systems and centralized, bursty star formation,
consistent with what we see in Leo~I.  The fact that population and/or
chemical gradients are commonly seen in other local dwarfs (Harbeck et
al. 2001; Tolstoy et al. 2004; Battaglia et al. 2006) suggests that
tidal stirring could be a common process.

If we adopt $M/L \geq 0.8$ for Leo~I and retain our interpretation of
$R_b$ as $R_{t,min}$, then $R_{apo} \leq 400$ kpc for any reasonable
Leo~I mass and present-day tangential velocity.  Defining the orbital
eccentricity as $e = (1 - R_{peri}/R_{apo})/(1 + R_{peri}/R_{apo})$
for perigalactic and apogalactic distances $R_{peri}$ and $R_{apo}$,
respectively, we conclude from the results in Table 8 and
Figure~\ref{figs:tidalperi} that $e \geq 0.74$ for $R_{apo} > 255$
kpc, the current distance of Leo~I.  For $M/L = 20$, $R_{peri} = 9$
kpc and $e \geq 0.93$.  To the extent that equilibrium models are
still valid near the core of Leo~I (Read et al. 2006b; Klimentowsky et
al. 2007), we can carry out a classic `core fitting' analysis
(Richstone and Tremaine 1986; M98) using the dispersion of only the
inner subsample of Leo~I members.  From this we derive $M_{Leo\ I} =
3.0 \times 10^7 M_\odot \times (\sigma_0/9.2)^2$, and infer $R_{peri}
\sim 18$ kpc (Figure~\ref{figs:tidalperi}) and $e \geq 0.87$.  These
large eccentricities are consistent with the lack of strong spatial
distortions in Leo~I (Section~4.3.2; IH95) since it implies we are
looking almost directly along the projected orbital path of the
galaxy.  If our interpretation of the fundamental physical origin of
$R_b$ is correct, it is difficult to avoid the conclusion that Leo~I
passed very close to the Galactic center about 1 Gyr ago, regardless
of its dark matter content.

This tidal history of Leo~I implies that stars with positive
velocities relative to the center of Leo~I are in its leading
(western) arm, while those with negative relative velocities are
trailing (to the east).  S07, whose model is broadly similar to the
one we propose, came to the opposite conclusion.  This appears to
reflect the fact that the S07 $n$-body models include the results of
{\it two} perigalactic passages.  Since we consider only the effects
of the last perigalactic passage in our description, we ignore stars
that may be projected onto Leo~I from portions of its orbit extremely
far ahead or behind the main body of the galaxy.  It may be that these
stars -- the ones from the perigalacticon some 7-9 Gyr ago -- are the
ones that contribute to the kinematic asymmetry S07 identify and
interpret in their dataset.

S07 adopt a static Milky Way potential during the entire timespan
between the present epoch and the last two perigalactica of Leo~I (7-9
Gyr ago).  It is conceivable that the Milky Way's gravitational potential may have
changed significantly over that timespan (Bullock et al. 2001; Taylor
and Babul 2004, 2005; Bell et al. 2006; though see Hammer et al. 2007)
undermining the reliability of models in which Leo~I orbits in a
static potential (e.g. S07). On the other hand, we have so far adopted the seemingly
{\it ad hoc} assumption that Leo~I has had only one perigalactic
passage with the Milky Way.  This implies that Leo~I was somehow injected into
its present orbit sometime between the time of 
its last perigalactic passage (about 1 Gyr ago) and the time of its
earlier putative perigalacticon 7-9 Gyr ago.  Is this plausible?

There are two aspects to this question: First, is such an orbital
change -- presumably the result of an interaction with a third body --
reasonably probable?  If so, could such an interaction alter Leo~I's
orbit significantly without destroying the galaxy?  For our purposes, 'significant'
implies any change that causes Leo~I to transition from an orbit with
$R_{peri} \geq 50$ kpc, to one that brings it in as close as 10 kpc or
so from the Galactic Center.  From the data in Table~8, we estimate
that the required change of orbital energy is as large as about 6\%.
In the impulse approximation (Binney and Tremaine 1987), the energy
change is approximately $\Delta E \sim G M_p/b$, for a perturber of
mass $M_p$ and an encounter impact parameters $b$.  Solving for $b$
and taking $\Delta E$ to be 6\%\ of the total energy of Leo~I for an
orbit that gives it $(v_r,v_t) = (180, 80)$ km/s at $R = 250$ kpc
(Table~8), we find that for $M_p = (10^8, 10^9, 10^{10}) M_\odot$, $b
\sim (0.15, 1.5, 15)$ kpc assuming that the energy change goes
entirely into altering Leo~I's orbit.  If we consider a population of
$N$ subhalos within the volume of the Milky Way's overall halo
($R_{MW} \sim 250$ kpc), then the instantaneous filling factor of
those subhalos is $f = N (R_{sh}/R_{MW})^3$.  For the
intermediate case above ($M_p = 10^9 M_\odot$ for which  $R_{sh} \sim b =
1.5$ kpc) and assuming $N = 100$, $f \sim 9 \times 10^{-6}$ implying a
very low probability of interaction.

But this calculation may be misleading.  In hierarchical models,
subhalos necessarily inhabit regions of comparatively high density
right from the start, so they invariably have considerably more
neighbors than in a uniform-density model.  Moreover, the interactions
may be 'slow' (relative velocities between subhalos comparable to the
circular velocities of individual subhalos), contrary to a basic
assumption of the impulse approximation where the relative velocity is
taken to be comparable to the (large) dispersion of the overall halo
in which the subhalos reside.  Indeed, subhalo interactions are a
common feature in CDM simulations of hierarchical structure formation
(see web sites for: The Center of Theoretical Physics, Univ. of
Zurich; The Center for Cosmological Physics, Univ. of Chicago; The
$n$-body Shop, Univ. of Washington, Seattle).  Qualitative inspection
of these simulations seem to reveal that subhalos with large outward
velocities become increasingly common, and that some of these cases
are due to interactions with other subhalos and not the most central,
parent halo.

Taylor and Babul (2004, 2005) explored this more quantitatively and
confirmed work by Tormen et al. (1998) and Knebe et al. (2004) that
showed that `significant' encounters between subhalos are indeed quite
common in systems forming hierarchically.  In about 5\%\ of the
subhalo encounters, these interactions are transformative, in the
sense that either or both halos disrupt or they merge together (Taylor
and Babul 2004, 2005).  But there is also a class of much weaker
interactions that alter orbits but not raise destructive tides ($x >
1$ in the nomenclature of Taylor and Babul, 2005, where $x$ is the
ratio $b/r_{c,p}$, and $r_{c,p}$ is the radius of the peak of the
rotation curve of a given subhalo).  These are much more common than
the transformative encounters, occurring at least once for 30-60\%\ of
all subhalos during the formation of a Milky Way-sized galaxy.  Most
of these encounters (about 70\%; see Figure 19 of Taylor and Babul,
2005) occur over the first half of the formation process of a massive
galaxy, compatible with our requirement that a third-body encounter
altered Leo~I's orbit up to 7-9 Gyr ago.  More recently, Sales et
al. (2007b) have confirmed these basic results from independent
hierarchical models.  They speculate that objects such as Leo~I and
some other odd Local Group galaxies (Cetus, Tucana) may have
experienced third-body encounters that could account for their unusual
orbital characteristics.

Zhao (1998) proposed a specific interaction between the Sgr dSph
galaxy and the Magellanic Clouds 2-3 Gyr ago to inject Sgr into its
current orbit.  This model provides an elegant solution to the puzzle
of the long-term survivability of Sgr (Zhao 1998; Bellazzini et
al. 2006), and is consistent with its detailed orbital characteristics
(Majewski et al. 2004).  Perhaps something similar has occurred to
Leo~I, boosting the star formation rate and initiating its
transformation into a spheroidal system (Mayer et al. 2001a,b) 7-9 Gyr
ago while simultaneously injecting it into its present orbit.  This
picture also addresses Leo~I's survival: two close encounters with the
Milky Way should have stripped most of its initial mass (Mayer et
al. 2001a,b; Taylor and Babul 2004, 2005; Read and Gilmore 2005),
making it difficult for the galaxy to have survived as relatively
unscathed as we observe today.

Detailed photometric studies of Leo~I reveal low-level star formation
for the first third of the galaxy's existence (Gallart et al. 1999b;
Hernandez et al. 2000; Dolphin 2002) when Leo~I may have resembled
dIrr system.  This was followed by an increase in star formation some
5-8 Gyr ago which ended, possibly after a final peak, about 1 Gyr
ago.  The event that first caused the star formation rate to rise
evidently did not clear the gas from Leo~I since star formation
continued after that epoch.  Nor did the event induce tidal features
that we can see today in the distribution of stars in and around Leo~I
or in its internal kinematics.  If it had, such features would by now
extend over a very long arc of Leo~I's present orbit and be seen as
obvious photometric depth that is not observed (Section 4.3.2; Held et
al. 2001).  In our model, we identify this event as an interaction
with another subhalo, specifically {\it not} with the Galactic Center.
We obviously require a better idea of the orbit of Leo~I before we can
hope to identify the third body that Leo~I may have scattered off of,
assuming of course that that body exists as an identifiable
entity today (Taylor and Babul 2004, 2005; Sales et al. 2007b).

The kinematic basis of the S07 model is their detection of a strong
asymmetry in the velocity distribution of Leo~I.  We do not confirm
this feature in our data.  The velocity distribution of all Leo~I
members (Figure~\ref{figs:velhist}; $N = 328$) has a skew of $0.08 \pm
0.14$ and a kurtosis of $-0.34 \pm 0.27$, both consistent with a
Gaussian distribution.  If we select stars from our sample that are
spatially distributed in the same manner as those in the S07 sample
(dashed histogram in Figure~\ref{figs:velhist}), we find no
significant skew or kurtosis ($0.10 \pm 0.21$ and $-0.23 \pm 0.41$,
respectively).  It is unclear to what extent the systematic errors of
the S07 velocities relative to our measurements (see Section~3.2)
contribute to their observation of an asymmetric velocity
distribution.  What is clear is that we see no such effect in our
larger sample of more precise velocities.  In the detailed $n$-body
models by S07, the asymmetry of the velocity distribution from their
model appears to result entirely from stars that have migrated along
the orbit since the penultimate perigalacticon (about 5-7 Gyr).  Based on
S07's reasoning, our data (Figure~\ref{figs:velhist}) argue that this
earlier perigalactic passage did not occur and adds weight to the idea
that Leo~I's orbit has evolved.

The extent of Leo~I along the line of sight offers another way of
distinguishing these models.  S07 predict that Leo~I should exhibit a
full depth of about 30\%, defined here as ${{\Delta D}\over{D}}$.  In
our description, this extent is much less, about 10-15\%, which we
estimate from the distance stars have traveled relative to the center
of Leo~I at a velocity of 10 km/s (characteristic of the internal
dispersion) since perigalacticon. Proper simulations are needed to
determine the tidal extent reliably, but we note that this rough
estimate agrees with distance range of stars extracted from Leo~I
during its last perigalactic passage in the S07 models.  Though the
horizontal branch of Leo~I is extended in luminosity, this seems to
result mostly from its unusual star-formation history and not a
distance spread (Gallart et al. 1999b).  As we summarized in Section
4.3.2, observations of RR~Lyr stars in Leo~I appear to rule out a
depth greater than about 15\%\ (Held et al. 2001), arguing against the
existence of significant tidal arms as predicted by S07.  Distance
tracers accurate to 5\% are needed if we hope to detect the much more
modest tidal arms hypothesized in our scenario.

We predict that Leo~I's proper motion should be from east to west in a
heliocentric reference frame; specific predictions are given in
Table~8.  These values differ significantly from the predicted proper
motion reported by S07 for several reasons.  First, S07 report their
prediction in units of marcsec/yr, though the values appear more
consistent with units of arcsec/yr.  Also, the predicted proper motion
in S07 is stated to be heliocentric, but the signs of the two
components are inconsistent with this and suggest that the values
reported by S07 correspond to a Galactostationary frame.  Note that
our predicted proper motions assume the orbital pole given by S07 and
rely on our interpretation that the leading side of Leo~I (west)
exhibits a net positive velocity relative to the systemic velocity of
the galaxy (see Figures~\ref{figs:gradient} and \ref{figs:xieta2}).

One final issue has to do with why we see streaming in the outer parts
of Leo~I (outside the break radius, $R_b = 400$ arcsec) while Koch et
al. (2007; K07) do not.  To explore this, we carried out simulations
in which the velocities of stars in our sample were assumed to have
(normal) errors 2.0 times as large as in Table 5, mimicking the mean
uncertainty of the K07 measurements.  For our full dataset, only 29\%\
of our simulations produce by chance a larger value of $\Delta v$ than
we observe (Figure~\ref{figs:deltav}).  When we select stars to
approximate the spatial distribution of stars in the K07 sample, 30\%\
of the simulations produce a stronger streaming signal by chance.
Interestingly, our data still reveal streaming outside $R_b \sim 400$
arcsec even when we double the velocity uncertainties or mimic the K07
spatial distribution.  It will be of useful to expand the size of the
Leo~I kinematic sample, particularly at and outside $R_b$, to explore
the nature of the streaming signal in Leo~I further.

\subsection{Halo Substructure?}

Figure~\ref{figs:rpa} reveals the presence of some stars with
velocities in the fairly narrow range $88$ to $105$ km/s. The mean
heliocentric velocity of this group (14 measurements of 6 stars) is
$95.8 \pm 2.4$ km/s with a `dispersion' of $5.8 \pm 1.9$ km/s.  These
stars appear uniformly distributed over the field and are
kinematically distinct from Leo~I.  The expected distribution of field
stars predicted by the Besancon Galaxy model (Robin et al. 2003)
toward Leo~I is shown in the Gaussian-smoothed histogram in the right
panel of Figure~\ref{figs:rpa}.  This model agrees reasonably well
with the distribution of velocities of other non-Leo~I stars plotted
in Figure~\ref{figs:rpa} and does very well in other dSph fields
(Walker et al. 2007b).  By counting stars in velocity intevals $< 85$
km/s and between 85-110 km/s in both our dataset and the model, we
find that these 6 stars represent a modest $\sim$2$\sigma$ excess
relative to this model, which predicts $1.2 \pm 0.2$ stars in this
velocity interval given the number of field stars we observe at lower
velocities.  Other kinematic studies based on relatively
high-resolution observations have identified possible evidence of cold
kinematic groups in other halo fields (Cote et al 1992; Ibata et
al. 1994; Odenkirchen et al. 2002; Mu\~noz et al. 2006).  Given the
complex and rich distribution of streams being discovered in
wide-field surveys (Belakurov et al. 2006; Grillmair and Dionatos
2006), serendipitous kinematic detection of streams may not be
surprising.  We may be seeing a similar feature near Leo~I, but the
statistics are obviously poor and further members of this putative
kinematic group need to be identified.

\section{Summary and Conclusions}

We have presented new kinematic results of stars located in and near
the Milky Way satellite dwarf spheroidal galaxy Leo~I.  Our sample
includes velocities of 328 likely Leo~I red giant members based on new
observations with the Hectochelle multi-object echelle spectrograph,
plus published kinematic results obtained by M98 obtained with HIRES
at the Keck Observatory.  These results are based on measurements of
spectra obtained around 5180\AA, a region virtually uncontaminated by
telluric emission or absorption features.

Repeat measurements of many stars in our sample allow us to estimate
the typical errors of the velocities of these stars to be 2.4 km/s.
Our results give a systemic heliocentric velocity for Leo~I of $282.9
\pm 0.5$ km/s, and a radial velocity dispersion for the full sample of
$9.2 \pm 0.4$ km/s, both in agreement with previous measurements.  The
large areal coverage and significant number of stars in the present
sample (see Figure~\ref{figs:rpa}) allow us to measure the radial
velocity dispersion profile to slightly beyond the formal King tidal
radius of Leo~I (IH95; Table 6).  As we find in other dSph systems
(Walker et al. 2007b), this profile is flat to large projected radius
(see Figures~\ref{figs:leoibin} and \ref{figs:leoibinfits}).

We have fit the dispersion profile to a variety of equilibrium
dynamical models.  The observed profile is strongly inconsistent with
an isotropic King model in which mass follows light, but can be fit
reasonably well with an isothermal sphere.  In this latter case, we
still infer that the mass distribution is much more extended than the
visible light.  We have also fit the dispersion profile two-component
Sersic+NFW model ({\L}okas 2002).  The isothermal model implies a mass
of $5.2 \pm 1.2 \times 10^7 M_\odot$ within a radius of $1040$ pc and
a central density of $\rho_0 = 0.23 \pm 0.04 M_\odot {\rm pc}^{-3}$
for a core radius equal to the King core radius.  The best fit to a
Sersic+NFW model gives a total mass of $7 \pm 1 \times 10^8 M_\odot$
to a virial radius of 18.3 kpc for $M_{tot}/M_{vis} = 129 \pm 45$, and
a tangentially anisotropic velocity distribution ($\beta = -1.5$).  An
isotropic Sersic+NFW model can be excluded at $> 95$\%\ confidence.
These results are summarized in Table 6 and illustrated in
Figure~\ref{figs:leoibinfits}.  All models that provide acceptable
fits to the dispersion profile of Leo~I demand that the DM profile is
much more extended than the visible matter.

One motivation to study Leo~I has been to determine the
characteristics of a plausibly isolated dark halo.  Ironically, our
observations reveal evidence that tidal effects may actually have
significantly affected the properties of the galaxy.  We find evidence
of a `break' radius, $R_b$, at about $R = 400$ arcsec (500 pc), where
the internal kinematics of Leo~I change from apparently isotropic
inside this radius, to a distribution consistent with rotation or
streaming along the major axis.  Monte-Carlo simulations reveal that
the statistical significance of the kinematic change is high
($> 97$\%).  We interpret this in the framework of a heuristic model in
which Leo~I passed very close to the center of the Milky Way about 1
Gyr ago.  The break radius corresponds to the instantaneous tidal
radius of Leo~I at perigalacticon (Table~8 and
Figure~\ref{figs:tidalperi}).

This simple model accounts for the observed kinematic and population
segregation in Leo~I, the mildly distorted structural properties of
the galaxy, the age and duration of the last prominent burst of star
formation, and the large outward radial velocity of Leo~I
relative to the Galactic Center.  The lack of gas today in the galaxy
presumably reflects the fact that the ISM was largely consumed by star
formation in the inner part of Leo~I (inside $R_b$), and stripped via
ram pressure near the Galactic disk outside that radius.  This is
consistent with our detection of population segregation in Leo~I, such
that younger stars predominate inside $R_b$, while older
populations are found at all radii.

Tidal arms would have formed in Leo~I during its perigalactic
passage(s).  Because of the high ellipticity of the galaxy's orbit and
its large distance, we predict that these arms are projected close to
the main body of the galaxy, and that they exhibit a full extent of
about 10-15\%\ the distance to Leo~I (25-35 kpc).  In this picture,
the leading (more distant) arm is associated with stars with positive
velocities relative to the center of Leo~I, corresponding to the west
side of the galaxy.  Existing observations of RR~Lyr stars in Leo~I
appear to rule out arms that extend 20-30\%\ of the distance to the
galaxy along the line of sight (Held et al. 2001).  Shoerter arms
cannot be excluded with existing observations, but could be detectable
with a distance indicator capable of distance resolution of $\sim$5\%,
such as dwarf Cepheids (Mateo et al. 1998b).  

The lack of long tidal arms is inconsistent with the simulations of
S07 who argue Leo~I has suffered at least two perigalactic passages in
its lifetime.  We speculate that Leo~I may have instead been injected
into its highly elliptical orbit via an interaction with a third body,
similar to a more specific model in which Sgr was injected into its
present orbit after interacting with the LMC (Zhao 1998).  Within the
context of hierarchical models, such an interaction for Leo~I is not
only possible, but probable (Taylor and Babul 2005, 2005; Sales et
al. 2007b).  This scattering event had to have occurred between about
2-9 Gyr ago to exclude as second close perigalactic passage of Leo~I.
The lack of asymmetry in the velocity distribution of our kinematic
sample of Leo~I members (Figure~\ref{figs:velhist}) is also consistent
with only one close perigalactic passage during Leo~I's
lifetime (S07).

It would be of interest to look for a similar effect in other
satellite systems.  Our kinematic observations of nearly 1000 stars in
Carina suggest that we may see evidence of a break radius in that
galaxy at comparable significance (97\% ; Walker et al. 2007c).  Other
dwarfs for which we have extensive kinematic data also show possible
break radii, but at lower significance than in Leo~I.  This may not be
entirely surprising.  Leo~I's exceptionally large radial velocity
makes the galaxy unique, and demands a highly elliptical orbit if it
is bound to the Milky Way (Byrd et al. 1994; Taylor et al. 2005).
Most other dSph systems appear to be on less extreme elliptical orbits
(Piatek et al. 2005, 2006, 2007; Dinescu et al. 2004), so the break
radius may not be as well-defined as in Leo~I.  Carina may have the
next most extreme orbital eccentricity after Leo~I ($e_{Car} = 0.67$;
Piatek et al. 2003).  It has long been discussed as a system with an
extended pseudo-stream of stars (Majewski et al. 2000; Mu\~noz et
al. 2006), and we may have already detected a break radius in its
kinematics.  Carina contains dwarf Cepheids (e.g. Mateo et al. 1998b)
which, if identified in greater numbers and over a larger fraction of
the galaxy, could be used to explore the galaxy's line-of-sight
extent.

Finally, it is worth noting here that the possibility that Leo~I has
been affected significantly by tides does not necessarily contradict
our conclusion from equilibrium models that the system is dominated by
dark matter.  For the larger masses we derive from the NFW models,
Leo~I would have had to pass within 10 kpc or so of the Galactic
Center.  Indeed, as long as we adopt Newtonian gravity, its survival
in this orbit {\it demands} the existence of dark matter.  If Leo~I
has no dark matter at all, then it could not have passed closer than
about 40 kpc of the Galactic Center and survive.  Such a large
perigalacticon may be problematic for models that account for the gas
loss via ram-pressure stripping and consumption in tidally-induced
star formation (Mayer et al. 2001a,b, 2005).  Measurements of precise
proper motions can help settle this issue, though Table 8 suggests it
may prove difficult to make fine distinctions within a broad range of
plausible Leo~I orbits.  In the meantime, our observations suggest
some obvious $n$-body simulations that could be designed to recreate
our detailed observations of the dispersion profile and break radius
in the enigmatic dwarf galaxy, Leo~I (Klimentowski et al. 2007).

\acknowledgements

We thank Carlton Pryor for calculating a range of orbits for Leo~I.
Nelson Caldwell, Gabor Furesz and John Roll helped immensely in
planning, scheduling and implementing our MMT observations, and we are
grateful for their efforts.  Andy Szentgyorgyi and Dan Fabricant
helped to address occasional problems with Hectochelle promptly and
expertly.  We are grateful to the entire Hectochelle team for their
impressive efforts in building and supporting this complex instrument.
At the telescope, we were expertly helped by Hectochelle robot
operators, Perry Berlind and Michael Calkins, and the MMT operators
Mike Alegria, John McAfee, and Alejandra Milone.  We thank them for
ensuring that our runs were successful and pleasureable.  We are
grateful to the referee for an insightful and thorough report.  This
work has been supported by NSF grants AST~02-06081 and AST~05-07453
(to MM) and AST~02-05790 and AST~05-07511 (to EO).

\clearpage

\begin{table}
\begin{center}
\caption{Log of Hectochelle Observations\tablenotemark{a}}
\vskip1em
\begin{tabular}{lccrccc}
\hline
     Configuration &  UT~Date & UT~Start/End &  \multicolumn{1}{c}{ET} &  \multicolumn{1}{c}{$N_{exp}$} 
& $\alpha_{J2000}$ & $\delta_{J2000}$ \\
                   &          &              &  \multicolumn{1}{c}{(sec)}    & & & \\
\hline
  SA57-n1   &     Mar 31, 2005   &   08:35/08:46     &      900   &    3  &   13:05:09.9 &   +30:06:32 \\
  HD~171232  &     Mar 31, 2005   &   12:05/12:11     &      180   &    3  &   18:32:35.9 &   +25:32:05 \\
  SA57-n3   &     Apr  2, 2005   &   07:34/07:40     &      600   &    2  &   13:05:09.9 &   +30:06:32 \\
  SA57-2006 &     Apr 19, 2006   &   08:12/08:28     &      900   &    2  &   13:05:09.9 &   +30:06:32 \\
  \\
  Leo~I/c1   &     Mar 31, 2005   &   03:03/07:18     &    16200   &    6  &   10:08:35.7 &   +12:16:49 \\
  Leo~I/c2   &     Apr  2, 2005   &   02:56/06:13     &    14400   &    4  &   10:08:15.5 &   +12:20:57 \\
  Leo~I/c3   &     Apr 19, 2006   &   04:57/07:45     &    10000   &    4  &   10:08:23.2 &   +12:21:32 \\
  Leo~I/c4   &     Apr 20, 2006   &   04:29/06:35     &     7500   &    3  &   10:08:40.0 &   +12:16:29 \\
  Leo~I/c5   &     Apr 24, 2006   &   03:34/05:52     &     8100   &    3  &   10:08:25.9 &   +12:18:36 \\
  Leo~I/c6   &     Mar 12, 2007   &   06:44/09:00     &     8100   &    3  &   10:08:24.5 &   +12:18:24 \\
  Leo~I/c7   &     Apr 22, 2007   &   03:53/06:20     &     8100   &    3  &   10:08:24.5 &   +12:18:24 \\
\hline
\end{tabular}

\tablenotetext{a}{ET is the total exposure time.  $N_{exp}$
is the number of individual exposures obtained for each target.}

\end{center}
\end{table}

\clearpage

\begin{table}
\begin{center}
\caption{Summary Standard-Star Observations\tablenotemark{a}}
\vskip1em

\begin{tabular}{lccccccccc}
\hline
\ \ \ Star\tablenotemark{b}   &  $v_{obs,n1}$  & $v_{obs,n3}$ &  $v_{obs,06}$ & $v_{obs,07}$ &  $v_h$   & \multicolumn{4}{c}{$\Delta v =  v_{obs} - v_h$} \\
                          &                &              &               &              &          &    n1   &   n3    &  06   &  07  \\
\hline
 W22942  &\hfill  --11.00 &\hfill  --14.23 &\hfill  --12.42 &       $\ldots$ &\hfill --16.35  & \hfill    5.35 &\hfill     2.12 &\hfill    3.93 &       $\ldots$ \\
 W23082  &\hfill  --17.51 &\hfill  --20.76 &       $\ldots$ &       $\ldots$ &\hfill --21.69  & \hfill    4.18 &\hfill     0.93 &      $\ldots$ &       $\ldots$ \\
 W23108  &\hfill  --10.91 &\hfill  --14.33 &\hfill  --12.25 &       $\ldots$ &\hfill --16.49  & \hfill    5.58 &\hfill     2.16 &\hfill    4.24 &       $\ldots$ \\
 W23131  &\hfill   --1.53 &\hfill   --4.80 &\hfill   --2.69 &       $\ldots$ &\hfill  --6.02  & \hfill    4.49 &\hfill     1.22 &\hfill    3.33 &       $\ldots$ \\
 W23833  &\hfill   --0.36 &\hfill   --1.93 &\hfill   --1.66 &       $\ldots$ &\hfill  --4.95  & \hfill    4.59 &\hfill     3.02 &\hfill    3.29 &       $\ldots$ \\
 W23870  &       $\ldots$ &\hfill  --11.60 &\hfill  --11.49 &       $\ldots$ &\hfill --14.89  &       $\ldots$ &\hfill     3.29 &\hfill    3.40 &       $\ldots$ \\
 W23961  &\hfill   --2.88 &\hfill   --4.42 &\hfill   --3.91 &       $\ldots$ &\hfill  --7.10  & \hfill    4.22 &\hfill     2.68 &\hfill    3.19 &       $\ldots$ \\
 W24128  &\hfill     8.16 &\hfill     4.32 &\hfill     5.93 &       $\ldots$ &\hfill    2.11  & \hfill    6.05 &\hfill     2.21 &\hfill    3.82 &       $\ldots$ \\
 W24226  &\hfill    16.58 &\hfill    14.86 &       $\ldots$ &       $\ldots$ &\hfill   11.62  & \hfill    4.96 &\hfill     3.24 &      $\ldots$ &       $\ldots$ \\
 W25209  &       $\ldots$ &       $\ldots$ &       $\ldots$ &\hfill --13.35  &\hfill  --16.73 &       $\ldots$ &       $\ldots$ &      $\ldots$ &\hfill     3.38 \\
 W33245  &\hfill  --16.36 &       $\ldots$ &\hfill  --17.68 &       $\ldots$ &\hfill --20.97  & \hfill    4.61 &       $\ldots$ &\hfill    3.29 &       $\ldots$ \\
 W50001  &       $\ldots$ &       $\ldots$ &       $\ldots$ &\hfill --68.71  &\hfill  --71.91 &       $\ldots$ &       $\ldots$ &      $\ldots$ &\hfill     3.20 \\
 W50325  &       $\ldots$ &       $\ldots$ &       $\ldots$ &\hfill   35.90  &\hfill    33.34 &       $\ldots$ &       $\ldots$ &      $\ldots$ &\hfill     2.56 \\
 W50747  &       $\ldots$ &       $\ldots$ &       $\ldots$ &\hfill   57.29  &\hfill    54.24 &       $\ldots$ &       $\ldots$ &      $\ldots$ &\hfill     3.05 \\
 W50806  &       $\ldots$ &       $\ldots$ &       $\ldots$ &\hfill --41.73  &\hfill  --45.37 &       $\ldots$ &       $\ldots$ &      $\ldots$ &\hfill     3.64 \\
 W50807  &       $\ldots$ &       $\ldots$ &       $\ldots$ &\hfill    2.36  &\hfill   --0.78 &       $\ldots$ &       $\ldots$ &      $\ldots$ &\hfill     3.14 \\
\\
Average\tablenotemark{c}  \hfill   &       &          &          &         &          &\hfill  4.89 &\hfill   2.22 &\hfill  3.56  &\hfill   3.16 \\
Corrected\tablenotemark{c}\hfill   &       &          &          &         &          &\hfill  1.49 &\hfill --1.18 &\hfill  0.16  &\hfill --0.24 \\
Stand. Dev.               \hfill   &       &          &          &         &          &\hfill  0.61 &\hfill   1.00 &\hfill  0.36  &\hfill   0.36 \\
\\
\hline
\end{tabular}
\tablenotetext{a}{The subscripts $n1$ and $n3$ refer to results
obtained on the first and third nights of our 2005 run.  The
subscripts `06' and `07' refer to SA57 observations during the 2006
and 2007 queue runs, respectively.  See Section 2 and Table 1 for
further details.}

\tablenotetext{b}{Star designations and heliocentric velocities,
$v_h$, are from Stefanik et al. (2006) for velocity standards near the
North Galactic Pole (SA57).}

\tablenotetext{c}{`Average' refers to the straight average $\Delta v$
for a given run/night.  `Corrected' refers to this average value minus
$3.4$ km/s to account for the zero-point offset we found for our
adopted template, HD171232 (see Section 3.1).}
\end{center}
\end{table}

\clearpage

\begin{table}
\begin{center}
\caption{Statistics of Leo I Hectochelle Fiber Configurations\tablenotemark{a}}
\vskip1em
\begin{tabular}{lccc}
\hline
     Configuration &  $N_{assigned}$ & $N_{2.8}$ & $N_{Leo I}$ \\
\hline
      c1/2005   &           109 & 82 & 54 \\
      c2/2005   & \phantom{0}99 & 78 & 58 \\
      c3/2006   &           105 & 87 & 71 \\
      c4/2006   &           115 & 65 & 51 \\
      c5/2006   &           107 & 68 & 57 \\
      c6/2007   &           114 & 72 & 63 \\
      c7/2007   &           100 & 91 & 86 \\
\\
   Total/2005   &           208 & 160 & 112 \\
   Total/2006   &           327 & 220 & 179 \\
   Total/2007   &           214 & 163 & 149 \\
     Total      &           749 & 543 & 440 \\
\\
   Efficiency/2005\tablenotemark{b}    &  $\ldots$ & 0.77  &  0.54 \\
   Efficiency/2006\tablenotemark{b}    &  $\ldots$ & 0.67  &  0.55 \\
   Efficiency/2007\tablenotemark{b}    &  $\ldots$ & 0.76  &  0.70 \\
   Total Efficiency\tablenotemark{b}   &  $\ldots$ & 0.72  &  0.59 \\
\hline
\\
\end{tabular}

\tablenotetext{a}{The counts in this table are, for each fiber
configuration, the total number of fibers assigned to a target
($N_{assigned}$), the total number of spectra that produce cross
correlations with $R_{TD} \geq 2.8$ ($N_{2.8}$), and the number of
Leo~I velocity members ($N_{Leo I}$, where membership is defined by
whether a star has a heliocentric velocity in the range $+250$ to
$+320$ km/s; Section 3.2 and Figure~\ref{figs:rpa}). Configuration
numbers are defined in Table 1.}

\tablenotetext{b}{Efficiencies are defined as $N_i/N_{assigned}$,
where $i$ refers to the various counts listed in this table.}

\end{center}
\end{table}

\clearpage

\begin{deluxetable}{rccccccccr}
\rotate
  \tablewidth{0pt}
  \tablecaption{Summary of Repeat Observations of Leo~I Targets\tablenotemark{a}\label{tab:repeats}}
  \tablehead{\colhead{ID} & 
    \colhead{$v_{h,1}$,\ $\sigma_1$} &
    \colhead{$v_{h,2}$,\ $\sigma_2$} &
    \colhead{$v_{h,3}$,\ $\sigma_3$} &
    \colhead{$v_{h,4}$,\ $\sigma_4$} &
    \colhead{$v_{h,5}$,\ $\sigma_5$} &
    \colhead{$v_{h,6}$,\ $\sigma_6$} &
    \colhead{$v_{h,7}$,\ $\sigma_7$} &
    \colhead{$v_{h,8}$,\ $\sigma_8$} &
    \colhead{$v_{avg}$,\ $\sigma_{avg}$} \\

  \colhead{} & 
  \colhead{(km/s)} &  
  \colhead{(km/s)} &  
  \colhead{(km/s)} &  
  \colhead{(km/s)} &  
  \colhead{(km/s)} &  
  \colhead{(km/s)} &  
  \colhead{(km/s)} &  
  \colhead{(km/s)} &  
  \colhead{(km/s)} }
\startdata
    5 &    286.8,   1.8 &    287.8,   1.7 &    288.6,   1.7 &    286.3,   1.7 &    $\ldots$     &    $\ldots$     &    $\ldots$     &    $\ldots$     &    287.4,   2.0 \\
   12 &    294.1,   1.8 &    $\ldots$     &    $\ldots$     &    298.9,   1.8 &    $\ldots$     &    $\ldots$     &    $\ldots$     &    295.3,   1.8 &    295.9,   2.8 \\
   16 &    $\ldots$     &    $\ldots$     &    $\ldots$     &    $\ldots$     &    284.7,   1.8 &    $\ldots$     &    $\ldots$     &    282.5,   1.8 &    283.6,   2.1 \\
   23 &    $\ldots$     &    281.4,   1.9 &    $\ldots$     &    281.1,   1.8 &    $\ldots$     &    $\ldots$     &    $\ldots$     &    $\ldots$     &    281.2,   1.9 \\
   25 &    $\ldots$     &    287.1,   1.7 &    284.6,   1.8 &    283.9,   1.9 &    $\ldots$     &    $\ldots$     &    $\ldots$     &    $\ldots$     &    285.2,   2.3 \\
   28 &    $\ldots$     &    $\ldots$     &    284.7,   1.8 &    282.9,   1.9 &    $\ldots$     &    $\ldots$     &    $\ldots$     &    284.2,   1.8 &    283.9,   2.0 \\
   31 &    287.4,   1.8 &    $\ldots$     &    291.1,   1.8 &    $\ldots$     &    $\ldots$     &    $\ldots$     &    $\ldots$     &    289.4,   1.8 &    289.3,   2.4 \\
   36 &    280.9,   2.0 &    280.4,   1.9 &    $\ldots$     &    282.2,   1.8 &    $\ldots$     &    $\ldots$     &    $\ldots$     &    282.6,   1.8 &    281.6,   2.1 \\
\enddata 
\tablenotetext{a}{The subscripts used in
the headings of columns 2-7 refer to the configurations listed in Table 1:
 1 = Leo~I/c1; 2 = Leo~I/c2; 3 = Leo~I/c3; 4 = Leo~I/c4; 5 =
Leo~I/c5; 6 = Leo~I/c6; 7 = Leo~I/c7; 8 = Keck velocities from M98.}
\end{deluxetable}

\clearpage

\begin{deluxetable}{rrrrccccclr}
  \tablewidth{0pt}
   \setlength{\tabcolsep}{0.065in}
   \tablecaption{Summary of Observations of Leo~I Targets\tablenotemark{a}\label{tab:alldata}}
\  \tablehead{\colhead{ID}\tablenotemark{b} & 
   \colhead{$R$\tablenotemark{c}} & 
   \colhead{$PA$\tablenotemark{c}} & 
   \colhead{$v_h$} & 
   \colhead{$\sigma_{v_h}$}  & 
   \colhead{$\alpha_{J2000}$} & 
   \colhead{$\delta_{J2000}$} &  
   \colhead{I}  &   
   \colhead{V--I}  &  
   \colhead{SRC\tablenotemark{d}} & 
   \colhead{M98} \\ 

   \colhead{} & 
   \colhead{($''$)} & 
   \colhead{(deg)} & 
   \colhead{(km/s)} & 
   \colhead{(km/s)} & 
   \colhead{} & 
   \colhead{} & 
   \colhead{} & 
   \colhead{} & 
   \colhead{} & 
   \colhead{No.\tablenotemark{e}} } \\ 

\startdata
   1 &      2.3 &    77.2 &   283.3 &    1.6 &  10:08:27.15 &  12:18:30.5 &   18.34 &   1.27 & 7        &     \\ 
   2 &      6.6 &   358.7 &   294.9 &    1.9 &  10:08:26.99 &  12:18:36.6 &   18.34 &   1.28 & 6        &     \\ 
   3 &     11.6 &   329.6 &   274.1 &    1.8 &  10:08:26.60 &  12:18:40.0 &   17.99 &   1.43 & 8        & 13  \\ 
   4 &     28.4 &    48.0 &   281.4 &    1.8 &  10:08:28.44 &  12:18:49.0 &   17.89 &   1.40 & 8        & 12  \\ 
   5 &     32.9 &    86.7 &   287.4 &    2.0 &  10:08:29.24 &  12:18:31.9 &   17.51 &   1.58 & 1234     &     \\ 
   6 &     35.4 &   158.2 &   264.8 &    1.9 &  10:08:27.90 &  12:17:57.1 &   18.19 &   1.50 & 6        &     \\ 
   7 &     36.7 &   160.1 &   287.8 &    2.0 &  10:08:27.85 &  12:17:55.5 &   17.89 &   1.63 & 7        &     \\ 
   8 &     43.3 &   143.7 &   288.4 &    2.0 &  10:08:28.75 &  12:17:55.1 &   18.55 &   1.26 & 5        &     \\
\enddata

\tablenotetext{a}{Stars with heliocentric velocities (column 4) in the
range 250-320 km/s are treated as members in our analysis.  Star 30
in this table was not identified in our new photometry because if fell
in a gap between 90Prime CCDs.  Its equatorial coordinates (in {\bf
boldface}) are taken directly from M98 and may be offset by 2-3 arcsec
relative to the coordinates of all other stars in the Table.}

\tablenotetext{b}{Star identification number, ordered by increasing
radial distance from the center of Leo~I ($R$, column 2).  IDs in {\it
italics} are possible `false positives' where we may have obtained the
velocity of the sky at the time of observation rather than an actual
target star (see Figure~\ref{figs:rpa} and Section 3.2). Star IDs in
{\bf boldface} denote possible binary stars based on velocity
variability (see Figure~\ref{figs:chi2} and Sections 3.4 and 4.1).}

\tablenotetext{c}{$R$ (distance from center in arcsec) and $PA$
(position angle relative to center in degrees with $PA$(North) = 0 deg
and $PA$(East) = 90 deg) are defined relative to the adopted center of
Leo~I: $(\alpha_{c,J2000},delta_{c,2000}) = $ (10:08:27,+12:18:30)
(Mateo 1998).  This center position differs somewhat from the one used
by M98 and from the center coordinate given by IH95.}

\tablenotetext{d}{The sources of the velocities used to estimate the
mean velocity listed in this table are coded as: 1 = Leo~I/c1, 2 =
Leo~I/c2; 3 = Leo~I/c3; 4 = Leo~I/c4; 5 = Leo~I/c5; 6 = Leo~I/c6; 7 =
Leo~I/c7; 8 = M98.  See Table 1 for details about each of these
configurations.}

\tablenotetext{e}{Star number from M98.}
\end{deluxetable}

\clearpage

\begin{deluxetable}{lccl}
\tablecolumns{4}
\tablewidth{0pt}
\tablecaption{Summary of Adopted and Derived Properties of Leo~I}
\tablehead{
\multicolumn{1}{c}{Quantity} & Value & \multicolumn{1}{c}{Units} & \multicolumn{1}{c}{Reference\tablenotemark{a}}
}
\startdata

\cutinhead{Adopted Parameters\tablenotemark{b}}\\

$\alpha_{2000}$ (center) & \phantom{$+$}10 08 27.0        &                      & M98R \\
$\delta_{2000}$ (center) & $+$12 18 30\phantom{.0}        &                      & M98R \\
Distance                 & $255 \pm 23$                   & kpc                  & C99; H01; B04\\
$R_{core}$               & $3.3 \pm 0.3$                  & arcmin               & IH95 \\
$R_{tidal}$              & $12.6 \pm 1.5$                 & arcmin               & IH95\\
$R_{core}$               & $245 \pm 35$                   & pc                   & IH95+Dist\\
$R_{tidal}$              & $935 \pm 140$                  & pc                   & IH95+Dist\\
\enddata

\tablenotetext{a}{References: M98R = Mateo 1998; C99 = Caputo et
al.  1999; H01 = Held et al. 2001; B04 = Bellazzini et al. 2004; IH95
= Irwin and Hatzidimitriou 1995; IH95+Dist refers to IH95 results
adjusted for the distance used in this paper; M98 = Mateo et al. 1998.
Figures and Section numbers refer to this paper.}

\tablenotetext{b}{$\Sigma_0$ and $S_0$ are the central surface
brightness of Leo~I in astronomical and physical units, respectively;
$I_0$ is the central luminosity density, and $\rho_{0,baryons}$ is the
central mass density for the assumed baryonic mass-to-light ratio,
$(M/L)_{V,baryons}$.  The quantities $S_0$ and $I_0$ are calculated as
described in M98.}

\tablenotetext{c}{`GS' refers to the Galactostationary reference frame
(see Section 4.3.1) based on an LSR motion of 220 km/s toward $(l,b) =
(90,0)$, and a solar peculiar motion relative to the LSR of 16.6 km/s
toward $(l,b) = (53,25)$.  The position angles ($PA$) and isophotal
ellipticities ($e$) are defined in Figure~\ref{figs:leoishape}. See
Section 4.3.1 for details.}

\tablenotetext{d}{$\rho_0$ is the total central mass density (DM +
baryons).}

\tablenotetext{e}{The Sersic index, $m_S$, and Sersic radius,
$r_{Sersic}$ are defined as $I_{Sersic} = I_0 \exp \left[
-(r/r_{Sersic})^{1/m_S} \right]$.  Values of $r_{Sersic}$ and $m_S$
were derived from fitting our stellar density profile for Leo~I
(Figure~\ref{figs:leoishape}).  The anisotropy parameter, $\beta$ is
defined such that $\beta = 0$ corresponds to an isotropic
distribution, $\beta = +1$ is a fully radially anisotropic
distribution, and $\beta = -\infty$ is a fully tangentially
anisotropic distribution (Binney and Tremaine 1987). Following
{\L}okas, we set the virial radius, $R_{vir}$, such that
$\rho_{NFW}(R_{vir}) = 200 \rho_{crit}$.}

\end{deluxetable}

\clearpage

\begin{table}
\begin{center}
\caption{Population Segregation in Leo~I}
\vskip1em
\begin{tabular}{ccccccc}
\hline 
\hline
\\ 
 Region\tablenotemark{a} & Region Name & $R$ (arcsec) & $N$ & $N_{vel}$ & $N_{mem}$ & $N_{nonmem}$ \\
\\
\hline
\\
   1 & RGB & $R \leq 400$         & 732 & 227 & 209 & 18 \\
   1 &     & $400 < R \leq 760$ & \phantom{0}91 &  \phantom{0}68 & \phantom{0}52 & 16 \\   
\\
   2 & AGB & $R \leq 400$         &  72 &  47 & 42 &  5 \\
   2 &     & $400 < R \leq 760$ &  18 &  10 &  \phantom{0}1 &  9 \\   
\\
   3 & RGB-Control & $R \leq 400$         &  25 & $\ldots$ & $\ldots$ & $\ldots$ \\
   3 &             & $400 < R \leq 760$ &  52 & $\ldots$ & $\ldots$ & $\ldots$ \\   
\\
   4 & Blue-Loop & $R \leq 400$         &  42 & $\ldots$ & $\ldots$ & $\ldots$ \\
   4 &           & $400 < R \leq 760$ &   \phantom{0}6 & $\ldots$ & $\ldots$ & $\ldots$ \\   
\\
\hline
\hline
\\
\end{tabular}
\tablenotetext{a}{The column labeled `Region' corresponds to the
regions of the Leo~I CMD identified in Figure~\ref{figs:cmd2}; `Region
Name' identifies the principal component in each region: RGB = Red
Giant Branch stars; AGB = Asymptotic Giant Branch stars; RGB-Control =
Field stars in a similar magnitude range as the RGB stars; Blue-Loop =
intermediate-age core He burning stars in a blue phase of their
post-main sequence evolution (Dohm-Palmer and Skillman 2002). The
other columns in this table are $R$, the radial extent fro the center
of Leo~I, $N$ the total number of stars in a given region, $N_{vel}$
the number of stars in a region with measured velocities, and
$N_{mem}$, $N_{nonmem}$, the number of kinematic members and
non-members in a given region, respectively.  The sole AGB star with
$R > 400$ arcsec is located at $R = 497$ arcsec (star 324 in Table
5).}
\end{center}
\end{table}

\newpage

\begin{table}
\begin{center}
\caption{Representative Orbital Parameters for Leo~I\tablenotemark{a}}
\vskip1em

\begin{tabular}{ccccccccc}
\hline
\hline
\\
 $V_{GS,rad}$ & $V_{GS,tan}$ & $R_{peri}$ & $R_{apo}$ & $T_{peri}$ & $P_{orb}$ & $e$ & $\mu_\alpha$ & $\mu_\delta$ \\
  (km/s)        &     (km/s)     &   (kpc)     &   (Gyr)      &   (kpc)    & &   (Gyr)    &   (marcsec/cent)  &  (marcsec/cent) \\ 
\\
\hline
\\
  181 &    \phantom{00}5 &     \phantom{0}2.5 &   399 &  0.91 &   \phantom{0}5.4 &  0.99 & \phantom{0}$-$4.1 &  $-$16.3  \\ 
  181 &   \phantom{0}13 &     \phantom{0}6.1 &   399 &  0.91 &   \phantom{0}5.4 &  0.97 & \phantom{0}$-$3.4 &  $-$15.9  \\ 
  180 &   \phantom{0}24 &    11.5 &   400 &  0.92 &   \phantom{0}5.4 &  0.94 & \phantom{0}$-$2.6 &  $-$15.4  \\ 
  180 &   \phantom{0}34 &    16.9 &   402 &  0.93 &   \phantom{0}5.5 &  0.92 & \phantom{0}$-$1.9 &  $-$15.0  \\ 
  180 &   \phantom{0}44 &    23.5 &   404 &  0.94 &   \phantom{0}5.6 &  0.89 & \phantom{0}$-$1.1 &  $-$14.6  \\ 
  180 &   \phantom{0}54 &    30.3 &   408 &  0.95 &   \phantom{0}5.7 &  0.86 & \phantom{0}$-$0.4 &  $-$14.1  \\ 
  177 &  207 &   159   &   656 &  0.80 &  10.2 &  0.61 &  +10.8 &  \phantom{0}$-$7.7  \\ 
\\
\hline
\hline
\\
\end{tabular}
\tablenotetext{a}{These results are derived for model orbits using the
Milky Way gravitational potential given by Johnston et al. (1995),
adopting an asymptotic halo circular velocity of 190 km/s.  The
results here were calculated for an orbit with an adopted pole of
$(l,b) = (130,0)$, slightly different than the value used to produce
the orbit shown in Figure~\ref{figs:leoiorbit}. The column headers
are: $V_{GS,rad}$ = Galactostationary radial velocity, $V_{GS,tan}$ =
Galactostationary tangential velocity, $R_{peri}$ = perigalactic
distance, $T_{peri}$ = time of last perigalactic passage, measured
backward in time from the present, $R_{apo}$ = apogalactic distance,
$P_{orb} =$ radial orbital period, and $e = (1 - (R_{peri}/R_{apo})/(1
+ (R_{peri}/R_{apo}))$ is the orbital eccentricity.  The columns
labeled $\mu_\alpha$ and $\mu_\delta$ are the observed heliocentric
proper motions (in RA and Dec, respectively) in units of
marcsec/century for the current epoch and assuming Leo~I is at its
present location and distance.  These orbital parameters are
essentially independent of the mass of Leo~I for the range of masses
explored in Figure~\ref{figs:tidalperi}.}
\end{center}
\end{table}

\clearpage

\clearpage

\clearpage

\clearpage

\begin{figure}
  \begin{center}
    \plotone{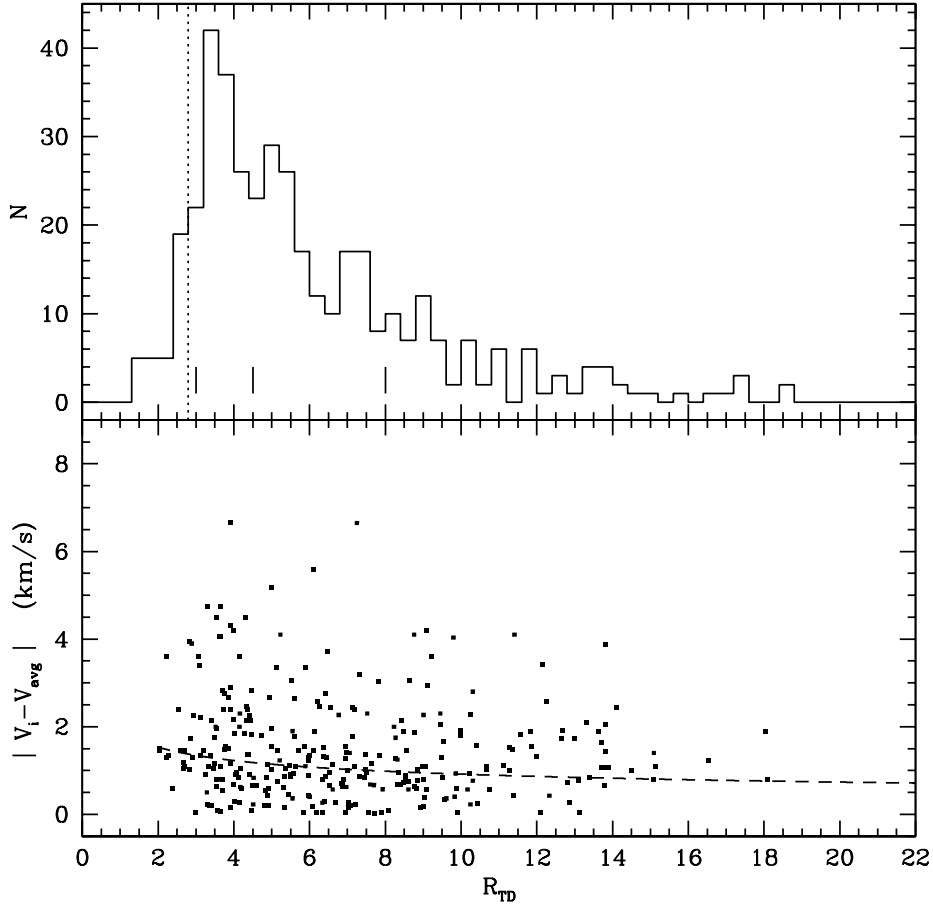}
      \caption{\label{figs:ccf2} {\it Top panel:} The distribution of
      the Tonry-Davis $R_{TD}$ values for all stellar spectra in the
      Leo~I sample (members and non-members included).  The small
      tickmarks near the lower x-axis indicate the $R_{TD}$ values of
      the three profiles shown in Figure~\ref{figs:ccf1}. The median
      value of $R_{TD}$ is 4.9. {\it Lower panel:} A plot of the
      absolute value of the velocity differences of individual
      measurements ($V_i$) and the weighted mean velocities
      ($V_{avg}$) for all stars in our sample with multiple velocity
      measurements.  The dashed line is our error model, $\sigma_i =
      \sigma(R_{TD,i}: \alpha, x, \sigma_0)$ (see Section 3.2).  The
      weights used to calculate $V_{avg}$ are $w_i =
      1/\sigma_{i,fx}^2$, where $\sigma_{i,fx}$ is the error estimate
      on $V_i$ from the IRAF cross-correlation routine {\bf fxcor}.
      The weighted means for multiply-observed stars listed in Tables
      4 and 5 are based on the values of $\sigma_i$ from the error
      model, so the means listed in the tables differ a bit from the
      values used here.}
  \end{center}
\end{figure}

\clearpage

\begin{figure}
  \begin{center}
    \plotone{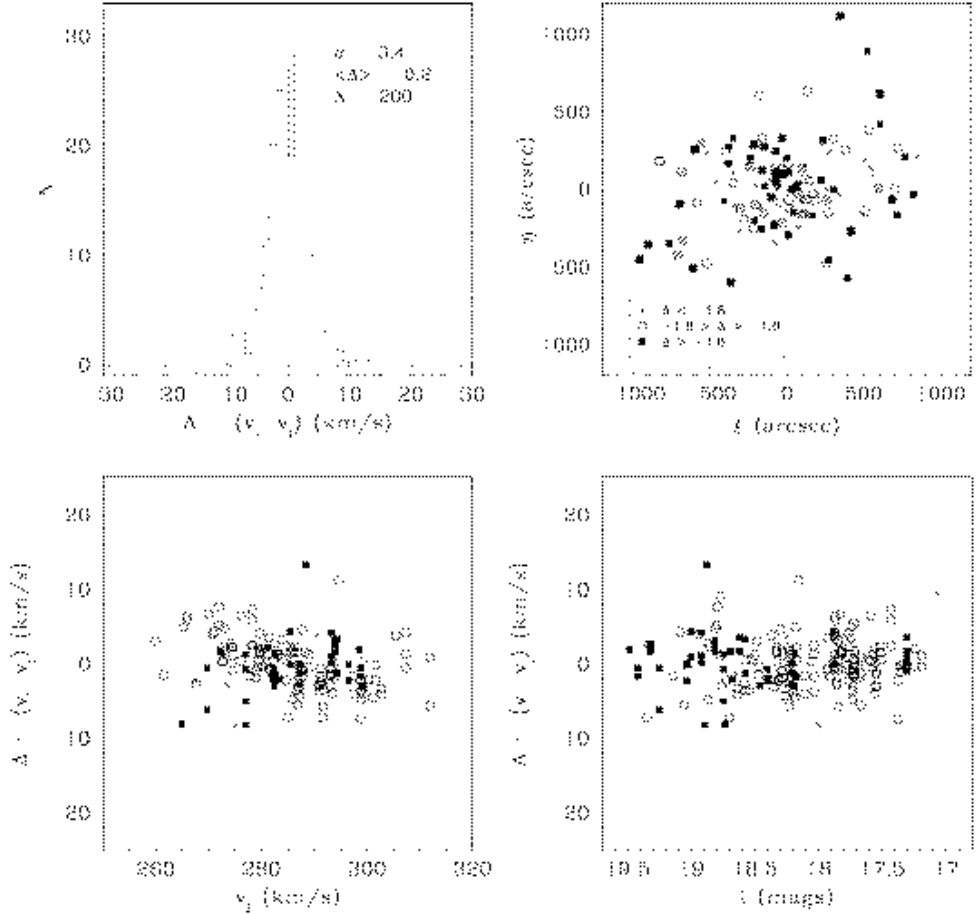}
      \caption{\label{figs:mmtdiffs} {\it Upper left: } The histogram
      of the velocity differences, $\Delta = v_i - v_j$ for stars with
      multiple velocity measurements.  The indices $i$ and $j$ refer
      to the 2005, 2006 or 2007 Hectochelle data (see Table 1), with
      the chronologically earlier observation labeled with index $j$.
      The standard deviation, $\sigma$, mean offset, $\langle
      \Delta\rangle$, and total number of $\Delta$ values in the
      histogram are shown.  {\it Upper right:} $\Delta$ as a function
      of standard coordinates, $\xi,\eta$, with different symbols
      denoting different ranges in $\Delta$ as noted in the legend.
      There is no evident trend of $\Delta$ with location on the sky
      apparent in this plot.  {\it Lower left:} The distribution of
      $\Delta = v_i - v_j$ as a function of $v_j$.  The crosses
      correspond to the case where $v_j = v_{2005}$; the open circles
      are for $v_j = v_{2006}$, and the filled squares are for $v_j =
      v_{2007}$.  This panel reveals no statistically significant
      dependence of $\Delta$ on fiber configuration or run.  {\it
      Lower right:} The distribution of $\Delta$ and $I$ magnitude.
      The symbols are the same as in the lower left plot.  We find no
      dependence of $\Delta$ with $I$ magnitude.}
  \end{center}
\end{figure}

\clearpage

\begin{figure}
  \epsscale{0.8}
  \begin{center}
    \plotone{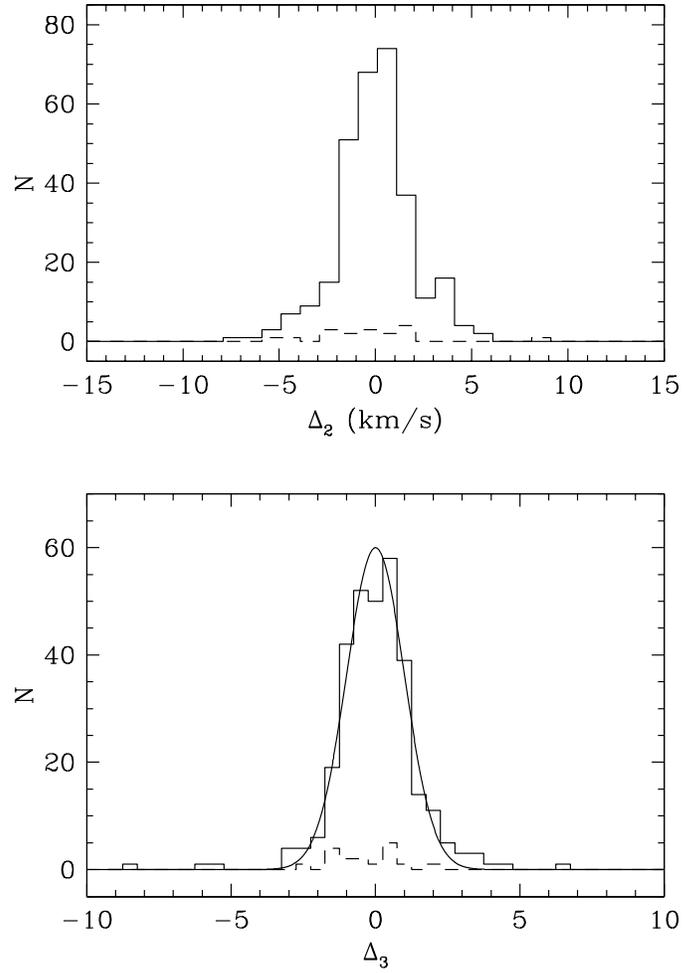}
    \caption{\label{figs:hists} {\it Upper plot:} Histogram of velocity
    differences relative to the mean velocity, $\Delta_2 \equiv v_i -
    \langle v\rangle$ ({\it not} the same $\Delta$ plotted in
    Figure~\ref{figs:mmtdiffs}), for all stars with repeat
    measurements within the MMT dataset only (solid line) and for
    stars with both Keck and MMT velocity measures (dashed line).
    {\it Lower plot:} The same as the upper plot except for the statistic
    $\Delta_3 \equiv \Delta v/\langle \sigma\rangle$.  The curve is a
    Gaussian with unit $\sigma$.}
  \end{center}
\end{figure}

\clearpage

\begin{figure}
  \begin{center}
    \plotone{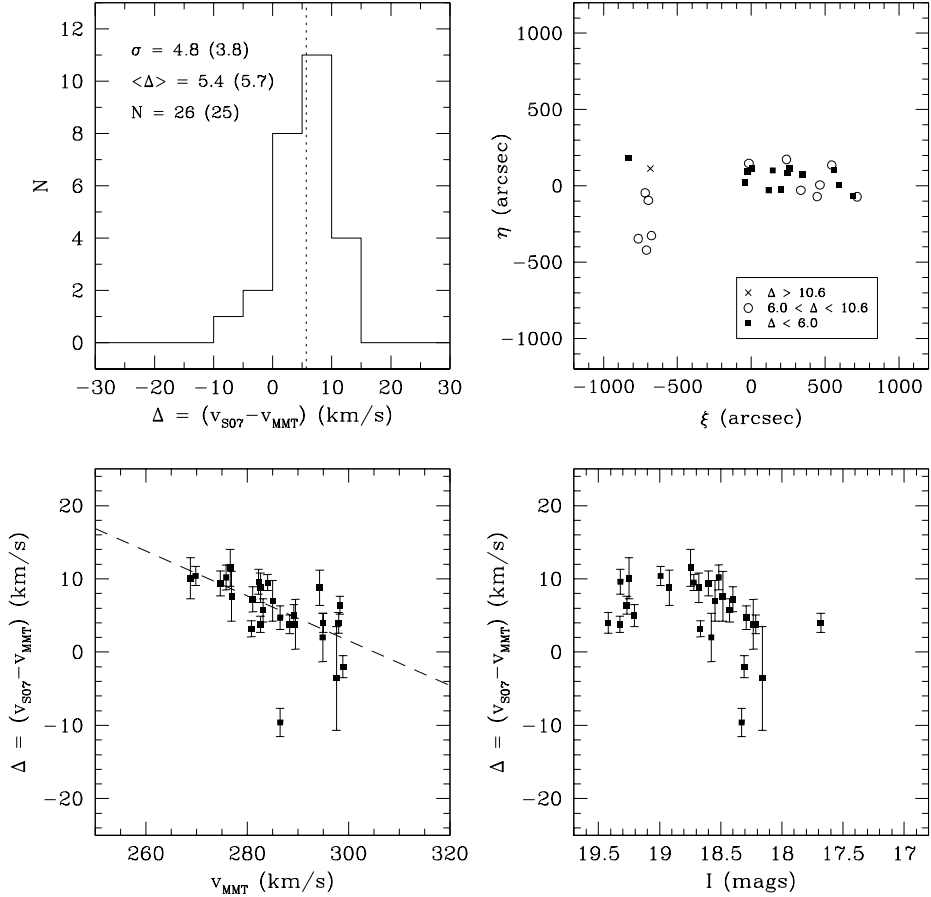}
      \caption{\label{figs:sohndiffs} {\it Upper left:} A histogram of
      the velocity differences $\Delta = v_{S07} - v_{MMT}$ where
      'S07' refers to the results from Sohn et al. (2007).  The
      standard deviation, $\sigma$, mean offset $\langle \Delta
      \rangle$, and number of $\Delta$ values are shown in the
      panel. Values in parentheses are for the sample that excludes
      the outlying point at $\Delta \sim -9.5$.  {\it Upper right:}
      The dependence of $\Delta$ on standard coordinates, $\xi, \eta$.
      We find no significant trend of position with $\Delta$. {\it
      Lower left:} The distribution of $\Delta$ vs $v_{MMT}$.  A
      strong trend is apparent.  The line corresponds to a
      least-squares fit to the data (excluding the one outlier at
      $\Delta \sim -10$) of the form $\Delta = -0.31(0.06) v_{MMT} +
      93.4 (18)$ (1-$\sigma$ parameter errors are in parentheses).
      The correlation coefficient for this fit is $|R| = 0.72$ for
      $N=25$, corresponding to a probability of 0.02\%\ of obtaining a
      better linear fit by chance.  {\it Lower right:} A plot of
      $\Delta$ vs. $I$-band magnitude.  A least-squares linear fit
      suggests that there is no significant correlation between
      $\Delta$ and $I$.}
  \end{center}
\end{figure}
 
\clearpage

\begin{figure}
  \begin{center}
    \plotone{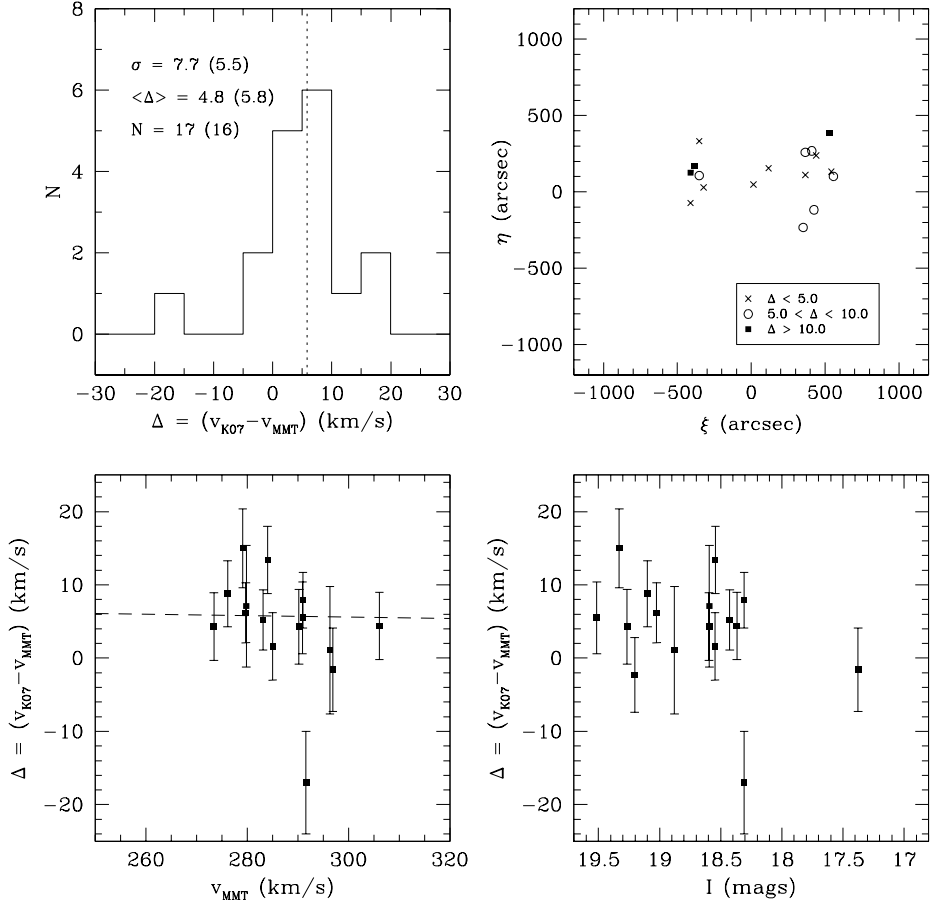}
      \caption{\label{figs:kochdiffs} {\it Upper left:} A histogram of
      the velocity differences $\Delta = v_{K07} - v_{MMT}$ where
      'K07' refers to the results from Koch et al. (2007).  The
      standard deviation, $\sigma$, mean offset $\langle \Delta
      \rangle$ (dotted line), and number of $\Delta$ values are
      plotted in the panel.  Values in parentheses exclude the outlier
      at $\Delta \sim -18$. {\it Upper right:} The dependence of
      $\Delta$ on standard coordinates, $\xi, \eta$.  We find no
      significant trend of position with $\Delta$. {\it Lower left:}
      The distribution of $\Delta$ vs $V_{MMT}$.  The dashed line is a
      least-squares linear fit to the data (excluding the one outlier
      at $\Delta \sim -17$ but is consistent with $\Delta = $\ {\it
      constant}\ $= 4.8$ km/s.  {\it Lower right:} A plot of $\Delta$
      vs. $I$-band magnitude.  Apart from the mean offset, we see no
      significant trend in this plot.}
  \end{center}
\end{figure}

\clearpage

\begin{figure}
  \begin{center}
    \plotone{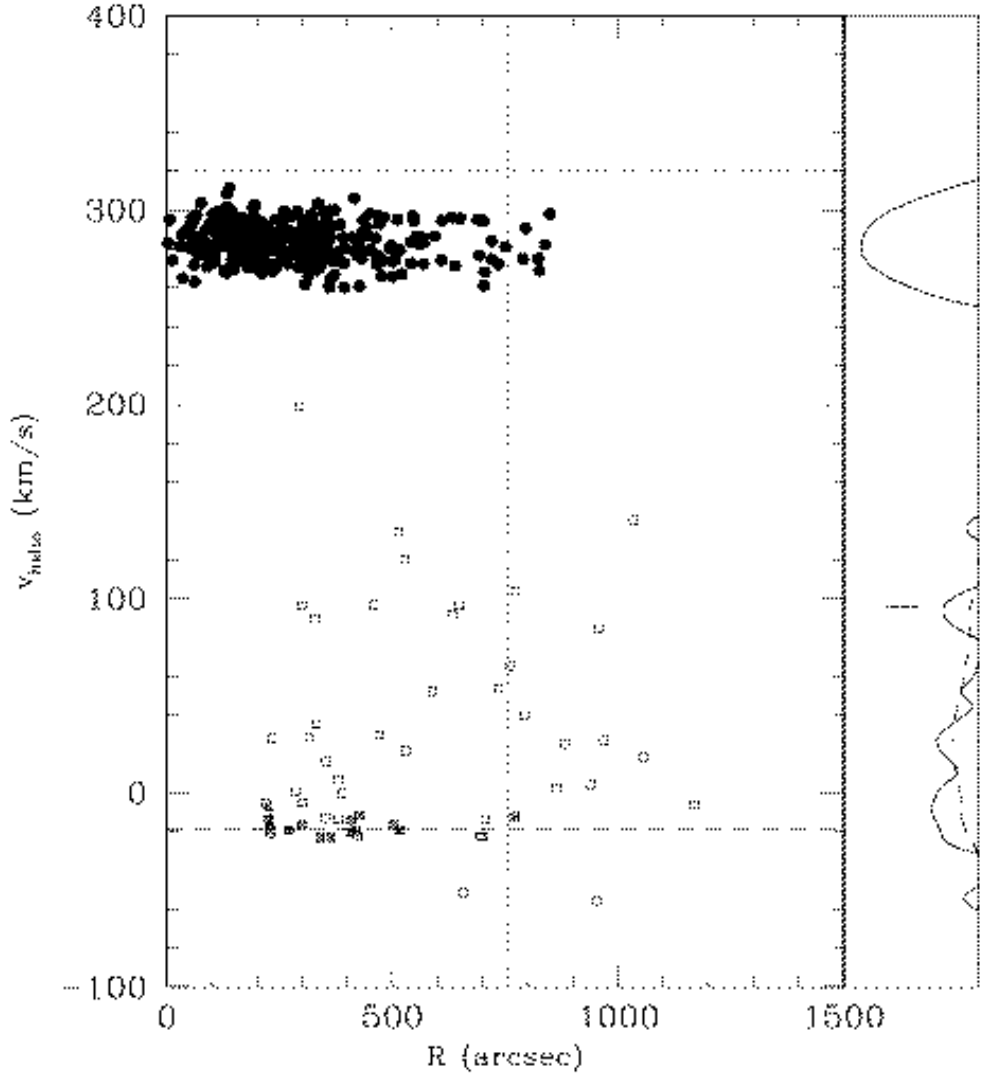}
      \caption{\label{figs:rpa} The radial distribution of
      heliocentric radial velocities in Leo~I for all targets with
      $R_{td} \geq 2.8$ (Table 5). The filled circles represent stars
      we take to be Leo~I kinematic members.  The dotted lines
      illustrate a crude Leo~I selection region between $v_{helio}$ =
      250 and 320 km/s, but it is clear that the identification of
      highly probable Leo~I members is not at all ambiguous in this
      sample.  The open squares denote likely non-members. The dashed
      horizontal line shows the typical heliocentric offset applied to
      the spectra; spectra from which we measured reflected
      sun/moonlight would have velocities near this line.  The open
      squares with crosses denoting 'false positives', cases where we
      have likely measured the sky velocity ($-29 \leq v_{helio} \leq
      -11$ km/s and $R_{td} \leq 4.2$) rather than the velocity of an
      actual star in non-sky fibers (see Section 3.2; Table 5).  The
      vertical dotted line marks the King tidal radius of Leo~I from
      IH95 (756 arcsec; Table 6). At the right is the logarithmic
      histogram of the velocity distribution where each star denoted
      as a solid symbol is represented by a Gaussian of unit area with
      $\sigma = 3.0$ km/s.  The dashed curve in the histogram is the
      predicted field star distribution from the Besancon Galaxy model
      (Robin et al. 2003).  A marginal ($2\sigma$) excess of stars
      relative to the model at $v_{helio} \sim 96$ km/s is marked (see
      Section 4.5).}
  \end{center}
\end{figure}

\clearpage

\begin{figure}
  \begin{center}
    \plotone{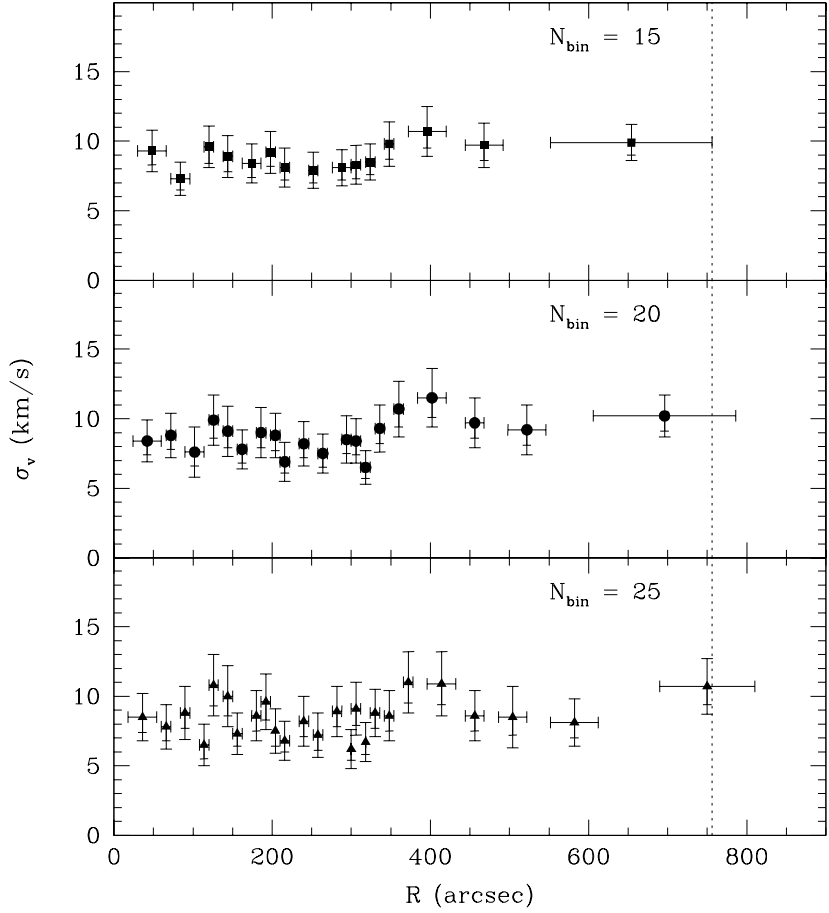}
      \caption{\label{figs:leoibin} The Leo~I velocity dispersion
      profile for all 328 members for three different binnings.  For a
      given profile, the bins have nearly equal numbers of stars
      approximately equal to $328/N_{bin}$.  The radius, $R$, of each
      bin is the mean radius for all stars in that bin.  The
      horizontal 'error bars' in $R$ is the standard deviation of the
      radius of the stars in each bin.  The dispersion values,
      $\sigma_v$, have two downward error bars.  The smaller errors
      are based on the method described by Kleyna et al. (2004), while
      larger errorbars correspond to the uncertainties calculated
      using the method described by Walker et al. (2006a, 2007a) and
      are equal to the upward error bars.  The vertical dotted line
      shows the location of the `tidal' radius of Leo~I (756 arcsec)
      from IH95.}
  \end{center}
\end{figure}

\clearpage

\begin{figure}
  \begin{center}
    \plotone{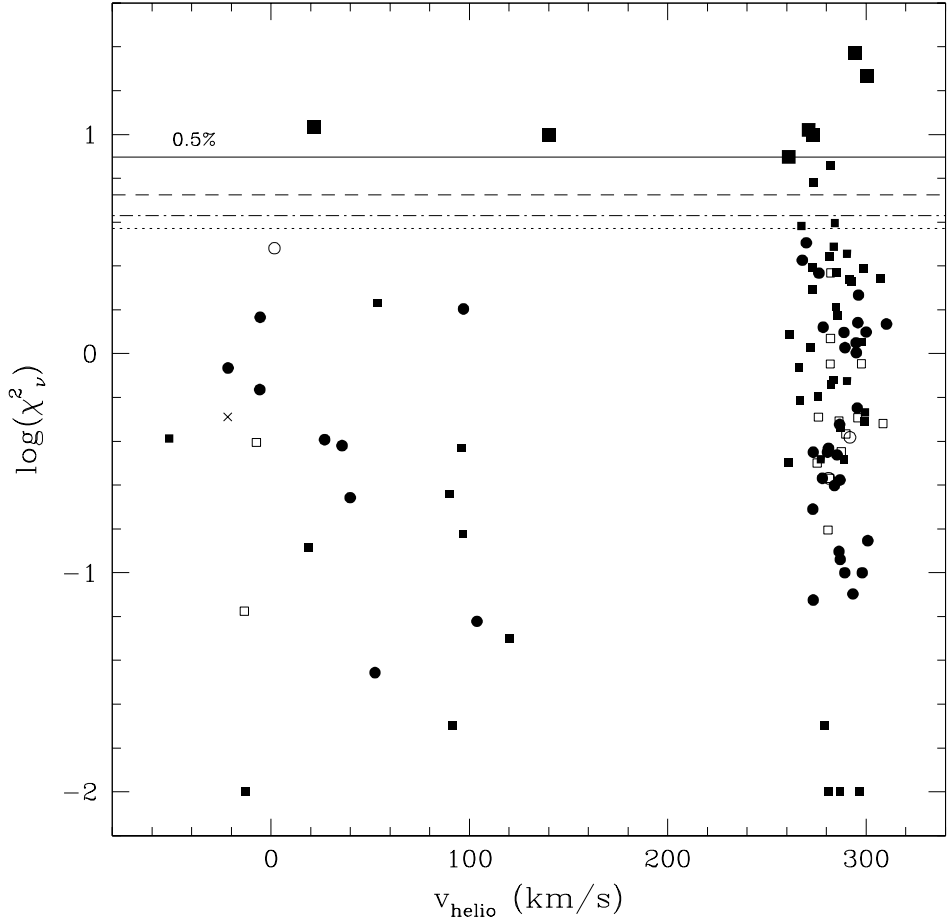}   
      \caption{\label{figs:chi2} Distribution of the reduced
        chi-squared statistic, $\chi_\nu^2$, as a function of
        heliocentric velocity, $v_h$.  The symbols denote the number
        of observations per star: $N = 2$, filled squares; $N=3$,
        filled circles; $N=4$, open squares; $N=5$, open circles;
        $N=6$, cross.  The lines denote the values of $\chi_\nu^2$ for
        which there is a 0.5\%\ chance to exceed $\chi_\nu^2$ by
        chance in a normal distribution for different degrees of
        freedom, $\nu = N - 1$.  From top to bottom, the lines
        correspond to $\nu = 1$ (solid line), 2 (dashed line), 3
        (dot-dash line) and 4 (dotted line). The $\nu = 5$ line is far
        above the sole 6-observation point and is not shown.  The
        large symbols denote the seven objects with $p(>\chi^2_\nu) <
        0.005$.  These seven stars are noted in Table 5.}
  \end{center}
\end{figure}

\clearpage

\begin{figure}
  \begin{center}
    \plotone{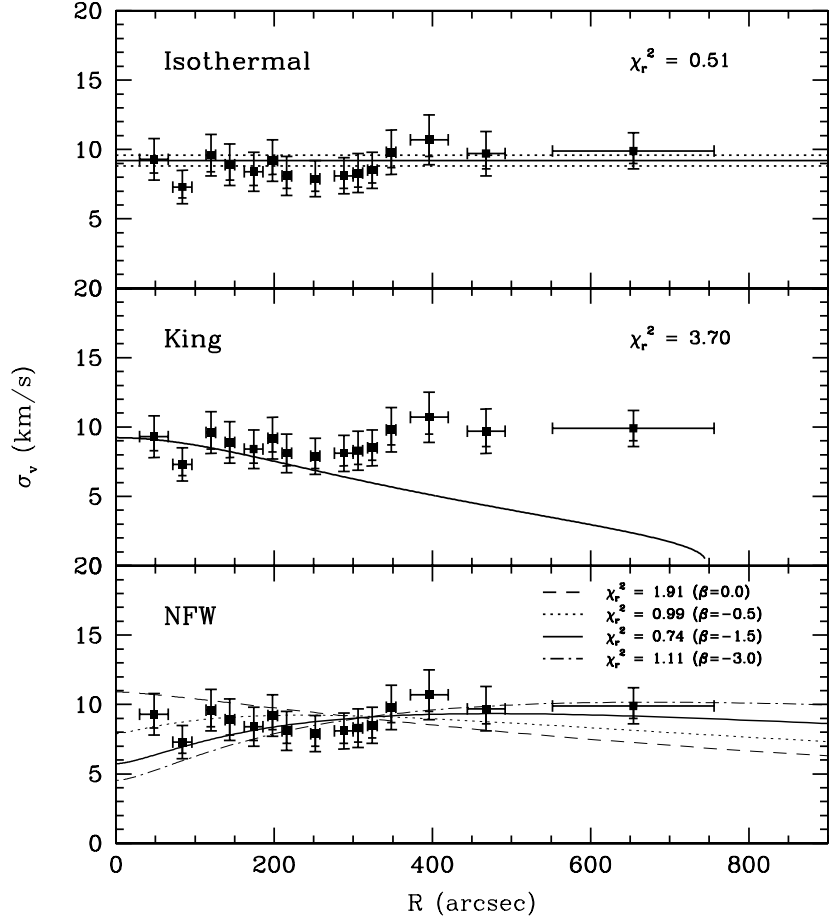}
     \caption{\label{figs:leoibinfits} Comparisons of the binned
     radial velocity dispersion profile for Leo~I ($N_{bin} = 15$; see
     Figure~\ref{figs:leoibin}) with equilibrium dynamical models.
     {\it Top panel:} The dispersion profile for an isothermal sphere
     set to have the dispersion equal to that of the entire sample
     (solid line; $\sigma = 9.2 \pm 0.4$ km/s).  The uncertainty in
     the mean dispersion is denoted by the dotted lines.  {\it Middle
     panel:} The dispersion profile for a King model with a central
     dispersion equal to the sample mean, and a concentration
     parameter of $c = 0.58$ (IH95).  {\it Bottom panel:} Dispersion
     profiles for NFW models assuming an isotropic velocity
     distribution ($\beta = 0$; solid line) and two tangentially
     anisotropic distributions ($\beta = -0.5$, dashed line, and
     $\beta = -1.0$., dotted line).  All of these fits include a
     visible and dark component, corresponding to a total mass of $9
     \times 10^8 M_\odot$.  Further model details are given in Section
     4.2.  Values of reduced $\chi^2_r$ are shown for each model 'fit'
     in each panel.}
  \end{center}
\end{figure}

\clearpage

\begin{figure}
  \begin{center}
    \plotone{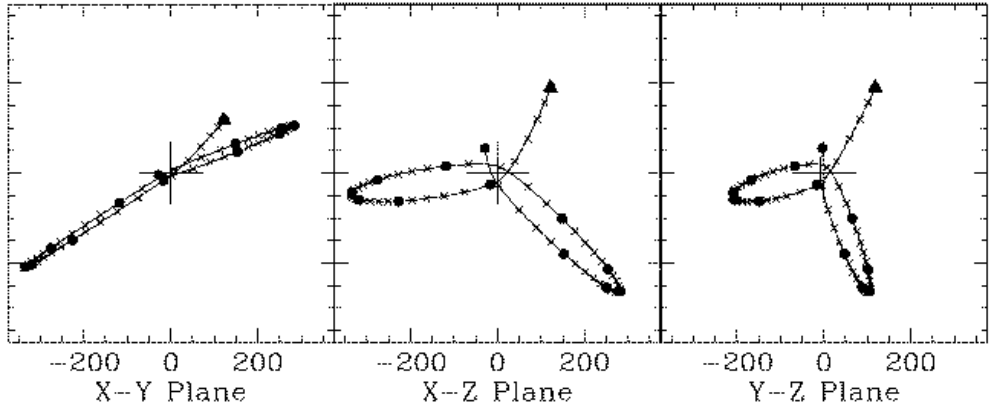}
     \caption{\label{figs:leoiorbit} Orthogonal projections of a
     plausible Leo~I orbit that has $R_{peri} = 15$ kpc and the
     preferred orbital pole of S07 [$(l,b) = (122,13)$] (similar to
     the orbits detailed in Table 8).  Crosses are separated by 200
     million years along the orbit, while filled circles represent
     locations along the orbit every 1 Gyr.  The large triangle
     denotes the present location of Leo~I.  The orbit is shown for
     the past 12 Gyr in an assumed static Galactic potential. The
     large cross in each panel denotes the Galactic Center.  These
     orbits were calculated using the static, multi-component
     potential of Johnston et al. (1995) to represent the Milky Way,
     and a halo with an asymptotic rotation velocity of 190 km/s. }
\end{center}
\end{figure}

\clearpage

\begin{figure}
  \begin{center}
    \plotone{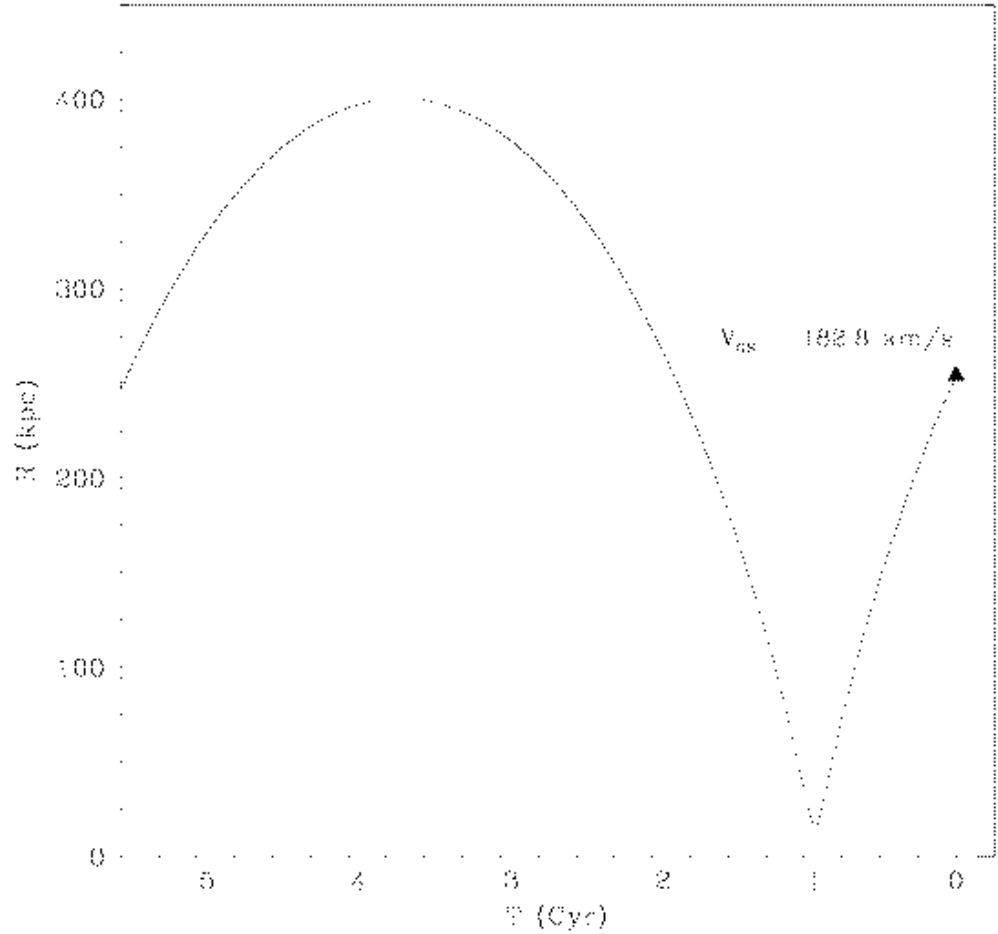}
     \caption{\label{figs:leoirorbit} The Galactocentric distance of
     Leo~I as a function of time (with $T = 0$ corresponding to the
     present) for the orbit described in Figure~\ref{figs:leoiorbit}.
     The triangle shows the (assumed) current location of Leo~I.  The
     galaxy's adopted Galactostationary radial velocity is noted.
     Only one orbital period, about 5.5 Gyr, is shown.  The assumed
     galactostionary velocity used for this calculation differs
     somewhat from the value we observe for Leo~I ($v_{GS} = 174.9$,
     but this has no significant impact on our analysis.}
\end{center}
\end{figure}

%

\clearpage

\begin{figure}
  \begin{center}
     \plotone{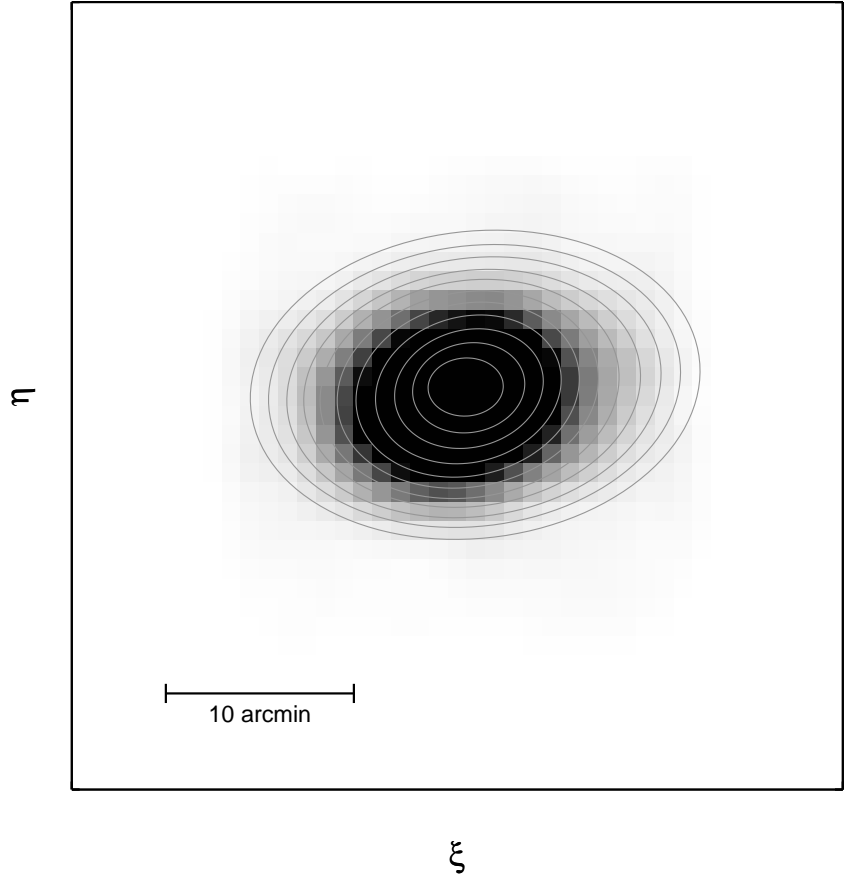}
     \caption{\label{figs:greycontour} A greyscale image of the
     smoothed star counts of the 12630 objects identified in the
     selection region defined in Figure~\ref{figs:cmd} and smoothed
     with a Guassian spatial filter with $\sigma = 2$ arcmin.
     Superimposed are fitted ellipses to the isopleths.  The ellipses
     have major axes ranging from 2 arcmin to 12 arcmin in 1 arcmin
     steps. This plot is oriented with North to the top, and East to
     the right.}
  \end{center}
\end{figure}

\clearpage

\begin{figure}
  \begin{center}
    \plotone{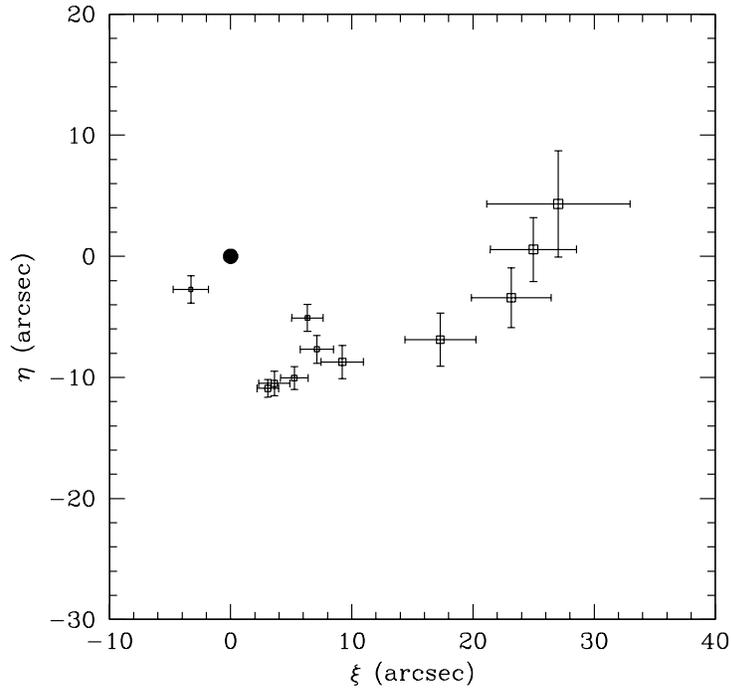}
     \caption{\label{figs:centers} The centers of the fitted ellipses
     shown in Figure~\ref{figs:greycontour}.  The increasing symbol
     sizes denote increasingly larger ellipses in
     Figure~\ref{figs:greycontour}.  Note that the coordinates of the
     center of the fitted ellipses remain close to the adopted center
     location of Leo~I (see Figure~\ref{figs:xieta}) for the smaller
     ellipses corresponding to the inner isopleths, but start to
     deviate to the NE for the 4-5 ellipses for the outermost
     isopleths.  The centroids of the innermost 2-3 ellipses may be
     systematically slightly affected by the presence of gaps in the
     CCDs apparent in Figure~\ref{figs:xieta}.}
  \end{center}
\end{figure}

\clearpage

\begin{figure}
  \begin{center}
    \plotone{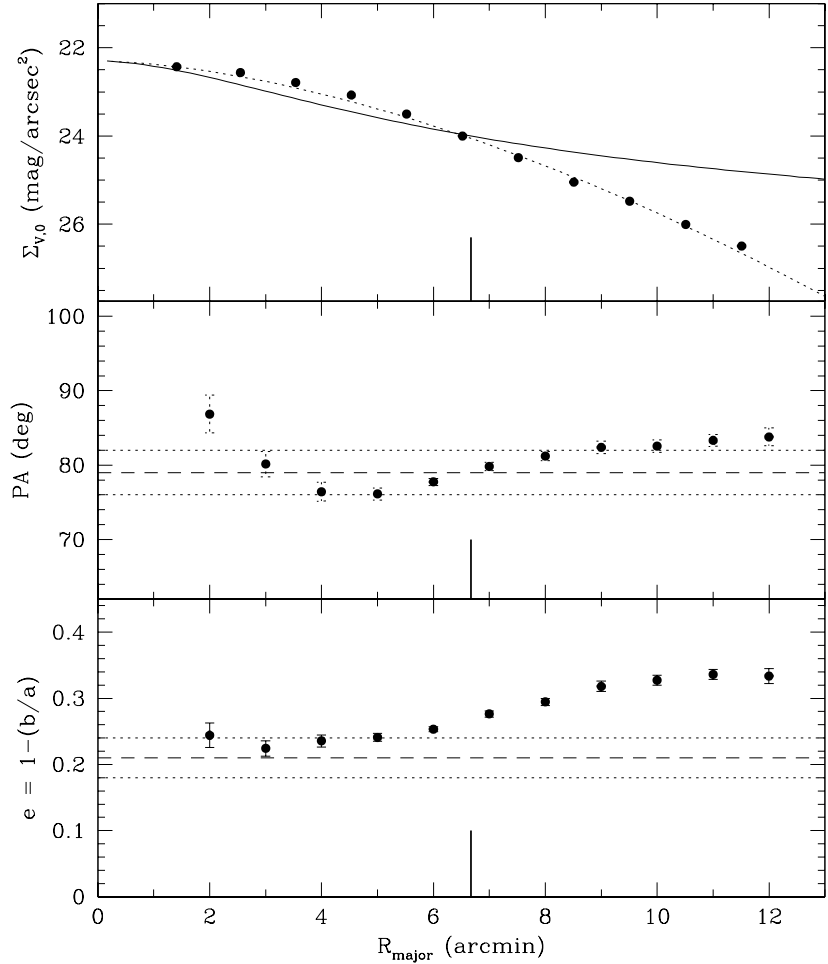}
     \caption{\label{figs:leoishape} {\it Top panel:} The surface
     brightness profile for Leo~I based on the isopleths in
     Figure~\ref{figs:greycontour} and measured in the elliptical
     annuli shown in that plot.  The radii are for the geometric mean
     major axes for each annulus.  This profile was produced assuming
     a constant baryonic $M/L$ ratio and the central surface
     brightness from IH95 (see Table 6).  The solid line is the
     projected density distribution for an isothermal sphere with a
     core radius of 270 arcsec (equal to the King core radius; see
     Table 6).  The dotted line is a Sersic profile with $m=0.6$ and a
     Sersic radius, $r_S$, of 300 arcsec.  {\it Middle panel:} The
     position angle of the major axes of the fitted ellipses in
     Figure~\ref{figs:greycontour} as a function of major axis.  {\it
     Lower panel:} The ellipticity of the fitted ellipses in
     Figure~\ref{figs:greycontour} as a function of major axis.  The
     ellipticity is defined as $e = 1 - b/a$, where $a$ and $b$ are
     the major and minor axes, respectively.  The horizontal lines in
     the lower two panels denote the position angle ($PA = 79 \pm 3$
     deg) and ellipticity ($e = 0.21 \pm 0.02$) from IH95, consistent
     with the inner values of the profiles shown here.  The short
     vertical lines in the panels are located at $R = 400$ arcsec, the
     radius at which we claim to see a change in the kinematic
     properties of the stars in Leo~I (Section 4.3.3).}
  \end{center}
\end{figure}

\clearpage

\begin{figure}
  \begin{center}
    \plotone{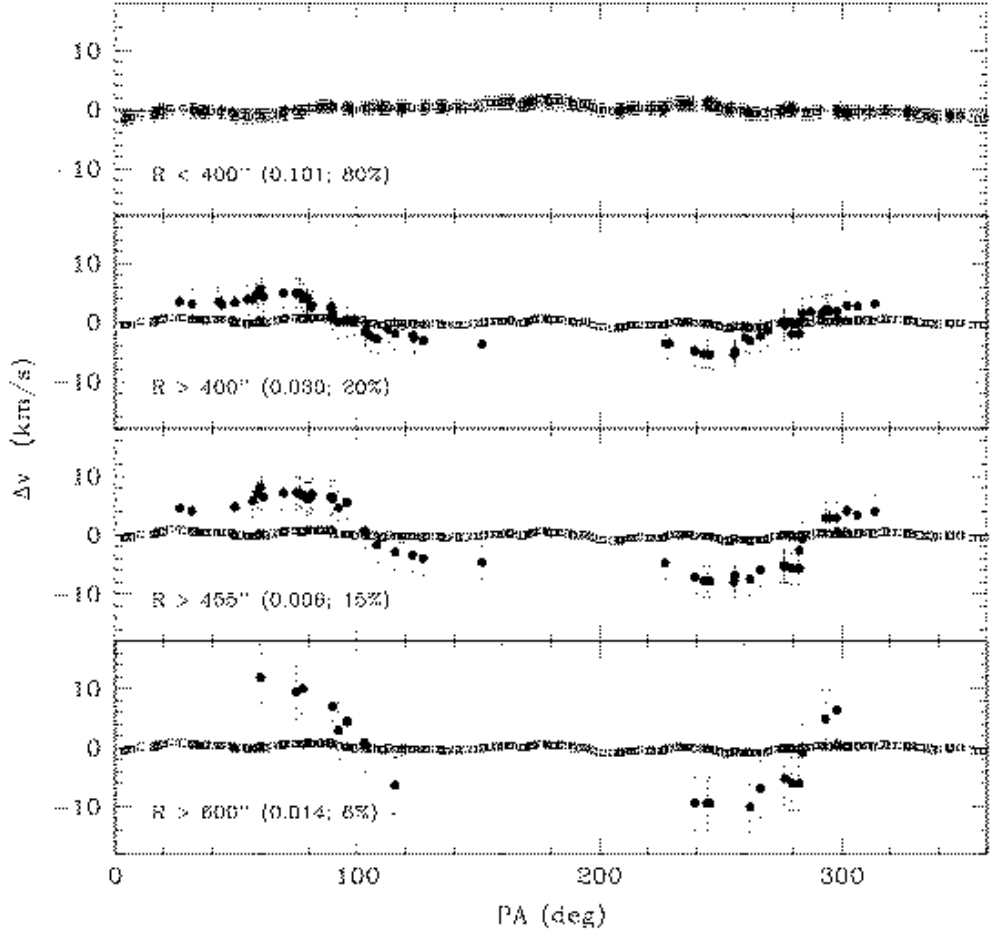}
    \caption{\label{figs:deltav} Plots of the bisector mean velocity
      difference, $\Delta v$, defined as the difference of the mean
      velocities of Leo~I members on both sides of a bisector oriented
      along a given position angle, $PA$ (see Figure~\ref{figs:xieta2}
      and Section 4.3.3 for further details). For the data plotted
      with error bars, the radial range of the dataset (from top: $R <
      400$ arcsec, $R > 400$ arcsec, $R > 455$ arcsec, and $R > 600$
      arcsec), the probability of exceeding $(\Delta v)_{max}$, and
      the fraction of all the Leo~I members used to produce each plot
      are given in each panel.  In the lower three panels the set of
      points (open squares) that lack error bars are for the full
      distribution of Leo~I members ($N = 328$, and for which the
      probability of exceeding $(\Delta v)_{max}$ is 31.3\%).  The
      line at $\Delta v = 0$ in the top panel is shown for reference.}
  \end{center}
\end{figure}

\clearpage

\begin{figure}
  \begin{center}
    \plotone{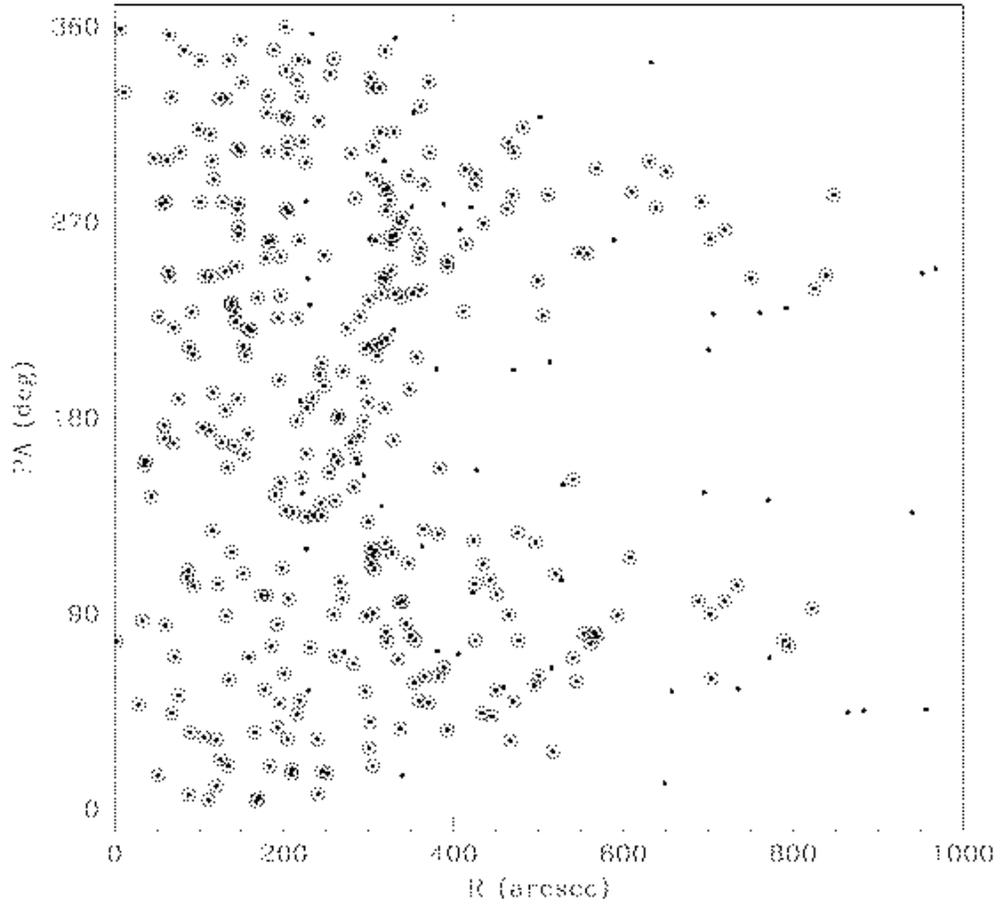}
     \caption{\label{figs:leoibreak} The position angle vs. radial
     distance from the center of Leo~I for all stars listed in Table 5
     ($R_{TD} \ge 2.8$).  The points enclosed in open circles are
     likely Leo~I members.  Note the transition from a fairly uniform
     distribution of members (in $PA$) for $R < 400$ arcsec (denoted
     by the vertical dashed line) to a strongly bimodal distribution
     for $R > 400$ arcsec.  The non-members (small dots) show a
     noticeably more uniform distribution in PA regardless of $R$.}
  \end{center}
\end{figure}

\clearpage

\begin{figure}
  \begin{center}
    \plotone{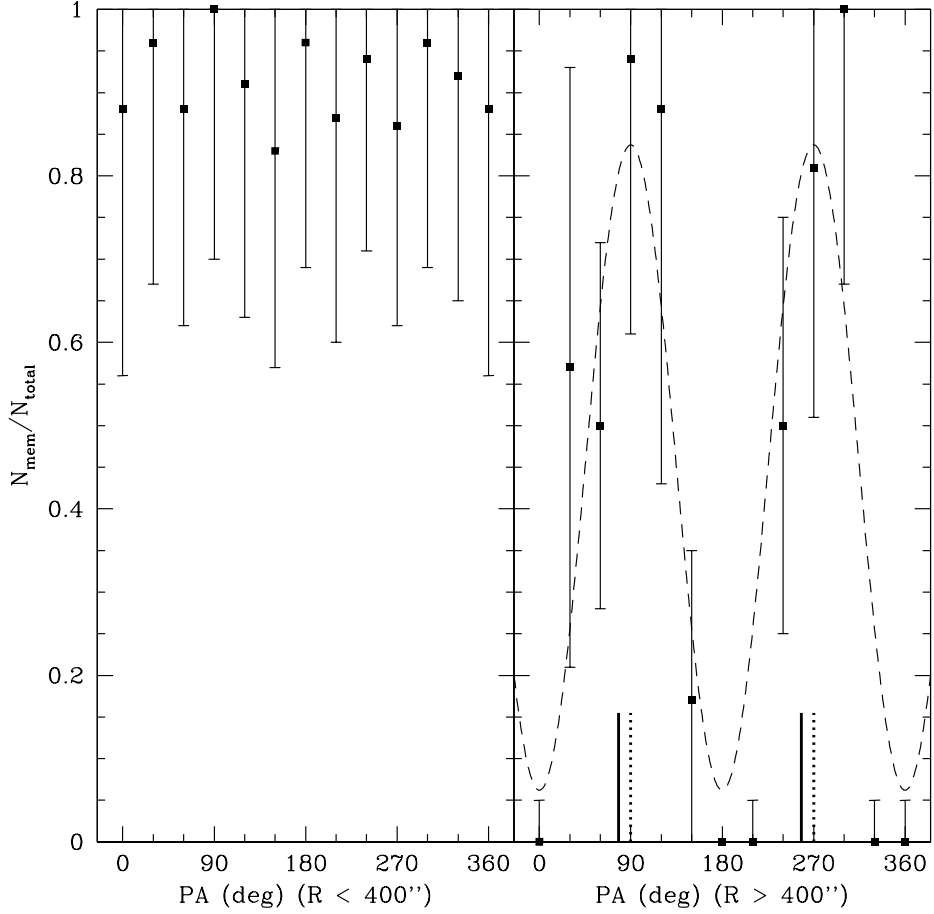}
     \caption{\label{figs:memfrac} Plots of the fraction of members
     divided by the total stars observed kinematically (data from
     Table~5) as a function of position angle for stars in the inner
     sample ($R < 400$ arcsec; {\it left panel}), and for stars of
     the outer sample ($R \geq 400$ arcsec; {\it right panel}).  The
     double sine curve is a fit to the outer sample data.  The maxima
     occur at $PA = 90$ and 270 degrees (short dotted vertical lines
     at the bottom of the plot).  The major axis position angle of the
     inner regions of Leo~I (see Figure~\ref{figs:leoishape}) is
     denoted by the short vertical solid lines at $PA = 78$ and 258
     degrees.  Bins in the right panel with values of
     $N_{mem}/N_{total} = 0$ and a small error bar contain no
     kinematic members but $N_{total} \geq 1$.  The one bin with
     $N_{mem}/N_{total} = 0$ and no error bar has $N_{total} = 0$.}
\end{center}
\end{figure}

\clearpage

\begin{figure}
  \begin{center}
    \plotone{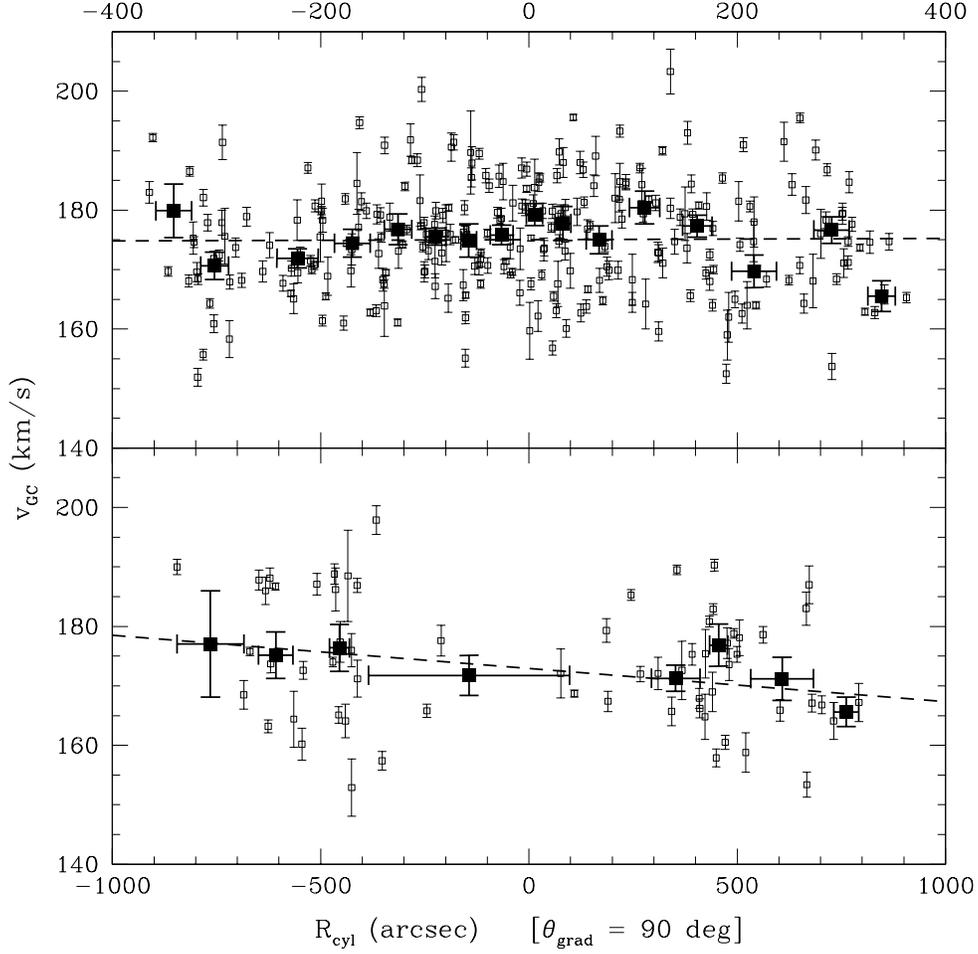}
     \caption{\label{figs:gradient} {\it Top panel:} A fit to the
     velocities corrected for Galactic rotation versus the $R_{cyl}$
     relative to a minor axis with position angle 108 deg (see text
     for details) for all stars with $R \leq 400$ arcsec (the inner
     sample).  {\it Lower panel:} The same for stars with $R > 400$
     arcsec. In both panels the open squares are the individual
     velocities with their uncertainties.  The larger filled squares
     are the binned mean velocities.  For these larger symbols, the
     vertical error bars are the standard deviations of the mean
     velocity in each bin, while the horizontal error bars are the
     standard deviations of the radial coordinates of the sample of
     stars in each bin.  The dashed lines show the least squared fit
     line to the respective binned mean velocities.  Note the change
     in scale in the horizontal axes of the two panels.  The slope of
     the fit in the lower panel is $-0.34 \pm 0.15$ km/s/arcmin, in
     the sense that the mean velocity becomes more negative toward
     {\it PA} = 90 deg.}
  \end{center}
\end{figure}

\clearpage

\begin{figure}
  \begin{center}
    \plotone{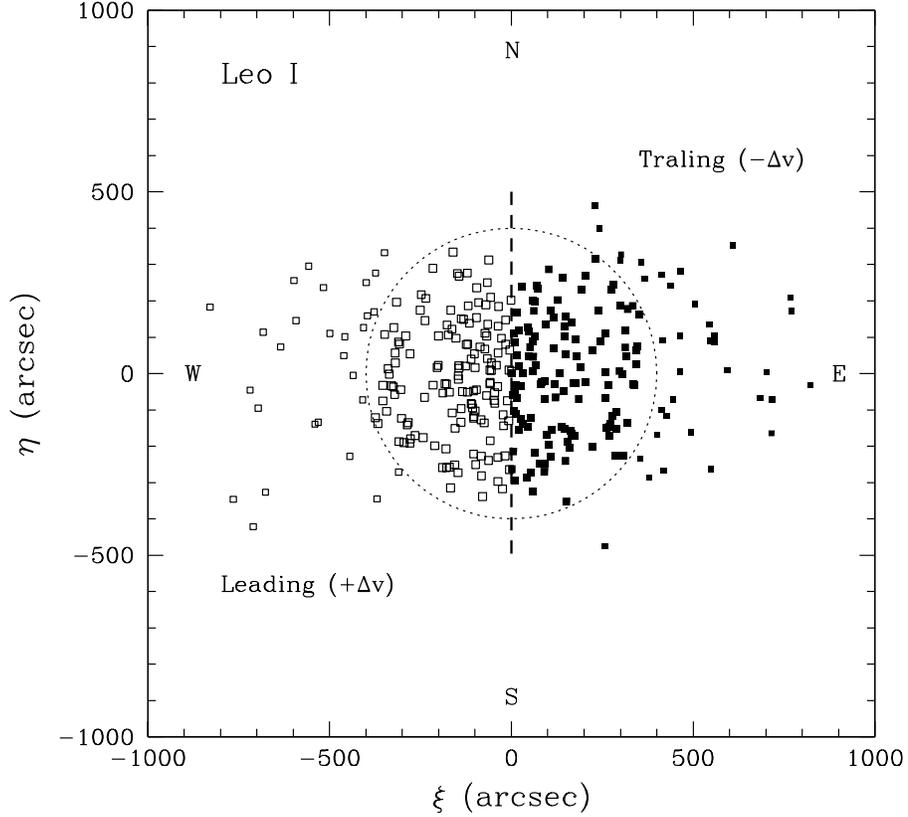}
    \caption{\label{figs:xieta2} The spatial distribution of Leo~I
     kinematic members.  The dashed straight line indicates the
     position angle of an axis perpendicular to the maximum velocity
     gradient observed in Leo~I.  The circle (dotted line) corresponds
     to $R = 400$ arcsec, the radius where we see a change in the
     internal kinematics in Leo~I.  This figure also illustrates how
     Figure~\ref{figs:deltav} was constructed.  The mean velocity of
     points on one side of the bisector (solid symbols) minus the mean
     velocity of points on the other side of the bisector (open
     symbols) represent the data that go into calculating $\Delta v$
     for $PA = 0$ degrees.  This orientation corresponds to the
     maximum velocity gradient in Leo~I along {\it PA} = 90 deg.
     Figure~\ref{figs:deltav} plots how $\Delta v$ changes as the
     bisector is rotated clockwise in position angle.}
\end{center}
\end{figure}

\clearpage

\begin{figure}
  \begin{center}
    \plotone{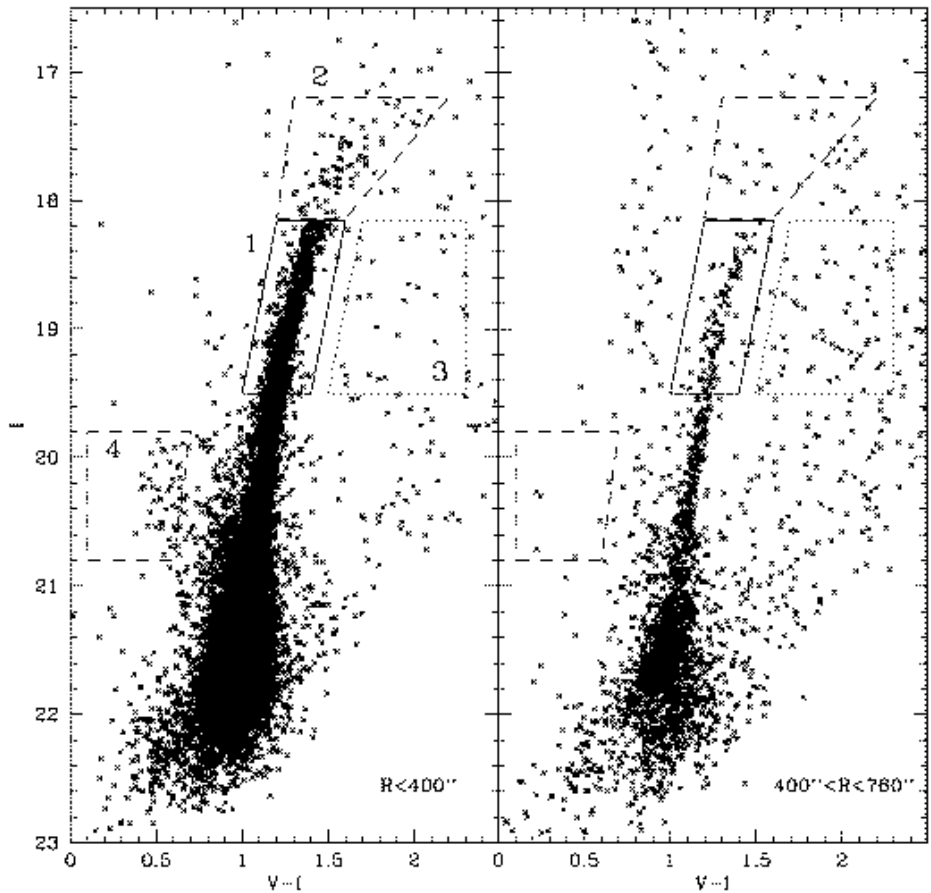}
     \caption{\label{figs:cmd2} {\it Left panel:} The Leo~I CMD
     for stars with $R_b \leq 400$ arcsec, where $R_b$ is the break
     radius' (Section 4.3.3).  The four regions correspond to RGB stars
     (1; solid line), AGB stars (2; dashed line), field stars (3;
     dotted line) and blue-loop stars (4; dashed line).  {\it Right
     panel:} The Leo~I CMD for stars with $400 < R \leq 760$ arcsec.
     The CMD regions are the same as in the left panel but the areal
     coverage in the right panel is 3.2 times that of the left panel.
     Star counts for the regions defined in these CMDs are listed in
     Table~7.}
  \end{center}
\end{figure}

\clearpage

\begin{figure}
  \begin{center}
    \plotone{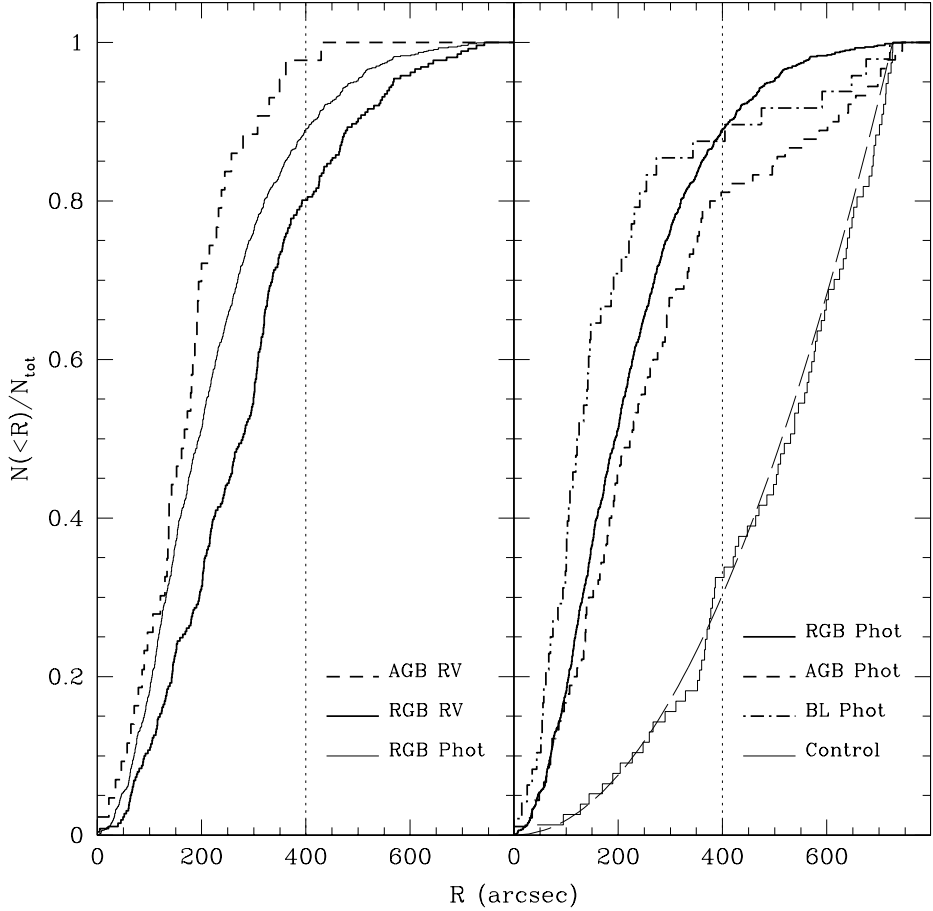}
     \caption{\label{figs:rcumulative} {\it Left panel:} The
     cumulative radial distributions of kinematically-confirmed member
     RGB (Region 1 of Figure~\ref{figs:cmd2}; thick solid line;
     $N_{tot} = 261$ stars) and AGB (Region 2; thick dashed line; 43
     stars) stars of Leo~I.  The thin solid line is the distribution
     of photometrically-selected RGB stars (from Region 1 of
     Figure~\ref{figs:cmd2}; 823 stars).  The location of the break
     radius, $R_b$, is denoted by the dotted vertical line at $R =
     400$ arcsec.  {\it Right Panel:} The cumulative radial
     distributions of photometrically-selected RGB (Region 1; thick
     solid line; 823 stars), AGB (Region 2; thick dashed line; 90
     stars), and blue-loop (Region 4; dot-dashed line; 48 stars) stars
     identified in the Leo~I CMD (Figure~\ref{figs:cmd2}).  Some idea
     of the completeness of our counts is given by the cumulative
     radial distribution of photometric non-members (Region 3; thin
     solid line; 77 stars).  This is compared to the parabola $f =
     N(<R)/N_{tot} = (R/R_{max})^2$ (thin, long-dashed line) expected
     of a constant surface density contaminating population.  The good
     fit suggests we are counting stars to fairly uniform completeness
     at all radii in Leo~I.  The break radius at $R = 400$ arcsec is
     denoted as the dotted vertical line.}
  \end{center}
\end{figure}

\clearpage

\begin{figure}
  \begin{center}
    \plotone{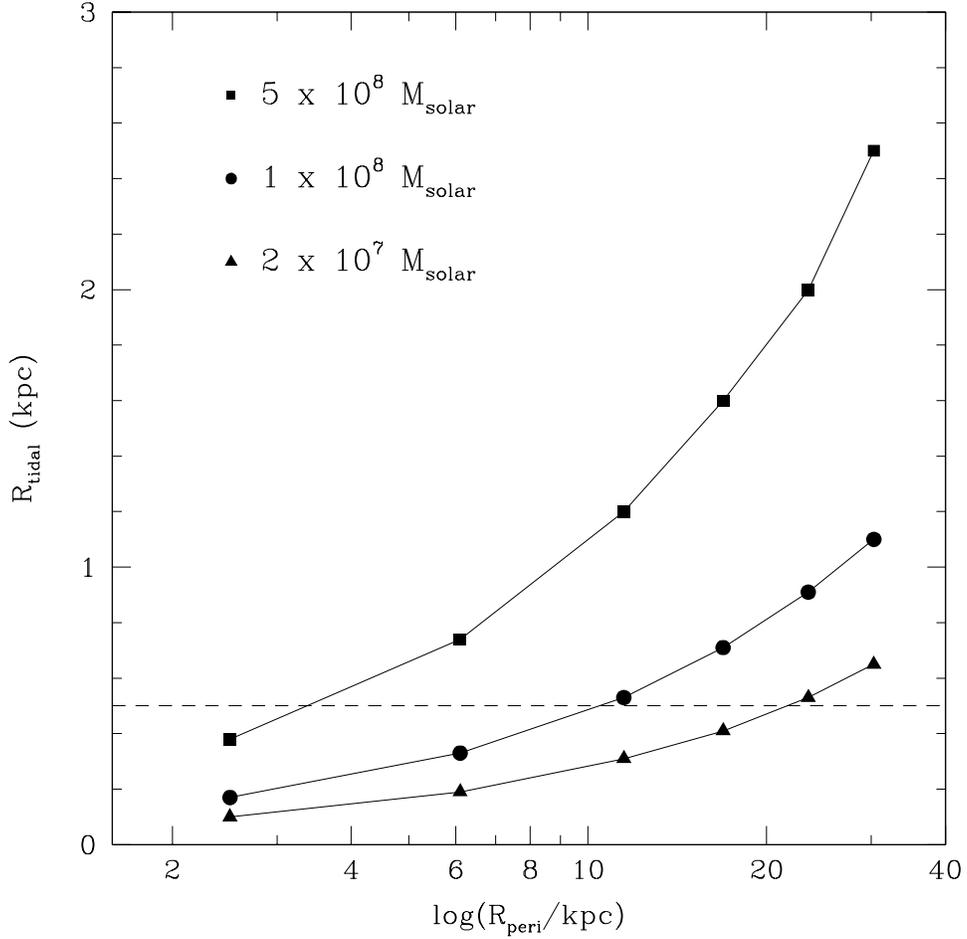}
     \caption{\label{figs:tidalperi} The instantaneous tidal radius
     ($R_{tidal}$) of Leo~I as a function of perigalactic distance for
     a range of assumed total masses (see Table 8 for more details of
     the orbits).  The horizontal dashed line corresponds to the
     observed break radius ($R_b \sim 400$ arcsec $ \sim 500$ pc) at
     which the internal kinematics of Leo~I begin to change from
     possible isotropy (interior to this radius) to streaming
     (exterior).  A representative orbit is shown in
     Figure~\ref{figs:leoiorbit}.}
\end{center}
\end{figure}

\clearpage

\begin{figure}
  \begin{center}
    \plotone{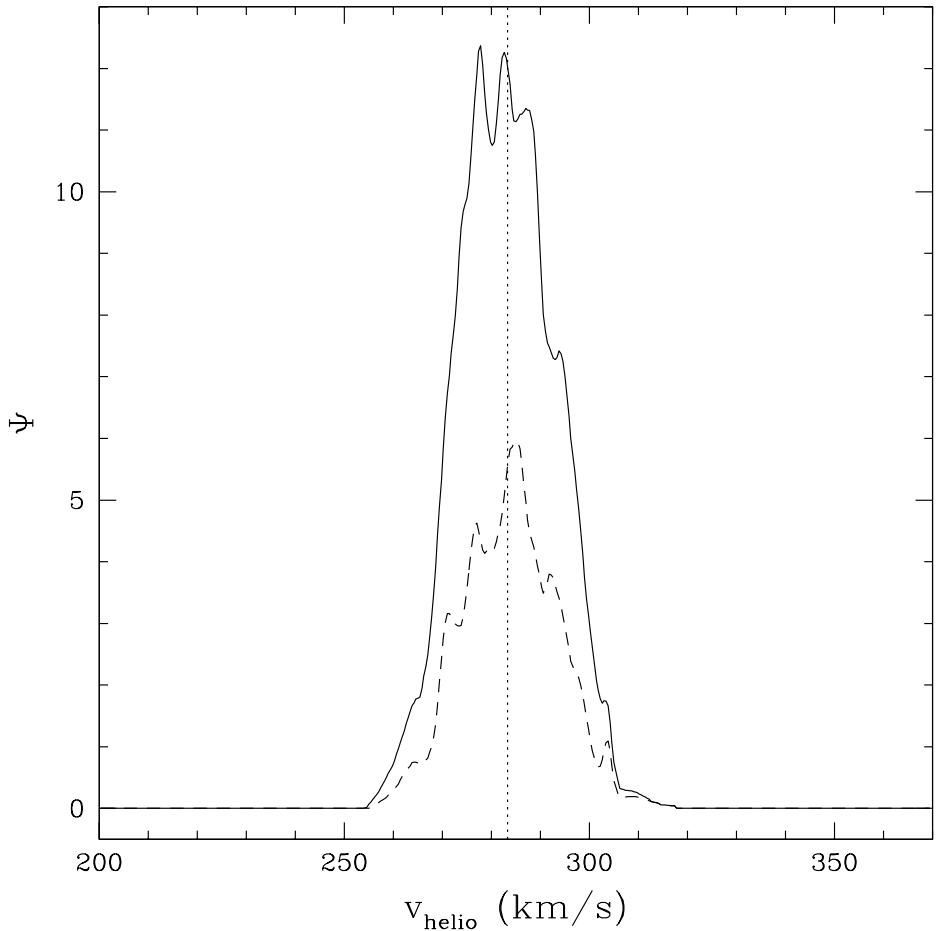}
     \caption{\label{figs:velhist} The velocity histogram of Leo~I
     members.  Each star is represented here as a normalized Gaussian
     centered at the star's heliocentric velocity and with $\sigma$
     equal to the observed 1-$\sigma$ errors listed in Table~5.  The
     function $\Psi$ plotted here is the sum of the individual
     Gaussians sampled every 0.4 km/s.  The vertical dashed line shows
     the mean heliocentric velocity of Leo~I (282.9 km/s; Section
     4.3.1).  The skew of this distribution is $0.08 \pm 0.14$ and the
     kurtosis is $-0.34 \pm 0.27$ (for $N = 328$), both consistent
     with a Gaussian distribution.  The dashed histogram is for stars
     from our kinematic sample selected from the regions on the sky
     observed by S07 (138 stars).  We see no significant skew ($0.10
     \pm 0.21$), kurtosis ($-0.23 \pm 0.41$), or velocity shift
     ($\langle v\rangle = 283.7 \pm 0.7$\ km/s) for this subsample.
     These histograms are for heliocentric velocities, but, apart from
     a shift by $\sim -108$ km/s, they would be essentially identical
     had we used Galactostationary velocities.}
\end{center}
\end{figure}

\end{document}